{
\documentclass[%
twocolumn,
 amsmath,amssymb,
 aps,
 pre,
]{revtex4}
\usepackage{dcolumn}
\usepackage{bm}
\usepackage{graphicx}
\usepackage{bbm}
\usepackage{comment}

\begin{document}


\title{Distribution of the Time Between Maximum and Minimum of Random Walks}

\author{Francesco Mori}
\address{LPTMS, CNRS, Universit\'e  Paris-Sud,  Universit\'e Paris-Saclay,  91405 Orsay,  France}

\author{Satya N. Majumdar}
\address{LPTMS, CNRS, Universit\'e  Paris-Sud,  Universit\'e Paris-Saclay,  91405 Orsay,  France}

\author{Gr\'egory Schehr}
\address{LPTMS, CNRS, Universit\'e  Paris-Sud,  Universit\'e Paris-Saclay,  91405 Orsay,  France}

\date{\today}

\begin{abstract}
We consider a one-dimensional Brownian motion of fixed duration $T$. Using a path-integral technique, we compute exactly the probability distribution of the difference $\tau=t_{\min}-t_{\max}$ between the time $t_{\min}$ of the global minimum and the time $t_{\max}$ of the global maximum. We extend this result to a Brownian bridge, i.e. a periodic Brownian motion of period $T$. In both cases, we compute analytically the first few moments of $\tau$, as well as the covariance of $t_{\max}$ and $t_{\min}$, showing that these times are anti-correlated. We demonstrate that the distribution of $\tau$ for Brownian motion is valid for discrete-time random walks with $n$ steps and with a finite jump variance, in the limit $n\to \infty$. In the case of L\'evy flights, which have a divergent jump variance, we numerically verify that the distribution of $\tau$ differs from the Brownian case. For random walks with continuous and symmetric jumps we numerically verify that the probability of the event ``$\tau = n$'' is exactly $1/(2n)$ for any finite $n$, independently of the jump distribution. Our results can be also applied to describe the distance between the maximal and minimal height of $(1+1)$-dimensional stationary-state Kardar-Parisi-Zhang interfaces growing over a substrate of finite size $L$. Our findings are confirmed by numerical simulations. Some of these results have been announced in a recent Letter [Phys. Rev. Lett. {\bf 123}, 200201 (2019)].
\end{abstract}

\maketitle

\newpage

\section{Introduction} \label{sec:intro}

The average global temperature over a century, the amount of rainfall in a given area throughout one year, or the price of a stock during a trading day are only few of several quantities whose maximal observed value within a fixed period of time $T$ plays a central role. Understanding extremal properties of the underlying stochastic processes is therefore of fundamental importance in a variety of disciplines -- for a recent review see \cite{EVS_review}. These include several applications in e.g., climate studies \cite{katz02,katz05,RP06,WK10,RC2011,Christ13,WHK14}, finance \cite{BP2000,yor01,embrechts13,Challet17, zou14} and computer science \cite{coffman98,KM2000,M05}. For instance, extreme natural events, such as  earthquakes, tsunamis, or hurricanes, have often devastating consequences. The statistics of records of such extremal events have been extensively studied in statistics and mathematics literature~\cite{Resnick_book,ABN_book,nevzorov_book}, and more recently in statistical physics~\cite{Krug07,MZ08,MSW12,GMS16,record_review,MBK19}. 

\begin{figure}[h]
  \centering
\includegraphics[width=1\linewidth]{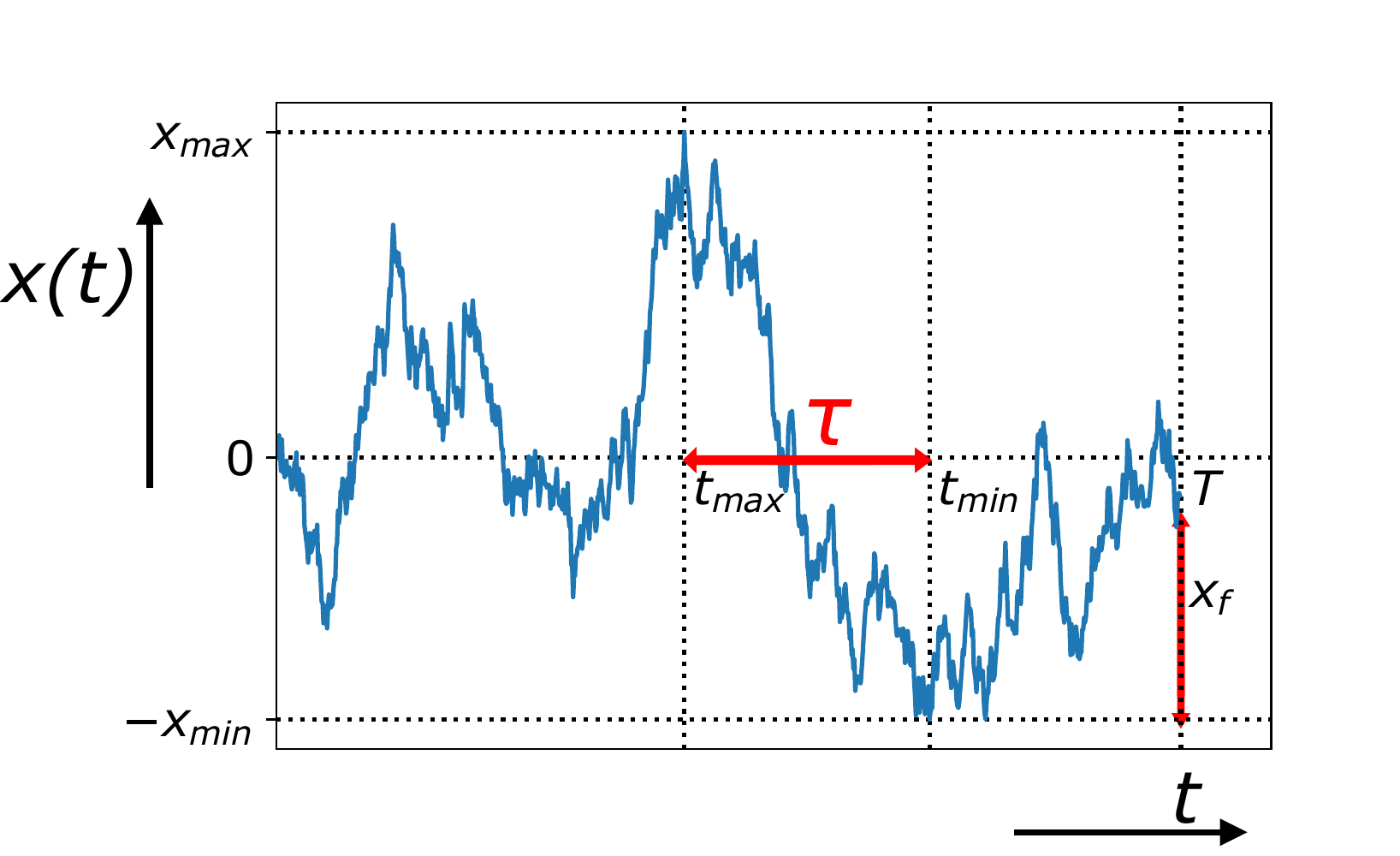}
\caption{Typical realisation of a Brownian motion $x(t)$ vs $t$ with time $t \in [0,T]$. The global maximum $x_{\max}$ is reached at time $t_{\max}$, while the global minimum $x_{\min}$ is attained at time $t_{\min}$. The time between the global maximum and the global minimum is $\tau=t_{\max}-t_{\min}$.}
\label{fig:brownian_1} 
\end{figure}

Thus, understanding statistical fluctuations of these calamities is a problem of great practical importance. Moreover, in many situations it is not only important to ask what is the intensity of the maximum value but also at what time $t_{\max}$ this maximum value is attained within the time interval $[0,T]$ (see Fig. \ref{fig:brownian_1}). This time $t_{\max}$ is extremely relevant in several applications. For instance, in finance it is important to estimate the time at which the price of a stock will reach the maximum value during a fixed period \cite{dale80,baz04}. Similarly, it is also natural to study the time $t_{\min}$ at which the global minimum is reached. 

The time $t_{\max}$ of the maximum has been studied for a variety of stochastic processes. For instance, the probability distribution $P(t_{\max}|T)$  of $t_{\max}$, for a given $T$, is exactly known for one-dimensional Brownian motion (BM), one of the most ubiquitous and paradigmatic stochastic processes. Indeed, if we consider a BM $x(t)$ of total duration $T$, then
~\cite{levy40,feller50,SA53,morters10} 
\begin{equation}\label{levy}
P(t_{\max}|T)=\frac{1}{\pi\sqrt{t_{\max}(T-t_{\max})}} \;,
\end{equation}
where $t_{\max}\in [0,T]$.
In the case of BM, $t_{\max}$ and $t_{\min}$ have, by symmetry, the same probability distribution (\ref{levy}). 
Moreover, the distribution of $t_{\max}$ has been computed for several generalisations of BM. For instance, 
$P(t_{\max}|T)$ is known for BM with additional constraints 
\cite{randon-furling07,randon-furling08,schehr10,PLDM03}, drifted BM \cite{majumdar08,drift}, fractional BM 
\cite{delorme16,tridib18} and Bessel processes \cite{schehr10}. The many body case has also been considered, 
both for $N$ independent BM's \cite{comtet10} and $N$ non-crossing Brownian walkers \cite{rambeau11}. The 
statistics of $t_{\max}$ has been investigated for more general stochastic processes such as L\'evy flights 
\cite{SA53,majumdar10}, random acceleration process \cite{rosso10}, and run-and-tumble particles 
\cite{anupam_rtp}. Moreover, $t_{\max}$ has been studied in a wide range of applications including disordered 
systems \cite{MRZ10}, stochastic thermodynamics \cite{barato18} and sports \cite{clauset15}.

Despite this increasing interest in the statistical properties of $t_{\max}$ or $t_{\min}$, their joint 
probability distribution $P(t_{\max},\,t_{\min}|T)$ was computed only recently, in the case of BM, in our 
recent Letter \cite{mori2019}. Indeed, due to the strong correlations between $t_{\max}$ and $t_{\min}$, this 
joint probability density function (PDF) is not just the product of the two marginals $P(t_{\max}|T)$ and 
$P(t_{\min}|T)$, each of which is given by the expression (\ref{levy}). 
To see that $t_{\max}$ and $t_{\min}$ are strongly correlated, it suffices to consider the following fact.
Since the process occurs in continuous time, 
it is clear that 
if the maximum occurs at a given time, it is highly unlikely that 
the minimum occurs immediately before or after the maximum. 
Thus $t_{\rm max}$ and $t_{\rm min}$ are strongly anti-correlated and the occurrence of one
forbids the occurrence of the other nearby in time.

Furthermore, computing $P(t_{\max},\,t_{\min}|T)$ is not 
only relevant to quantify the anti-correlations of $t_{\max}$ and $t_{\min}$ but also to investigate observables 
that depend on both $t_{\max}$ and $t_{\min}$. One relevant example is the time difference between maximum and 
minimum: $\tau=t_{\min}-t_{\max}$ (see Fig. \ref{fig:brownian_1}). Note that $\tau\in[-T,T]$ can be positive or 
negative. This quantity $\tau$ has a natural application in finance. Let us consider the price of a stock in a 
period of time $T$. Then if $t_{\max}<t_{\min}$, as in Fig. \ref{fig:brownian_1}, an agent would try to sell 
his/her shares at time $t_{\max}$ when the price is the highest and then wait up to time $t_{\min}$ to re-buy at 
the best price. Thus, $\tau=t_{\min}-t_{\max}$ represents the time the agent has to wait before re-buying his/her 
shares in order to maximise the gain. Notably, similar quantities, for instance the time between a local maximum 
and a local minimum, have been empirically investigated for stock market data \cite{zou14}. Consequently, 
computing $P(t_{\max},t_{\min}|T)$ and $P(\tau|T)$ is a problem of fundamental relevance, with broad 
interdisciplinary applications. 

The main goal of this paper is to compute these probabilities 
$P(t_{\max},t_{\min}|T)$ and $P(\tau|T)$ for different one-dimensional stochastic processes. We will first 
consider BM, finding an exact solution for the PDF of $\tau$. Then we will consider more general stochastic 
processes and possible applications of our results. Some of the main results presented here were announced in 
our previous Letter \cite{mori2019}. However, the details of the calculations, which are rather involved, were 
not presented in \cite{mori2019}. Here we provide a detailed description of these techniques, which we
hope would be useful in other problems. In addition, we present new results for 
discrete-time random walks including L\'evy flights, which can not be obtained directly by applying the 
techniques used for the Brownian motion. They require different methods that are presented in detail in this
paper. We will see that our analysis of $t_{\max}$ and $t_{\min}$ for discrete-time random walks 
also raises interesting mathematical questions that would be of interest to the probability theory community 
in mathematics.\\

The rest of the paper is organised as follows. In Section \ref{sec:main_results}, we briefly present our main 
results. In Section \ref{sec:BM} we use a path-integral technique to compute the PDF of the time $\tau$ between 
the maximum and the minimum of a BM. In Section \ref{sec:BB} we present two alternative derivations for the PDF 
of $\tau$ in the case of a Brownian Bridge (BB), which is a periodic BM of period $T$. The first derivation is 
based on a path-integral method, while the second exploits a mapping, known as Vervaat's construction, between 
the BB and the Brownian excursion, i.e. a BB constrained to remain positive between the initial and final 
positions. In Section \ref{sec:RW} we study $\tau$ in the case of discrete-time random walks (RWs). In 
particular, we perform an exact computation in the cases of double-exponential jumps and of lattice walks. We 
also present the results of numerical simulations for other jump distributions such as L\'evy flights. 
In Section \ref{sec:universal}, we study the probability of the event ``$\tau = n$'' for discrete-time random 
walks of $n$ steps and conjecture that it is universal, i.e., independent
of jump distributions as long as it is symmetric and continuous.
In Section \ref{sec:fluctuating} we apply our results to $(1+1)-$dimensional fluctuating interfaces. Finally, in 
Section \ref{sec:conclusions} we conclude with a summary and related open problems. Some details of computations 
are relegated to the appendices.

\section{Summary of the main results}
\label{sec:main_results}

\begin{figure*}[t] 
   \includegraphics[angle=0,width=\linewidth]{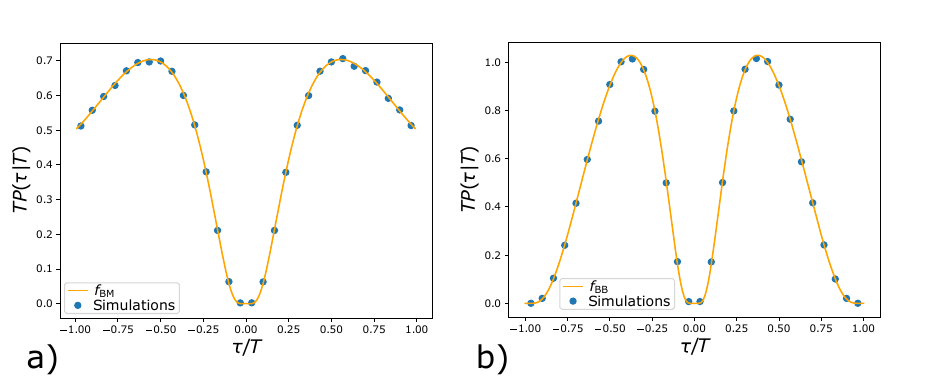}
    \caption{ {\bf a)} The scaled distribution $T\, P(\tau|T)$ plotted as a function of $\tau/T$ for the BM (the solid line corresponds to the exact scaling function $f_{\rm BM}(y)$ in Eq. (\ref{eq:f_bm}), while the filled dots are the results of simulations). 
{\bf b)} The same scaled distribution for the  Brownian bridge where the exact scaling function $f_{\rm BB}(y)$ is given in Eq. (\ref{eq:f_bb}). Numerical simulations are obtained by averaging over $10^7$ samples.}
    \label{fig:numeric}
\end{figure*}
Since this paper is rather long, it would be convenient and handy to the reader to have a summary
of our main results--this section does precisely that. This section is rather self-contained and readers not interested in details
may just read this section only.\\
\\

\noindent{\bf Probability distribution of $\tau=t_{\min}-t_{\max}$ for Brownian motion:}
We consider a BM $x(t)$, starting from some fixed initial position $x_0$ and evolving over a time interval $[0,T]$. By using a path-integral method we show that the PDF of $\tau=t_{\min}-t_{\max}$ has a scaling form for any $\tau$ and $T$:
\begin{equation}
P(\tau|T)=\frac{1}{T}f_{\rm BM}\left(\frac{\tau}{T}\right)\,,
\end{equation}
where the scaling function $f_{\rm BM}(y)$ is given by
\begin{eqnarray}\label{eq:f_bm}
f_{\rm BM}(y)=\frac{1}{|y| }\sum_{n=1}^{\infty}(-1)^{n+1}\tanh^2\left(\frac{n\pi}{2}\sqrt{\frac{|y|}{1-|y|}}\right)\,,
\end{eqnarray}
where $-1\leq y\leq 1$. The function $f_{\rm BM}(y)$ is symmetric around $y=0$ and is non-monotonic as a function of $y$ (see Fig. \ref{fig:numeric}a).
Moreover, $f_{\rm BM}(y)$ has the asymptotic behaviours 
\begin{eqnarray}\label{asymptote_bm}
f_{\rm BM}(y) \underset{y \to 0^+}{\approx} \frac{8}{y^2} e^{-\frac{\pi}{\sqrt{y}}} \;\;\; , \;\;\; f_{\rm BM}(y) \underset{y\to 1}{\approx} \frac{1}{2}  \;.
\end{eqnarray}
We also show that the scaling function $f_{\rm BM}(y)$ satisfies the following integral relation:
\begin{equation}\label{eq:integral_BM}
\int_{0}^{1}dy\frac{f_{\rm BM}(y)}{1+uy}=
\int_{0}^{\infty}dz \frac{1}{\sinh(z)}\tanh^2\left(\frac{z}{2\sqrt{1+u}}\right)\, .
\end{equation}
This integral identity turns out to be very useful to compute the moments
of $\tau$ explicitly in an efficient way. For example, for BM, the first few moments of $\tau$ are given 
explicitly as :
\begin{eqnarray}
\langle |\tau|\rangle &=&\frac{4\log(2)-1}{3} \; T = (0.5908\ldots)\;T\,,\\ \nonumber \langle \tau^2\rangle &=& \frac{7\zeta(3)-2}{16}\; T^2 = (0.4009\ldots) \;T^2\,,\\ \nonumber 
\langle |\tau|^3 \rangle &=& \frac{147\zeta(3)-34}{480}\; T^3 = (0.2972\ldots)\; T^3\,,\\ \nonumber \langle \tau^4\rangle &=&\frac{1701\zeta(3)-930\zeta(5)-182}{3840}\;T^4\\&=  &(0.2339\ldots)\;T^4\,,\nonumber
\end{eqnarray}
where $\zeta(z)$ is the Riemann zeta function.

\vspace*{0.3cm}
\noindent{\bf Probability distribution of $\tau=t_{\min}-t_{\max}$ for Brownian bridge:}
For a BB, which is a periodic BM of period $T$, we show that the PDF of $\tau$ has a scaling form for any value of $\tau$ and $T$:
\begin{equation}
P(\tau|T)=\frac{1}{T}f_{\rm BB}\left(\frac{\tau}{T}\right)\, ,
\end{equation}
where the scaling function is
\begin{equation}\label{eq:f_bb}
f_{\rm BB}(y)=3\,(1-|y|)\sum_{m,n=1}^{\infty}\frac{(-1)^{m+n}m^2n^2}{\left[m^2|y|+n^2(1-|y|)\right]^{5/2}} \, .
\end{equation}
This scaling function is again symmetric around $y=0$ (see Fig. \ref{fig:numeric}b) and it has the asymptotic behaviours \footnote{Note that there were typos in the exponents of the pre-exponential factors in Eq. (5) of Ref. \cite{mori2019}.}:
\begin{eqnarray}\label{asymptote_bb}
f_{\rm BB}(y)\underset{y \to 0^+}{\approx}\frac{\sqrt{2}\pi^2}{y^{\frac{9}{4}}}e^{-\frac{\pi}{\sqrt{y}}} \;, \; f_{\rm BB}(y) \underset{y\to1}\approx\frac{\sqrt{2}\,\pi^2}{(1-y)^{\frac{5}{4}}}e^{-\frac{\pi}{\sqrt{1-y}}} .
\end{eqnarray}
Moreover, we show that the scaling function $f_{\rm BB}(y)$ satisfies the integral equation:
\begin{equation}\label{eq:integral_BB}
\int_{0}^{1}dy\, \frac{f_{\rm BB}(y)}{\sqrt{1+ u y}}=
\int_{0}^{\infty}dz  
\frac{\frac{z}{\sqrt{1+u}}\coth\left(\frac{z}{\sqrt{1+u}}\right)-1}{\sinh(z)\sinh 
\left(\frac{z}{\sqrt{1+u}}\right)} \, .
\end{equation}
This integral relation for BB is the counterpart of Eq. (\ref{eq:integral_BM}) for BM. As in the case of BM,
the integral relation in Eq. (\ref{eq:integral_BB}) can be used to compute the moments of $\tau$ for~BB:
\begin{eqnarray}
\langle |\tau|\rangle &=&\frac{\pi^2-6}{9}\;T = (0.4299\ldots)\;T\,,\\ \nonumber \langle \tau^2\rangle &=& \frac{\pi^2-6}{18}\; T^2 = (0.2149\ldots)\;T^2\,,\\ \nonumber 
\langle |\tau|^3 \rangle &=& \frac{375\pi^2-14\pi^4-1530}{6750}\; T^3 = (0.1196\ldots)\; T^3\,,\\ \nonumber \langle \tau^4\rangle &=&\frac{125\pi^2-7\pi^4-390}{2250}\;T^4 = (0.0719\ldots)\; T^4\,.\nonumber
\end{eqnarray} 
\vspace*{0.3cm}

\noindent{\bf Covariance of $t_{\min}$ and $t_{\max}$:}
In the two cases of BM and BB we quantify the anti-correlation of $t_{\min}$ and $t_{\max}$. Indeed, we compute exactly the covariance function 
\begin{equation}\label{eq:cov}
\operatorname{cov}(t_{\min},t_{\max}) = \langle t_{\min} t_{\max} \rangle - \langle t_{\min} \rangle \langle t_{\max}\rangle \, .
\end{equation}
In the case of BM we find that
\begin{eqnarray}
\operatorname{cov}_{\rm BM}(t_{\min},t_{\max})&=&-\frac{7\zeta(3)-6}{32} \;T^2 \\ & =  & (-0.0754 \ldots)\textit{   }  T^2\,, \nonumber \label{cov_BM_summ} 
\end{eqnarray}
where $\zeta(z)$ is the Riemann zeta function. While for BB, we get
\begin{eqnarray}
\operatorname{cov}_{\rm BB}(t_{\min},t_{\max})&=&-\frac{\pi^2-9}{36}\; T^2 \\ & =  &  (-0.0241 \ldots)\textit{   } T^2 \nonumber \label{cov_BB_summ} \;.
\end{eqnarray}
\\

\noindent{\bf Discrete-time random walks:}
We show that the result in (\ref{eq:f_bm}) is universal in the sense of the Central Limit Theorem. Indeed, let us consider a discrete-time stochastic process of $n$ steps, generated by the position of the~RW
\begin{equation}
x_k=x_{k-1}+\eta_k\,, 
\end{equation}
where $x_0=0$ and $\eta_k$ are independent and identically distributed (IID) jumps drawn from the symmetric probability distribution $p(\eta)$. We show that in the limit of large $n$ the probability distribution of $\tau=t_{\min}-t_{\max}$ has a scaling form
\begin{equation}
P(\tau|n)\underset{n \to \infty}{\longrightarrow}\frac{1}{n}f\left(\frac{\tau}{n}\right)\, ,
\end{equation}
where the scaling function $f(y)$ depends only on the tail behaviour of the 
jump distribution $p(\eta)$. Moreover, we show that if the jump variance 
$\sigma^2=\int_{-\infty}^\infty d\eta\,\eta^2\,p(\eta)$ is finite, $f(y)$ is given by the Brownian scaling 
function (\ref{eq:f_bm}): $f(y)=f_{\rm BM}(y)$. We numerically verify this result for several choices of the 
distribution $p(\eta)$. Moreover, we demonstrate exactly that $f(y)=f_{\rm BM}(y)$ for two particular 
distributions $p(\eta)$ with a finite variance: the double-exponential distribution $p(\eta)=(1/2)e^{-|\eta|}$ and 
the discrete distribution $p(\eta)=(1/2)\delta(|\eta|-1)$, which corresponds to lattice walks. 
On the contrary, for L\'evy flights with divergent jump variance, i.e. for $p(\eta)\sim 1/|\eta|^{\mu+1}$ for 
large $\eta$ with $0<\mu<2$, we verify numerically that the scaling function does depend on the 
L\'evy exponent $\mu$ of $p(\eta)$, i.e. $f(y)=f_{\mu} (y)$. 

Moreover, we uncover a very interesting fact. For an $n$-step discrete-time random walk on a line with IID increments,
we studied the event ``$\tau=t_{\rm min}-t_{\rm max}=n$'', i.e., the time between
the minimum and the maximum has the maximal possible value $n$. This corresponds to the event
that the maximum occurs at time $0$ and the minimum at step $n$, i.e., at the end of the interval.
Thus the event counts the probability that the walk, starting at the origin, stays below $0$ up to step
$n$ and that the last position at step $n$ is the global minimum.
For the double-exponential jump distribution, we proved that for any finite $n$, this probability
is given exactly by
\begin{equation}\label{eq:univ_prob}
P(\tau=n|n)=\frac{1}{2n}\,.
\end{equation}
Indeed, our numerical simulations suggest that this result, for any $n$, seems to 
be valid for arbitrary symmetric and continuous jump distribution, and thus include L\'evy flights also. 
This leads us to conjecture
that the result in Eq. (\ref{eq:univ_prob}) is actually super-universal, i.e, independent
of jump distributions as long as they are continuous and symmetric.  
We could prove this conjecture for $n\le 3$ (the proof is given in the Appendix E) and
verified it numerically for higher $n$.
We suspect that there must be an elegant proof of this conjecture using some
generalisation of the Sparre Andersen theorem---but the mathematical proof of this conjecture for $n> 3$
has eluded us so far (rather frustratingly !). This remains an outstanding open problem.\\

\noindent{\bf Fluctuating interfaces:} 
Finally, we show that our results can be directly applied to study similar quantities in the context of 
fluctuating interfaces. We consider a one-dimensional Kardar-Parisi-Zhang (KPZ)~\cite{kpz86} or 
Edwards-Wilkinson (EW)~\cite{edwads82} interface evolving over a substrate of finite length $L$. 
Let $H(x,t)$ be the height of the interface at position $x$ and at time $t$. 
We consider both free boundary conditions (FBC), where the endpoints $H(0,t)$ and $H(L,t)$ evolve freely, 
and periodic boundary conditions (PBC), i.e. imposing $H(0,t)=H(L,T)$.

In order to use our results to study this system, we exploit a useful mapping between the fluctuating interface 
in the stationary state and a BM, in the case of FBC, or a BB, in the case of PBC. More precisely, we identify 
space with time, i.e. $x \Leftrightarrow t$, the height of the interface with the position of the Brownian 
particle, i.e. $H\Leftrightarrow x$, and the substrate length with the total duration of the BM/BB, i.e. $L 
\Leftrightarrow T$. The statistical properties of most observables of the interface height and of the BM/BB are 
in general quite different, due to a specific constraint which will be discussed later. For instance, the 
distribution of the maximal height of the interface with FBC/PBC is known to be different from the distribution 
of the maximum value of an usual BM/BB \cite{majumdar04,comtet05}. Nevertheless, we show that, in the stationary 
state, the position difference between the maximal and minimal height has the same PDF as $\tau$ for BM, in the 
case of FBC, and for BB, in the case of PBC. In the case of KPZ interfaces with FBC, this result 
is valid only in the large $L$ limit.

\section{Derivation of the distribution of $\tau = t_{\min} - t_{\max}$ for Brownian motion} \label{sec:BM}

\begin{figure}[t]
  \centering
\includegraphics[width=1\linewidth]{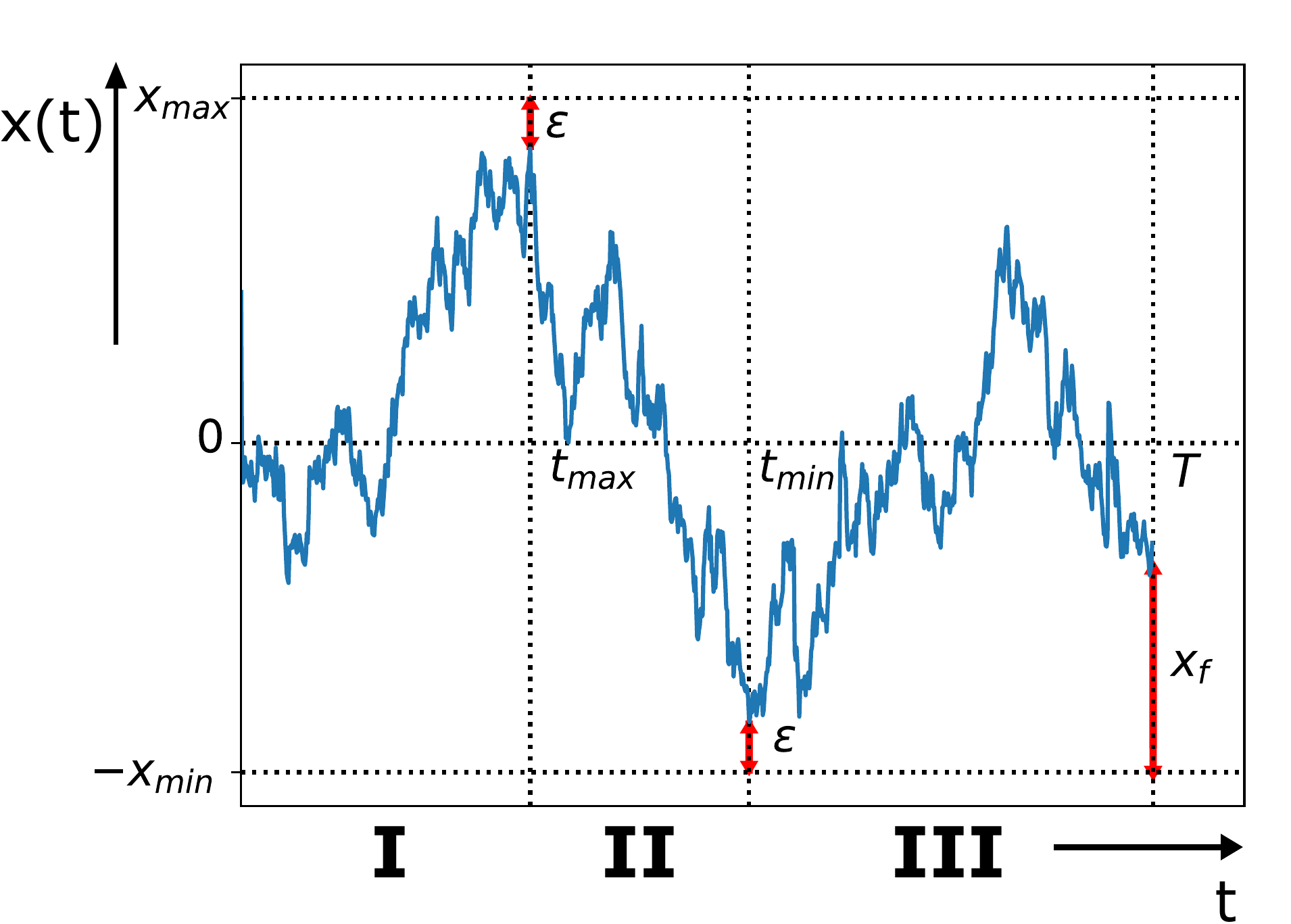}
\caption{Brownian motion $x(t)$ during the time interval $[0,T]$, starting from $x(0)=0$. The value of the global maximum is $x_{\max} - \epsilon$ (with $\epsilon>0$) and the value of the global minimum $-x_{\min} + \epsilon$, where $\epsilon$ is the cut-off needed to enforce absorbing boundary conditions at $x_{\max}$ and $x_{\min}$ (as explained in the text). The time at which the maximum (respectively the minimum) occurs is denoted by $t_{\max}$ (respectively $t_{\min}$). The final position $x(T)$, measured with respect to $-x_{\min}$ is denoted by $x_{\rm f}$. The total time interval $[0,T]$ is divided into three segments: $[0,t_{\max}]$ (I), $[t_{\max}, t_{\min}]$ (II) and $[t_{\min}, T]$ (III), for the case $t_{\min}>t_{\max}$.}
\label{fig:brownian} 
\end{figure}

In this section, we derive the exact expression for the PDF of the time $\tau=t_{\min}-t_{\max}$ between the maximum and the minimum of a one-dimensional BM. We consider a BM $x(t)$, starting from position $x(0)=x_0$ at initial time and evolving according to
\begin{equation}
\dot{x}(t)=\eta(t),
\end{equation}
where $\eta(t)$ is a Gaussian white noise with zero mean $\left\langle \eta(t) \right\rangle=0$ and correlator 
$\left\langle\eta(t)\eta(t')\right\rangle=2D\,\delta\left(t-t'\right)$. For simplicity, we assume $D=1/2 $. The 
probability distribution of the time difference $\tau$ will not depend on the initial position $x_0$. Indeed, 
changing $x_0$ corresponds to a global shift in position but not in time. Hence, without loss of generality we 
set $x_0=0$ in the rest of the paper. Let us define the amplitude of the maximum in $[0,T]$ as 
$x_{\max}=x\left(t_{\max}\right)$ and the amplitude of the minimum as $x_{\min}=-x\left(t_{\min}\right)$ (see 
Fig. \ref{fig:brownian}). Let $x_{\rm f}=x(T)+x_{\min}$ be the final position measured with respect to the 
global minimum. First, we will compute the joint distribution $P(x_{\min},x_{\max},t_{\min},t_{\max}|T)$ of 
these four random variables and then integrate out $x_{\max}$ and $x_{\min}$ to obtain the joint PDF 
$P(t_{\min}, t_{\max}|T)$. Finally, from this joint PDF $P(t_{\min}, t_{\max}|T)$ we will obtain the marginal 
PDF of $\tau$. Note that, integrating out $t_{\min}$ and $t_{\max}$ in 
$P(x_{\min},x_{\max},t_{\min},t_{\max}|T)$, one could also obtain the joint distribution of $x_{\max}$ and 
$x_{\min}$ that is already known in the literature \cite{KMS13,MSS16}. Notably, the joint distribution of the 
extrema of a BM also plays an important role in disordered systems, such as the Sinai model, where the energy 
landscape is modelled by a Brownian motion in space. In that context, the joint distribution of $x_{\min}$ and $x_{\max}$
was computed by using real space renormalisation group methods in Refs. \cite{PLDM03,schehr10}. 
Indeed, the position and the depth of the energy minima as well as the height of the 
energy barriers are quantities of fundamental importance in the description of disordered systems.

\subsection{Computation of the joint distribution of $t_{\max}$ and $t_{\min}$}

For simplicity, we assume that $t_{\max}<t_{\min}$ (the complementary case $t_{\min} < t_{\max}$ can be studied in the same way).    
Exploiting the Markovian property of the process, we can write the total probability $P(x_{\min},x_{\max},t_{\min},t_{\max}|T)$ as the product of the probabilities of the three individual time segments (see Fig. \ref{fig:brownian}): (I) $[0,t_{\max}]$, (II) $[t_{\max}, t_{\min}]$ and (III) $[t_{\min},T]$. We denote by $P_{\rm I}$, $P_{\rm II}$, and $P_{\rm III}$ the probability of the trajectory in the segments (I), (II), and (III), respectively. First of all, in each segment the trajectory has to stay inside the space interval $[-x_{\min},x_{\max}]$ because, by definition, it can never go above the global maximum $x_{\max}$ nor below the global minimum $-x_{\min}$. In the first segment (I), the particle goes from the origin at time $t=0$ to position $x_{\max}$ at time $t_{\max}$ (see Fig. \ref{fig:brownian}). In the second segment (II), the trajectory starts at $x_{\max}$ at time $t_{\max}$ and arrives at $-x_{\min}$ at time $t_{\min}$. Finally, in the third segment (III), the trajectory starts at $-x_{\min}$ at time $t_{\min}$ and arrives at $x(T)$ at time $T$. We finally integrate over all possible values of $x(T)\in[-x_{\min}, x_{\max}]$. It is useful to define the amplitude of the space interval $[-x_{\min},x_{\max}]$
\begin{equation}
M = x_{\min} + x_{\max} \geq 0.
\label{eq:M}
\end{equation} 
To avoid that the trajectory goes outside the interval $[-x_{\min},x_{\max}]$ we impose absorbing boundary 
conditions at both $-x_{\min}$ and $x_{\max}$. However, since we are considering a continuous-time BM, we cannot 
impose that the trajectory arrives exactly at the absorbing boundary at a certain time. Indeed, due to the 
continuous nature of the process, if $x(t)$ arrives at position $x_{\max}$ at time $t_{\max}$ it will go above 
$x_{\max}$ infinitely many times in the time interval $[t_{\max},t_{\max}+\delta]$ for any $\delta>0$ 
\cite{feller50}. Thus, it is impossible to satisfy the constraint $x(t)<x_{\max}$ while imposing 
$x(t_{\max})=x_{\max}$. Analogously, one cannot impose that the BM arrives at position $-x_{\min}$ at time 
$t_{\min}$ in the presence of an absorbing boundary at $-x_{\min}$. A way to avoid this issue is to introduce a 
cut-off $\epsilon$ such that the actual values of $x(t)$ at $t_{\max}$ and $t_{\min}$ are respectively $x_{\max} - 
\epsilon$ and $-x_{\min}+\epsilon$ \cite{randon-furling08,M05}, as in Fig. \ref{fig:brownian}. In this way, we 
can compute the probability $P(x_{\min},x_{\max},t_{\min},t_{\max}|T)$ for a fixed $\epsilon$ and then take the 
limit $\epsilon \to 0$ at the very end of the computation. \\

To compute the grand joint PDF $P(x_{\min},x_{\max},t_{\min},t_{\max}|T)$, we need, as a basic building block, 
the Green's function $G_M(x,t|x_0,t_0)$ denoting the probability density for a BM, starting from $x_0$ at time 
$t_0$, to arrive at $x$ at time $t$, while staying inside the box $[0,M]$ during the interval $[t_0,t]$. An explicit expression for this Green's function can be easily computed solving the 
diffusion equation
\begin{equation} \label{eq:diffusion}
\partial_t G_M(x,t|x_0,t_0)=\frac{1}{2}\partial_x^2 G_M(x,t|x_0,t_0)\,,
\end{equation}
 with absorbing boundary conditions both at $x=0$ and $x=M$. The solution of Eq. (\ref{eq:diffusion}) is given by (see e.g. \cite{Risken, Redner,bray13}):
 \begin{equation}\label{eq:g} 
G_M=\frac{2}{M}\sum_{n=1}^{\infty}\sin\left(\frac{n\pi x}{M}\right)\sin\left(\frac{n \pi x_0}{M}\right)e^{-\frac{n^2\pi^2}{2M^2}(t-t_0)}\;.
\end{equation}
To make use of this building-block, we first shift the origin in Fig. \ref{fig:brownian} to $-x_{\min}$. We start with the segment (I), where the probability $P_{\rm I}$ is just proportional to 
\begin{eqnarray} \label{PI_1}
P_{\rm I} \propto G_{M}(M-\epsilon,t_{\max}|x_{\min},0) \, .
\end{eqnarray}
Using the expression (\ref{eq:g}), after shifting the origin to $-x_{\min}$, and expanding to leading order in $\epsilon$, we get
\begin{equation}\label{PI_2}
P_{\rm I} \propto -\frac{2 \pi \epsilon}{M^2} \sum_{n_1=1}^\infty (-1)^{n_1} n_1 \sin{\left(\frac{n_1 \pi \, x_{\min}}{M} \right)} \,e^{-\frac{n_1^2 \pi^2}{2M^2}\, t_{\max}}\;.
\end{equation}
We next consider the segment (II). Here the probability $P_{\rm II}$ is proportional to 
\begin{eqnarray} \label{PII_1}
P_{\rm II} \propto G_{M}(\epsilon,t_{\min}|M-\epsilon,t_{\max})  \;.
\end{eqnarray} 
Using again Eq. (\ref{eq:g}), and expanding to leading order for small $\epsilon$, we obtain
\begin{eqnarray} \label{PII_2}
P_{\rm II} \propto -\frac{2 \pi^2\,\epsilon^2}{M^3} \sum_{n_2=1}^\infty (-1)^{n_2} n_2^2 \, e^{-\frac{n_2^2 \pi^2}{2M^2}(t_{\min}-t_{\max})} \;.
\end{eqnarray}
Finally, for the third time segment (III), we obtain
\begin{eqnarray} \label{PIII_1}
P_{\rm III} \propto \int_0^M G_{M}(x_{\rm f},T|\epsilon,t_{\min}) \, dx_{\rm f} \;,
\end{eqnarray}
after integrating over the final position $x_{\rm f} \in [0,M]$.  
Using Eq. (\ref{eq:g}) and expanding for small $\epsilon$ we obtain (after integration over $x_{\rm f}$)
\begin{eqnarray} \label{PIII_2}
P_{\rm III} \propto \frac{2\epsilon}{M} \sum_{n_3=1}^\infty \left[1 - (-1)^{n_3} \right] \, e^{-\frac{n_3^2 \pi^2}{2M^2} (T-t_{\min})} \;.
\end{eqnarray}
Taking the product of the three segments (\ref{PI_2}), (\ref{PII_2}) and (\ref{PIII_2}), we obtain that the total probability of the trajectory is proportional to 
\begin{widetext}
\begin{eqnarray}\label{produc_1}
P(x_{\min},x_{\max},t_{\min},t_{\max}|T) &\propto & P_{\rm I} P_{\rm II} P_{\rm III}\propto  \frac{\epsilon^4}{M^6} \sum_{n_1=1}^\infty (-1)^{n_1} n_1 \sin{\left(\frac{n_1 \pi \, x_{\min}}{M} \right)} \,e^{-\frac{n_1^2 \pi^2}{2M^2}\, t_{\max}}  \\ & \times & \sum_{n_2=1}^\infty (-1)^{n_2} n_2^2 \, e^{-\frac{n_2^2 \pi^2}{2M^2}(t_{\min}-t_{\max})} \sum_{n_3=1}^\infty \left[1 - (-1)^{n_3} \right] \, e^{-\frac{n_3^2 \pi^2}{2M^2} (T-t_{\min})} \,,\nonumber
\end{eqnarray}
\end{widetext}
where $M=x_{\min}+x_{\max}$. As pointed out earlier, the result in Eq. (\ref{produc_1}) was derived in Ref.~\cite{schehr10} using real-space renormalisation group (RSRG) method. The method used here, using directly the constrained propagator, is rather different from the RSRG method used in Ref.~\cite{schehr10}.  
Note that Eq. (\ref{produc_1}) is valid in the case $t_{\min}>t_{\max}$ and that by 
`$\propto$', we have omitted the explicit dependence on the volume factors of the variables, i.e. $dt_{\max}, 
dt_{\min}, dx_{\max}$ and $dx_{\min}$, since $P(x_{\min},x_{\max},t_{\min},t_{\max}|T)$ is a probability 
density, and not a probability. We now want to integrate $x_{\min}$ and $x_{\max}$ over $[0,+\infty)$, in order 
to obtain the joint PDF $P(t_{\min}, t_{\max}|T)$ for $t_{\min} > t_{\max}$
\begin{eqnarray} \label{Pminmax}
&& P(t_{\min}, t_{\max}|T) \\
&=& \int_0^\infty dx_{\min} \, \int_0^\infty dx_{\max} P(x_{\min},x_{\max},t_{\min},t_{\max}|T)  \;.\nonumber
\end{eqnarray}
Plugging the expression for $P(x_{\min},x_{\max},t_{\min},t_{\max}|T)$ given in Eq. (\ref{produc_1}) into Eq. (\ref{Pminmax}) and performing the integral over $x_{\max}$, we get
\begin{eqnarray}\label{Pminmax1}
&& P(t_{\min}, t_{\max}|T)\\ &\propto & \epsilon^4 \sum_{n_1,n_2,n_3=1}^{\infty}(-1)^{n_1+n_2}\left[1-(-1)^{n_3}\right]n_1 n_2^2 \,J(\alpha,\beta)\,,\nonumber
\end{eqnarray}
where we have defined the double integral
\begin{eqnarray}\label{Iab}
J(\alpha, \beta)& =&\int_0^\infty dx_{\min}  \int_0^\infty dx_{\max} e^{-\frac{\beta}{(x_{\min} + x_{\max})^2}}\\ &\times & \frac{1}{(x_{\min} + x_{\max})^6}\sin \left(\frac{\alpha \, x_{\min}}{x_{\min}  + x_{\max}} \right)  \, . \nonumber
\end{eqnarray}
with
\begin{eqnarray}
\alpha = n_1 \pi\, ,
\end{eqnarray}
and
\begin{equation}
\beta = \frac{\pi^2}{2} \Big( n_1^2\, t_{\max} + n_2^2\left(t_{\min} - t_{\max}\right) + n_3^2 \left(T-t_{\min}\right)  \Big)\, .
\end{equation}
The integral (\ref{Iab}) can be explicitly evaluated (see Appendix \ref{app:Iab}) and we get
\begin{eqnarray} \label{Iab_2}  
J(\alpha, \beta) = \frac{1-\cos \alpha}{2 \alpha \, \beta^2 } \;.
\end{eqnarray} 
Thus, using this result (\ref{Iab_2}) in Eq. (\ref{Pminmax1}), one obtains (for $t_{\max} < t_{\min}$)
\begin{eqnarray}\label{Pminmax2}
&& P_<(t_{\min}, t_{\max}|T)  = A_< \; \theta(t_{\min} - t_{\max}) \, \sum_{n_1, n_2, n_3 = 1}^\infty \nonumber \\ &\times &\frac{(-1)^{n_2+1} n_2^2  [1-(-1)^{n_1}] [1-(-1)^{n_3}]}{\left[n_1^2 t_{\max} + n_2^2 (t_{\min}-t_{\max})+ n_3^2 (T-t_{\min})\right]^2}  \, ,
\end{eqnarray}
where we used $dt_{\max} dt_{\min} \propto \epsilon^4$ and the subscript `$<$' indicates $t_{\max} < t_{\min}$. Here $\theta(x)$ is the Heaviside theta function. In arriving at this final form (\ref{Pminmax2}), we have used that $(-1)^{n_1} = -1$ since only the odd values of $n_1$ contribute to the sum. The overall proportionality constant $A_<$ has to be fixed from the normalisation condition
\begin{eqnarray}\label{eq:normalization<}
\int_0^T dt_{\min} \int_0^T dt_{\max} P_<(t_{\min}, t_{\max}|T)\\ =P(t_{\max}<t_{\min}|T)=\frac{1}{2}\,.\nonumber
\end{eqnarray}
Note that, for the complementary case $t_{\max} > t_{\min}$, one can perform a similar computation and one obtains 
\begin{eqnarray} \label{symmetry}
&&P_>(t_{\min},t_{\max}|T) = A_> \; \theta(t_{\max} - t_{\min})  \, \sum_{n_1, n_2, n_3 = 1}^\infty \nonumber\\ &\times & \frac{(-1)^{n_2+1} n_2^2[1-(-1)^{n_1}] [1-(-1)^{n_3}]}{\left[n_1^2 t_{\min} + n_2^2 (t_{\max}-t_{\min})+ n_3^2 (T-t_{\max})\right]^2} \,,
\end{eqnarray}
where $A_>$ is again a proportionality constant. This constant $A_>$ can be fixed from a normalisation condition similar to the one in Eq. (\ref{eq:normalization<}) but with $P_<(t_{\min}, t_{\max}|T)$ replaced by $P_>(t_{\min}, t_{\max}|T)$.
Indeed, it is easy to see that $A_>$ and $A_<$ have to satisfy the same condition. This implies that
\begin{equation}
A_<=A_>=A\,.
\end{equation}
However, computing the exact value of $A$ from condition (\ref{eq:normalization<}) appears to be non-trivial. We will see later that the normalisation constant $A$ is given exactly by
\begin{eqnarray} \label{A_1}
A = \frac{4}{\pi^2} \;.
\end{eqnarray}     
Moreover, one sees the symmetry 
\begin{eqnarray} \label{symmetry2}
P_>(t_{\min},t_{\max}|T) = P_<(t_{\max}, t_{\min}|T)  \;.
\end{eqnarray} 
This non-trivial symmetry can be traced back to the fact that the BM is symmetric under the reflection $x \to -x$.\\

\subsection{Computation of the PDF of $\tau = t_{\min} - t_{\max}$} 

To compute the PDF $P(\tau|T)$ of $\tau = t_{\min} - t_{\max}$, we focus on the case $t_{\min}>t_{\max}$, i.e. $\tau >0$. The complementary case $\tau<0$ is simply determined from the symmetry $P(-\tau|T) = P(\tau|T)$, obtained from exchanging $t_{\max}$ and $t_{\min}$ and using Eq. (\ref{symmetry2}). For $\tau >0$, one has
\begin{eqnarray}\label{Ptau_1}
&&P(\tau|T) = \int_0^T dt_{\max} \int_0^T dt_{\min} P_<(t_{\min}, t_{\max}|T)\nonumber \\ &\times &\delta(t_{\min} - t_{\max}-\tau) \;, 
\end{eqnarray}
where $P_<(t_{\min}, t_{\max},|T)$ is given in Eq. (\ref{Pminmax2}). Integrating over $t_{\min}$ gives
\begin{widetext}
\begin{eqnarray}
\label{Ptau_2}
P(\tau|T) &=& \int_0^{T-\tau} dt_{\max} P_<(t_{\max} + \tau, t_{\max}|T)\\  &=& A \sum_{n_1, n_2, n_3 =1}^\infty (-1)^{n_2+1} n_2^2 (1-(-1)^{n_1})(1-(-1)^{n_3})  \int_0^{T-\tau} dt_{\max}\, \frac{1}{\left( n_1^2 t_{\max} + n_2^2 \tau + n_3^2 (T-t_{\max}-\tau)\right)^2} \nonumber \\&=& A \; (T - \tau)     \sum_{n_1, n_2, n_3 =1}^\infty   (-1)^{n_2+1} n_2^2 \frac{(1-(-1)^{n_1})(1-(-1)^{n_3}) }{(n_1^2 (T-\tau) + n_2^2 \tau)(n_3^2 (T-\tau) + n_2^2 \tau)} \;.\nonumber
\end{eqnarray}
\end{widetext} 
Remarkably, the sums over $n_1$ and $n_3$ get decoupled and each yields exactly the same contribution. Hence we get
\begin{eqnarray}\label{PTau_5}
P(\tau |T) &=& A \; (T-\tau) \sum_{n_2=1}^\infty (-1)^{n_2+1}n_2^2\\ & \times &  \left[ \sum_{n=1}^\infty \frac{1-(-1)^n}{n^2(T-\tau)+ n_2^2 \tau}\right]^2 \;.\nonumber
\end{eqnarray}   
This sum over $n$ inside the parenthesis can be performed using the identity \cite{prudnikov}
\begin{eqnarray}\label{identity_Ptau}
\sum_{n=1}^\infty \frac{1-(-1)^ n}{b+ n^2} = \frac{\pi}{2 \sqrt{b}} {\rm tanh}\left( \frac{\pi}{2} \sqrt{b}\right) \;.
\end{eqnarray}  
Using this identity (\ref{identity_Ptau}) into Eq. (\ref{PTau_5}) one obtains, for $\tau >0$, 
\begin{eqnarray}\label{PTau_6}
P(\tau|T) = \frac{1}{T} f_{\rm BM}\left( \frac{\tau}{T}\right) \;, 
\end{eqnarray}
where
\begin{eqnarray}\label{eq:f_BM_A}
f_{\rm BM}(y) = A \; \frac{\pi^2}{4y} \, \sum_{n=1}^\infty (-1)^{n+1} {\rm \tanh}^2 \left( \frac{n\pi}{2}  \sqrt{\frac{y}{1-y}}\right) \;,
\end{eqnarray}
which is only valid for $y>0$. However, the symmetry $\tau\to -\tau$ implies that $f_{\rm BM}(y)=f_{\rm BM}(-y)$ and hence
\begin{equation}\label{eq:f_BM_A2}
f_{\rm BM}(y) = A \; \frac{\pi^2}{4|y|} \, \sum_{n=1}^\infty (-1)^{n+1} {\rm \tanh}^2 \left( \frac{n\pi}{2}  \sqrt{\frac{|y|}{1-|y|}}\right) \;,
\end{equation}
which is valid for $-1\leq y\leq 1$.
The constant $A$ can be determined from the normalisation condition
\begin{equation}\label{eq:normalization3}
\int_{-T}^{T}d\tau \, P(\tau|T)=1\,.
\end{equation}
Using the scaling form (\ref{PTau_6}) and changing variable $\tau\to y=\tau/T$, we get the equivalent condition on $f_{\rm BM}(y)$
\begin{equation}
\int_{-1}^{1}dy \, f_{\rm BM}(y)=1\,,
\end{equation}
which, using the symmetry $f_{\rm BM}(y)=f_{\rm BM}(-y)$, becomes
\begin{equation}
\int_{0}^{1}dy \, f_{\rm BM}(y)=\frac{1}{2}\,.
\end{equation}
Using the expression for $f_{\rm BM}(y)$ in Eq. (\ref{eq:f_BM_A2}), we get the following condition for $A$:
\begin{equation}\label{eq:normalization_for_A}
A\frac{\pi^2}{4} \int_{0}^{1} \frac{dy}{y} \;  \, \sum_{n=1}^\infty (-1)^{n+1} {\rm \tanh}^2 \left( \frac{n\pi}{2}  \sqrt{\frac{y}{1-y}}\right)=\frac{1}{2}\,.
\end{equation}
It turns out that the integral on the left-hand side of Eq. (\ref{eq:normalization_for_A}) can be 
computed exactly (see Appendix \ref{app:integral}), yielding
\begin{equation}\label{eq:integral_1/2}
\int_{0}^{1} \frac{dy}{y} \;  \, \sum_{n=1}^\infty (-1)^{n+1} {\rm \tanh}^2 \left( \frac{n\pi}{2} 
\sqrt{\frac{y}{1-y}}\right)=\frac{1}{2}\,.
\end{equation}
Thus, using Eqs. (\ref{eq:normalization_for_A}) and (\ref{eq:integral_1/2}) 
we get that $A=4/\pi^2$. Using this exact value of $A$ in Eq. (\ref{eq:f_BM_A}) gives us our
complete result for the scaling function
\begin{eqnarray}\label{eq:f_bm_2}
f_{\rm BM}(y)=\frac{1}{|y| }\sum_{n=1}^{\infty}(-1)^{n+1}\tanh^2\left(\frac{n\pi}{2}\sqrt{\frac{|y|}{1-|y|}}\right) \;,
\end{eqnarray}
for $-1<y<1$, as given in Eq. (\ref{eq:f_bm}). A plot of the scaling function $f_{\rm BM}(y)$ is shown in Fig. \ref{fig:numeric}a, where we also compare it with numerical simulations, finding an excellent agreement. The scaling function $f_{\rm BM}(y)$ is symmetric around $y=0$ and it is non-monotonic as a function of $y$. We numerically identify the values
\begin{equation}
y^*=\pm (0.5563\ldots)
\end{equation}
at which $f_{\rm BM}(y)$ is maximal. This non-trivial value has a nice application in finance. Indeed, let us consider again the situation described in Section \ref{sec:intro} and let $x(t)$ represent the price of a stock during some fixed time window of duration $T$. Assume that $t_{\min}>t_{\max}$ and that an agent has sold her or his stock at time $t_{\max}$. In order to re-buy the stock at the best price, i.e. at time $t_{\min}$, then, the optimal time that the agent has to wait between selling and buying is exactly $y^* \,T$, which is the value of $\tau=t_{\min}-t_{\max}$ with the highest probability density.

\subsection{Asymptotic analysis of $f_{\rm BM}(y)$}

We consider the function $f_{\rm BM}(y)$ given explicitly in Eq. (\ref{eq:f_bm}) where $y \in [-1,1]$. Using the symmetry $f_{\rm BM}(-y) = f_{\rm BM}(y)$ it is sufficient to consider the case $y>0$. We first study the limit when $y \to 1$ (or equivalently $y \to -1$). In this limit, the term $\tanh^2\left(\frac{n\pi}{2}\sqrt{\frac{y}{1-y}}\right)$ can be expanded as
\begin{equation}
\tanh^2\left(\frac{n\pi}{2}\sqrt{\frac{y}{1-y}}\right)\simeq \frac{1+e^{-\frac{n\pi}{\sqrt{1-y}}}}{1-e^{-\frac{n\pi}{\sqrt{1-y}}}}\simeq 1+2e^{-n\pi/\sqrt{1-y}}\,.
\end{equation}
Thus, the scaling function $f_{\rm BM}(y)$ can be expanded, as $y\to 1$ as
\begin{eqnarray}
&&f_{\rm BM}(y)\simeq \left(1+(1-y)\right)\sum_{n=1}^{\infty}(-1)^{n+1}\left(1+2e^{-n\pi/\sqrt{1-y}}\right) \nonumber \\
&&=  \left(1+(1-y)\right)\left(\sum_{n=1}^{\infty}(-1)^{n+1}+ \frac{2}{1+e^{\pi/\sqrt{1-y}}}\right)\nonumber \\
&& \simeq \left(1+(1-y)\right)\left(\frac{1}{2}+ 2\, e^{-\pi/\sqrt{1-y}}\right)\simeq
\frac{1}{2}+\frac{1-y}{2}\,.
\end{eqnarray}
In going to the second to the third line above, we have used the equality
\begin{equation}
\sum_{n=1}^{\infty}(-1)^n=\frac{1}{2}\,.
\end{equation} 
Of course, this sum is not convergent. However, one can interpret it in a regularised sense as follows \cite{randon-furling08}
\begin{eqnarray}\label{yto1_2}
\lim_{\alpha \to -1} \sum_{n=1}^\infty \alpha^{n+1} = \lim_{\alpha \to -1} \frac{\alpha^2}{1-\alpha} = \frac{1}{2} \;.
\end{eqnarray}
Thus, in the limit $y\to 1$, we have verified that $f_{\rm BM}(y)$ goes to the value $1/2$. Moreover, in the vicinity of $y=1$, $f_{\rm BM}(y)$ is linear with negative slope $-1/2$.
Next, we consider the limit $y \to 0^+$. In order to investigate this limit, it turns out that the representation given in Eq. (\ref{eq:f_bm}) is not convenient, since the series diverges strongly if one naively takes the limit $y\to 0^+$. Hence, it is convenient to derive an alternative representation of $f_{\rm BM}(y)$ which will allow us to obtain the $y \to 0^+$ behaviour correctly. To proceed, we use the Poisson summation formula. Consider the sum
\begin{eqnarray}\label{s_of_a}
s(a)&=&\sum_{n=0}^{\infty}(-1)^{n+1}\tanh^2(na)\\&=&-\sum_{n=0}^{\infty}e^{i \pi n}\tanh^2(na)=-\frac{1}{2}\sum_{n=-\infty}^{\infty}e^{i\pi n}\tanh^2(na) \;.\nonumber
\end{eqnarray}
This sum can be re-written, using the Poisson summation formula, as
\begin{eqnarray} \label{Poisson1}
s(a) = - \frac{1}{2} \sum_{m=-\infty}^\infty \hat F(2\pi m) \;,  
\end{eqnarray}
where
\begin{eqnarray}
\hat F(2\pi m) = \int_{-\infty}^\infty e^{i 2 \pi m x} e^{i \pi x} \tanh^2(a x)\, dx \;.
\end{eqnarray}
This integral can be performed explicitly, using the identity \cite{gradshteyn} 
\begin{eqnarray}\label{Poisson2}
\int_{0}^\infty \cos{(b y)}\,  \tanh^2(y) \, dy = - \frac{\pi}{2} \frac{b}{\sinh\left(\frac{\pi b}{2} \right)} \;.
\end{eqnarray}
Using this, we get 
\begin{eqnarray}\label{Poisson3}
s(a) = \frac{\pi^2}{2a^2} \sum_{m=-\infty}^{\infty}\frac{2m+1}{\sinh\left(\frac{\pi^2 (2m+1)}{2a}\right)}.
\end{eqnarray}
Using $a = \frac{\pi}{2} \sqrt{|y|/(1-|y|)}$ in Eq. (\ref{eq:f_bm_2}) and Eq. (\ref{Poisson3}), we obtain an exact alternative representation of $f_{\rm BM}(y)$ as
\begin{equation}\label{Poisson4}
f_{\rm BM}(y) =  \frac{2(1-|y|)}{|y|^2} \sum_{m = -\infty}^\infty \frac{2m+1}{\sinh \left( (2m+1) \pi \sqrt{\frac{1-|y|}{|y|}} \right)}  \;.
\end{equation}
One can now take the $y \to 0^+$ limit in the last expression, where the terms $m=0$ and $m=-1$ dominate in this limit. This yields, to leading order,  
\begin{equation}\label{asympt_y0}
f_{\rm BM}(y) \approx \frac{8}{y^2}e^{-{\pi}/{\sqrt{y}}} \;, 
\end{equation}
Including also higher order corrections, one obtains
\begin{equation}
f_{\rm BM}(y) \approx \frac{8}{y^2}e^{-{\pi}/{\sqrt{y}}} -\frac{8}{y}e^{-{\pi}/{\sqrt{y}}} \;, \quad {\rm as} \quad y \to 0^+ \;.
\end{equation}
Hence, using the $y\to -y$ symmetry, the asymptotic behaviours of $f_{\rm BM}(y)$ can be summarised as
\begin{eqnarray}\label{summary_asymptotics}
f_{\rm BM}(y) \approx
\begin{cases}
&\dfrac{1}{2}+\dfrac{1-|y|}{2} \quad \; \quad \quad\quad \quad \quad \quad \; {\rm as} \quad y \to \pm 1 \\
& \\
& \dfrac{8}{y^2}\, e^{-{\pi}/{\sqrt{|y|}}} -\dfrac{8}{|y|}\, e^{-{\pi}/{\sqrt{|y|}}} \quad {\rm as} \quad y \to 0	\,.
\end{cases}
\end{eqnarray}

\subsection{Moments of $\tau$ for BM} \label{sec:moments}

Since the distribution $P(\tau|T)$ is symmetric in $\tau\in[-T,T]$, the odd moments of $\tau$ vanish by symmetry and
only the even moments are nonzero. Hence, it is more appropriate to compute the moments of the absolute value of $\tau$ 
\begin{equation}
\label{absolute_moment.BM}
\langle|\tau|^k\rangle=\int_{-T}^{T}d\tau\,P(\tau|T)\,|\tau|^k\, \quad k\ge 0\, .
\end{equation}
Using the scaling form $P(\tau|T)= (1/T)\, f_{\rm BM}(\tau/T)$, one obtains from Eq. (\ref{absolute_moment.BM})
\begin{equation}\label{eq:tau_k_1}
\langle|\tau|^k\rangle=T^k\,\int_{-1}^{1}dy\,f_{\rm BM}(y)\,|y|^k\,,=2T^k\,\int_{0}^{1}dy\,f_{\rm BM}(y)\,|y|^k\, ,
\end{equation}
where the scaling function $f_{\rm BM}(y)$ is given in Eq. (\ref{eq:f_bm_2}).
However, evaluating the integral on the right hand side of Eq. (\ref{eq:tau_k_1}) directly with the form of $f_{\rm BM}(y)$
in Eq. (\ref{eq:f_bm_2}) seems rather hard. Here we found an alternative way to evaluate the moments explicitly.

In fact, we found an integral identity satisfied by the scaling function $f_{\rm BM}(y)$, namely,
\begin{equation}\label{eq:int_BM}
\int_{0}^{1}dy\frac{f_{\rm BM}(y)}{1+uy}=\int_{0}^{\infty}dz \frac{1}{\sinh(z)}\tanh^2\left(\frac{z}{2\sqrt{1+u}}\right)\, .
\end{equation}
The proof of this identity is provided later in Section~V.A.1 in the context of discrete-time
random walks with exponential jump distribution, as it emerges quite naturally there. 
It turns out that this identity in Eq. (\ref{eq:int_BM}) plays the role of a generating function of moments
and moments can be simply extracted by expanding both sides of this identity in powers of $u$. Let us just quote here
first four moments of $\tau$
\begin{eqnarray}\label{eq:moments_BM}
\langle |\tau|\rangle &=&\frac{4\log(2)-1}{3}\,T=(0.5908\ldots)\,T\,,\\ \nonumber 
\langle \tau^2\rangle &=& \frac{7\zeta(3)-2}{16}\,T^2=(0.4009\ldots)\,T^2\,,\\ \nonumber 
\langle |\tau|^3 \rangle &=& \frac{147\zeta(3)-34}{480}\,T^3=(0.2972\ldots)\,T^3\,,\\ 
\nonumber \langle \tau^4\rangle &=&\frac{1701\zeta(3)-930\zeta(5)-182}{3840}\,T^4\\&=&(0.2339\ldots)\,T^4\,,\nonumber
\end{eqnarray}
which are fully consistent with the estimates from simulation.\\

\noindent {\bf{Covariance of $t_{\min}$ and $t_{\max}$:}} As an application of these results on the moments of $\tau$, 
we show now that the covariance
of $t_{\rm min}$ and $t_{\rm max}$ can also be computed explicitly using the known second moment $\langle \tau^2\rangle $ above.
By definition, the covariance function is given by
\begin{equation} \label{def_cov}
\operatorname{cov}\left(t_{\min},t_{\max}\right)= \langle t_{\min} t_{\max}\rangle  -
\langle t_{\min}\rangle  \langle t_{\max}\rangle  \;.
\end{equation}
While, in principle, one can evaluate the covariance exactly knowing the joint distribution
$P(t_{\min},t_{\max}|T)$ in Eq. (\ref{Pminmax2}), it turns out to be rather cumbersome.
A more elegant and much shorter method consists in using the moments of $\tau$ as we show now.
Since $\tau=t_{\rm min}-t_{\rm max}$, it follows that
\begin{eqnarray}\label{id_tau}
\langle \tau^2 \rangle = \langle t_{\min}^2 \rangle + 
\langle t_{\max}^2 \rangle - 2 \langle t_{\min} \, t_{\max} \rangle  \;.
\end{eqnarray}
We can now eliminate $\langle t_{\min} \, t_{\max} \rangle$ from Eqs. (\ref{def_cov}) and (\ref{id_tau})
and express the covariance in terms of $\tau^2$
\begin{eqnarray}\label{formula_covariance}
&&\operatorname{cov}\left(t_{\min},t_{\max}\right) \\ &=&\frac{1}{2}\left(\langle t_{\min}^2\rangle +
\langle t_{\max}^2\rangle -\langle \tau^2\rangle \right)-\langle t_{\min}\rangle \langle t_{\max}\rangle  \;. \nonumber
\end{eqnarray}
Thus, we just need the first two moments of $t_{\min}$, $t_{\max}$ and $\tau$. In the case of the BM the marginal PDFs of
$t_{\min}$ and $t_{\max}$ are given by the expression in Eq. (\ref{levy}):
\begin{eqnarray} \label{arcsine_min}
&&P(t_{\min}|T)=\frac{1}{\pi\sqrt{t_{\min}(T-t_{\min})}} \;,
\end{eqnarray}
for $ 0\leq t_{\min} \leq T$, and similarly
\begin{eqnarray}\label{arcsine_max}
&&P(t_{\max}|T)=\frac{1}{\pi\sqrt{t_{\max}(T-t_{\max})}}\;,
\end{eqnarray}
for $0\leq t_{\max} \leq T $. 
From Eqs. (\ref{arcsine_min}) and  (\ref{arcsine_max}) we get
\begin{eqnarray}
&&\langle t_{\min} \rangle = \langle t_{\max} \rangle = \frac{T}{2} \label{av_tmin} \;, \\
&&\langle t_{\min}^2 \rangle = \langle t_{\max}^2 \rangle = \frac{3}{8}T^2 \;. \label{var_tmin}
\end{eqnarray}
The second moment of $\tau$ is computed in Eq. (\ref{eq:moments_BM}) above. Thus,
substituting the results from Eqs. (\ref{av_tmin}), (\ref{var_tmin}) and (\ref{eq:moments_BM}) in
Eq. (\ref{formula_covariance}), we get
\begin{equation} \label{cov}
\operatorname{cov}_{\rm BM}(t_{\min},t_{\max})=-\frac{7\zeta(3)-6}{32}\,T^2=-(0.0754\ldots) \, T^2.
\end{equation}

\section{Derivation of the distribution of $\tau = t_{\min} - t_{\max}$ for Brownian bridge} \label{sec:BB}

Here, we study the statistics of $\tau=t_{\min}-t_{\max}$ in the case of a BB, i.e. a periodic BM of fixed period $T$. The PDF of $\tau$ for BB will be directly applicable to study KPZ/EW interfaces with PBC in space (see Section \ref{sec:fluctuating}). In the case of the BB, $t_{\max}$ is uniformly distributed over the interval $[0,T]$ \cite{feller50}
\begin{equation}\label{eq:t_max_BB}
P(t_{\max}|T)=1/T\,.
\end{equation} 
By the symmetry of the BB, $t_{\min}$ has the same distribution (\ref{eq:t_max_BB}). However, as explained below, the PDF of the time difference $\tau=t_{\min}-t_{\max}$ for the BB has a scaling form $P(\tau|T)=(1/T)f_{\rm BB}(\tau/T)$ where $f_{\rm BB}(y)$ is the non-trivial scaling function given in Eq. (\ref{eq:f_bb}). In the next two sections we present two alternative derivations for the PDF $P(\tau | T)$. The first is based on a path-integral technique analogous to the one presented in Section \ref{sec:BM}, while the second is based on a useful mapping between the BB and the Brownian excursion, namely the Veervat construction.

\subsection{Derivation 1: path-integral method}\label{sec:BB_1}

\begin{figure}[t]
  \centering
\includegraphics[width=1\linewidth]{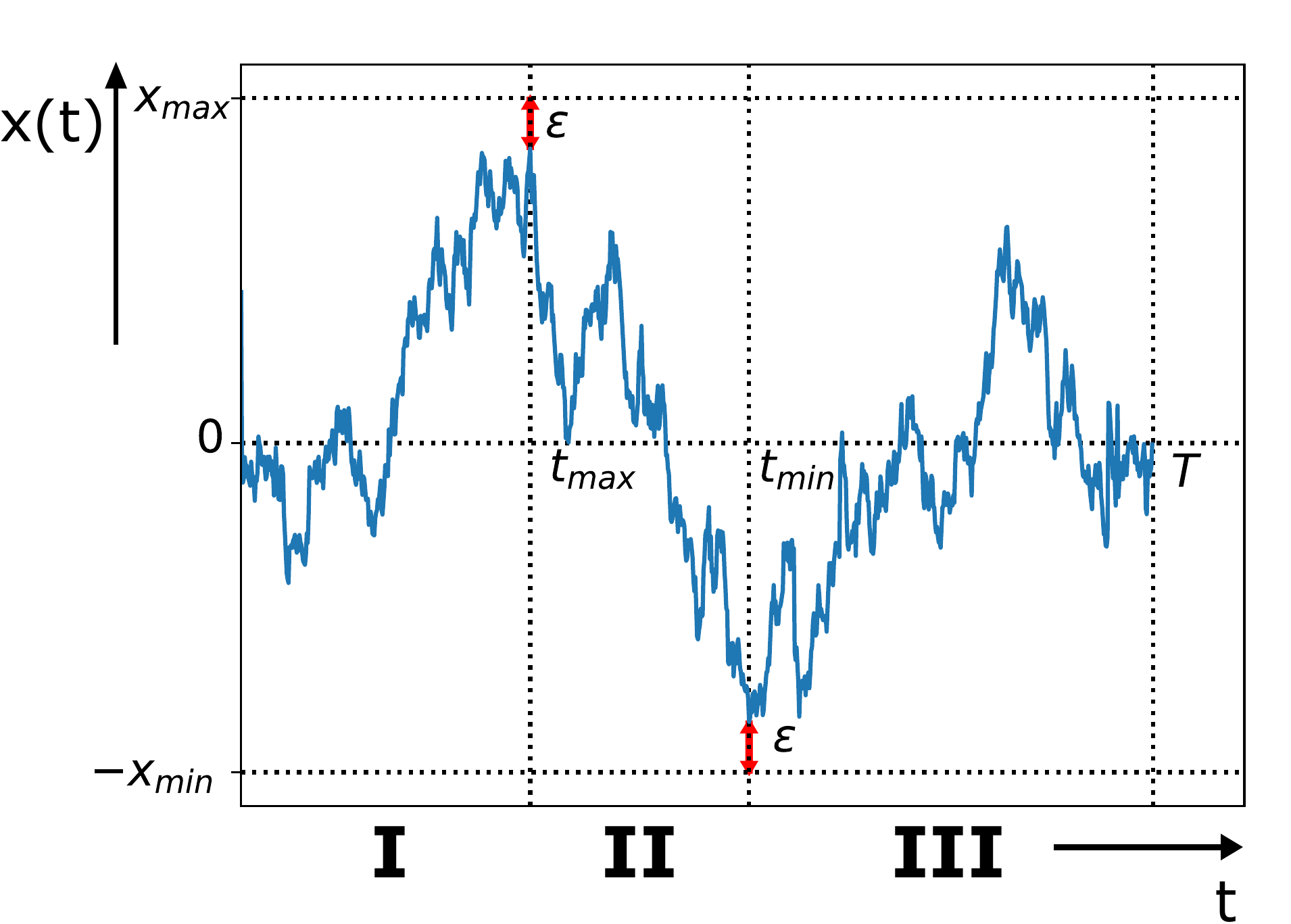}
\caption{A typical trajectory of a Brownian bridge $x(t)$ during the time interval $[0,T]$, starting from $x(0)=0$ and ending at $x(T)=0$. The value of the global maximum is $x_{\max} - \epsilon$ (with $\epsilon>0$) and the value of the global minimum $-x_{\min} + \epsilon$, where $\epsilon$ is the cut-off needed to enforce absorbing boundary conditions at $x_{\max}$ and $x_{\min}$ (as explained in the text). The time at which the maximum (respectively the minimum) occurs is denoted by $t_{\max}$ (respectively $t_{\min}$). The total time interval $[0,T]$ is divided into three segments: $[0,t_{\max}]$ (I), $[t_{\max}, t_{\min}]$ (II) and $[t_{\min}, T]$ (III), for the case $t_{\min}>t_{\max}$.}
\label{fig:bridge} 
\end{figure}

The derivation in the case of BB follows more or less the same steps as in the BM case (Section \ref{sec:BM}). We consider a 
typical trajectory going from $x_0=0$ at time $t=0$ to the final position $x(T)=0$ at time $T$, as in Fig. \ref{fig:bridge}.

Let $t_{\min}$ and $t_{\max}$ denote the time of occurrences 
of the minimum and the maximum respectively. The actual values of the minimum and the maximum are again denoted by $-x_{\min}$ and 
$x_{\max}$. As in the BM case, we first compute the grand joint PDF $P(x_{\min}, x_{\max}, t_{\min}, t_{\max} | T)$ by decomposing the interval $[0,T]$ into three segments I, II, III. While the probabilities $P_{\rm I}$ and $P_{\rm II}$ for the first two segments are exactly identical as in the BM case,
the probability for the last segment $P_{\rm III}$ is different, due to the bridge constraint $x(T)=0$. 
Once again, in terms of the Green's equation	
defined in Eq. (\ref{eq:g}), with the origin shifted to $-x_{\min}$ as in the BM case, this grand PDF can be written as
\begin{eqnarray} \label{jointPDF_BB}
&& P(x_{\min}, x_{\max}, t_{\min}, t_{\max} | T)\\& \propto & G_M(M-\epsilon, t_{\max}|x_{\min},0) \, G_M(\epsilon, t_{\min}|M-\epsilon, t_{\max}) \, \nonumber\\ &\times & G_M(x_{\min}, T|\epsilon, t_{\min}) \;,\nonumber
\end{eqnarray}     
where we have again used the cut-off $\epsilon$, as explained in Section \ref{sec:BM} and $M=x_{\min}+x_{\max}$ as before. We recall that in Eq. (\ref{jointPDF_BB}) we measure all positions with respect to the global minimum $-x_{\min}$. Taking the $\epsilon \to 0$ limit and integrating over $x_{\min}$ and $x_{\max}$, we can obtain the joint PDF $P(t_{\min}, t_{\max} | T)$. The intermediate steps leading to the final result are very similar to the BM case, hence we do not repeat them here and just quote the final result. For $t_{\min} > t_{\max}$, we get
\begin{eqnarray}\label{P>BB}
&& P_<(t_{\min}, t_{\max}|T) = B \sqrt{T} \, \theta(t_{\min}-t_{\max}) \\ &\times & \sum_{n,m=1}^\infty \frac{(-1)^{m+n}\,  m^2 n^2}{\left[n^2 T+ (m^2-n^2) (t_{\min}-t_{\max})\right]^{5/2}}  \;, \nonumber
\end{eqnarray}
where the constant $B$ can be fixed from the overall normalisation. The factor $\sqrt{T}$ in Eq. (\ref{P>BB}) comes from the fact that, since we are considering a BB, we are implicitly conditioning on the event ``$x(T) = 0$''. Thus, after integrating out the variables $x_{\min}$ and $x_{\max}$ from the joint distribution in Eq. (\ref{jointPDF_BB}), one has also to divide by the probability 
\begin{equation}
P(x(T)=0|T)=\frac{1}{\sqrt{2\pi T}}\,,
\end{equation}
which gives the additional factor $\sqrt{T}$ in Eq. (\ref{P>BB}).
We recall that the subscript '$<$' in Eq. (\ref{P>BB}) indicates $t_{\max} < t_{\min}$. To compute the PDF $P(\tau|T)$ of $\tau = t_{\min} - t_{\max}$, we focus on the case $t_{\max}<t_{\min}$, i.e. $\tau >0$. The complementary case $\tau<0$ is simply determined from the symmetry $P(-\tau|T) = P(\tau|T)$, as in the BM case. For $\tau >0$, one has
\begin{eqnarray}\label{Ptau_BB1}
&& P(\tau|T) = \int_0^T dt_{\max} \int_0^T dt_{\min} P_<(t_{\min}, t_{\max}|T) \; \nonumber\\ &\times & \delta(t_{\min} - t_{\max}-\tau) \; ,
\end{eqnarray}
where $P_<(t_{\min}, t_{\max},|T)$ is given in Eq. (\ref{P>BB}). Noting that $P_<(t_{\min}, t_{\max},|T)$ depends only on the difference $\tau = t_{\min} - t_{\max}$, we can first carry out the integral over $t_{\min}$ in Eq. (\ref{Ptau_BB1}) keeping $\tau$ fixed. This gives an additional factor $(T-\tau)$ and we get, for $\tau > 0$
\begin{eqnarray}\label{Ptau_BB2}
P(\tau|T) = \frac{1}{T} f_{\rm BB}\left( \frac{\tau}{T}\right)\,,
\end{eqnarray}
where the scaling function $f_{\rm BB}(y)$, for $0 \leq y \leq 1$, is given by
\begin{eqnarray}\label{fBB_SM1}
f_{\rm BB}(y) = B\;(1-y) \sum_{m,n=1}^{\infty}\frac{(-1)^{m+n}m^2n^2}{\left[m^2 \,y+n^2(1-y)\right]^{5/2}}.
\end{eqnarray}
For $\tau < 0$, using the symmetry $P(-\tau|T) = P(\tau|T)$, it follows that $P(\tau|T)$ takes exactly the same scaling form as in Eq. (\ref{Ptau_BB2}), with $y$ replaced by $-y$. Thus for all $-T \leq \tau \leq T$, $P(\tau|T) = (1/T) f_{\rm BB}(\tau/T)$ where $f_{\rm BB}(y)$, for all $-1 \leq y \leq 1$, is given by
\begin{equation} \label{fBB_SM2}
f_{\rm BB}(y) = B (1-|y|) \sum_{m,n=1}^{\infty}\frac{(-1)^{m+n}m^2n^2}{\left[m^2 \,|y|+n^2(1-|y|)\right]^{5/2}} \;, 
\end{equation}
The prefactor $B$ can, in principle, be fixed from the normalisation condition 
\begin{equation}
\int_{-1}^1 f_{\rm BB}(y) dy = 1 \, .
\end{equation}
However, computing explicitly this integral appears to be challenging. An alternative way to obtain the constant $B$ is presented in the next section, where we show that 
\begin{equation}
B=3\,.
\end{equation}
The scaling function $f_{\rm BB}(y)$, shown in Fig. \ref{fig:numeric}b, is symmetric around $y=0$ and it is non-monotonic as a function of $y$. We numerically find that at the two points $y^*=\pm (0.3749\ldots )$ the scaling function $f_{\rm BB}(y)$ reaches its maximal value.
To confirm this result (\ref{fBB_SM2}) we have performed numerical simulations, using a simple algorithm for generating Brownian bridges \cite{majumdar15}. The results of simulations (see Fig. \ref{fig:numeric}b) are in excellent agreement with the scaling function in Eq. (\ref{fBB_SM2}). 

\subsection{Derivation 2: mapping to Brownian excursion}\label{sec:BB_2}

\begin{figure*}
\includegraphics[scale=0.5]{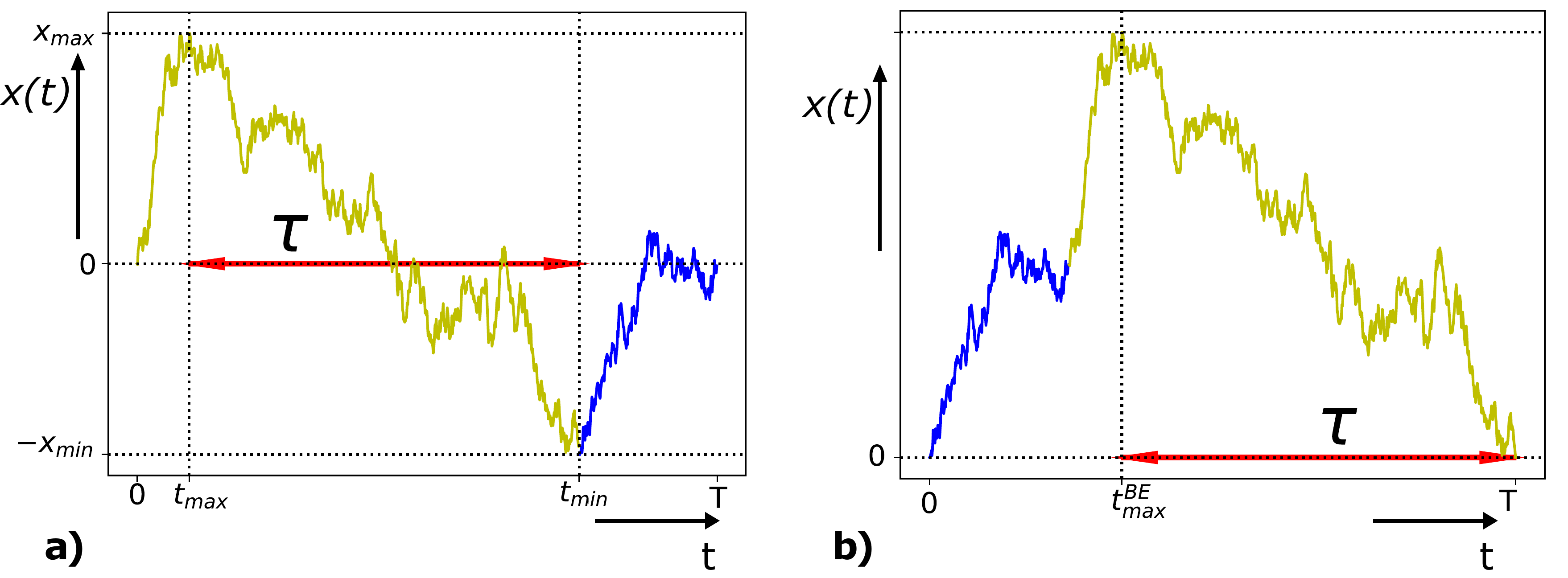}
\caption{Schematic representation of the Vervaat transformation from a Brownian bridge $x(t)$ in panel a) to a Brownian excursion in panel b). On the left panel a), we have a Brownian bridge going from $x(0) = 0$ at time $t=0$ to the final position $x(T) = 0$ at time $t=T$. We first locate the time $t_{\min}$ at which the minimum of the bridge occurs with value $-x_{\min}$. We decompose the trajectory into two parts: the left of $t_{\min}$ (shown in yellow) and the right of $t_{\min}$ (shown in blue). Keeping the blue part of the trajectory fixed, we first slide forward the yellow part of the trajectory by an interval $T$ and glue this yellow part of the trajectory to the right end of the blue part. Next we shift the origin of the space to $-x_{\min}$. After these two transformations, the new trajectory on the right panel b) corresponds to a  Brownian excursion path. Note that the time difference $\tau = t_{\min} - t_{\max}$ in the bridge configuration on the left (indicated by a double-arrowed red line) corresponds exactly to the time at which the maximum of the excursion occurs (measured from the right end of the interval) on the right panel (also shown by a double arrowed red line).}
\label{fig:vervaat} 
\end{figure*}

We can derive the result for the BB in Eq. (\ref{fBB_SM2}) by using an alternative method based on a mapping between a BB trajectory and a Brownian excursion (BE) trajectory -- known as Vervaat construction in probability theory \cite{vervaat79} (see also \cite{majumdar15}). Let us first recall that a BE on the time interval $[0,T]$ is a BB with the additional constraint that the path remains positive at all intermediate times between $0$ and $T$ (for a typical trajectory of BE, see Fig. \ref{fig:vervaat}b). From any BB configuration, one can obtain a BE configuration by sliding and fusing as explained in the caption of Fig. \ref{fig:vervaat}. Additionally, Vervaat proved that the configurations of BE generated by this construction from a BB configuration occur with the correct statistical weight. Clearly, under this mapping, as also explained in the caption of Fig. \ref{fig:vervaat}, the time difference $\tau = t_{\min} - t_{\max}$ for a BB gets mapped onto $t^{\rm BE}_{\max}$ of a BE, measured from the right end of the interval $[0,T]$, where $t^{\rm BE}_{\max}$ denotes the time at which the maximum of a Brownian excursion occurs. This is a random variable, and let us denote its PDF by
\begin{eqnarray} \label{def_PBE}
P_{\rm BE}(\tau|T) = {\rm Prob.}\left(t_{\max}^{\rm BE} = \tau|T\right) \;.
\end{eqnarray}
Note that this mapping is one-to-one only if we fix the value of $t_{\min}$. Thus, focusing on the case where $\tau >0$, i.e. $t_{\min} > t_{\max}$ for BB, the Vervaat construction provides the exact identity
\begin{equation}\label{eq:BB_BE}
P_{\rm BB}(t_{\min}-t_{\max}|t_{\min},T)  = P_{\rm BE}(t_{\min}-t_{\max}|T)   \;,
\end{equation}
where the left-hand side denotes the PDF of $t_{\min}-t_{\max}$ for a BB, conditioned on $t_{\min}$ and on the total time $T$. For the BE, the PDF $P_{\rm BE}(\tau|T)$ was computed exactly in \cite{randon-furling08} and it reads
\begin{eqnarray}\label{PBE_explicit}
P_{\rm BE}(\tau|T)=3\, T^{3/2} \,\sum_{m,n=1}^{\infty} \frac{(-1)^{m+n}m^2n^2 }{\left[m^2 \tau+n^2(T-\tau)\right]^{5/2}} \;.
\end{eqnarray}
The joint PDF of $t_{\max}$ and $t_{\min}$ for a BB can be written as
\begin{eqnarray}\label{eq:joint_BB}
P_{\rm BB}(t_{\max},t_{\min}|T)&=& P_{\rm BB}(t_{\max}-t_{\min}|t_{\min},T)P_{\rm BB}(t_{\min}|T)\nonumber\\
&=&P_{\rm BE}(t_{\min}-t_{\max}|T) P_{\rm BB}(t_{\min}|T)\,,
\end{eqnarray}
where we have used Eq. (\ref{eq:BB_BE}) in going from the first to the second line above. The distribution of the time $t_{\min}$ of the minimum of a BB is uniform over $[0,T]$ \cite{morters10}
\begin{equation}\label{eq:prob_tmin}
P_{\rm BB}(t_{\min}|T)=\frac1T\,,
\end{equation}
Thus, plugging the expressions for $P_{\rm BE}(t_{\min}-t_{\max}|T)$ and $P_{\rm BB}(t_{\min}|T)$, given in Eqs. (\ref{PBE_explicit}) and (\ref{eq:prob_tmin}), into Eq. (\ref{eq:joint_BB}), we obtain
\begin{eqnarray}\label{eq:P_tmax_tmin_BB}
&& P_{\rm BB}(t_{\max},t_{\min}|T)=
3\, \sqrt{T} \,\\ &\times & \sum_{m,n=1}^{\infty} \frac{(-1)^{m+n}m^2n^2 }{\left[m^2 (t_{\min}-t_{\max})+n^2(T-t_{\min}+t_{\max})\right]^{5/2}}\,,\nonumber
\end{eqnarray}
Finally, integrating Eq. (\ref{eq:P_tmax_tmin_BB}) over $t_{\max}$ and $t_{\min}$, keeping $\tau=t_{\min}-t_{\max}$ fixed, we obtain that the PDF of the time $\tau$ between the maximum and the minimum of a BB is given by
\begin{equation}
P(\tau|T)=\frac{1}{T}f_{\rm BB}(\frac{\tau}{T})
\end{equation}
where
\begin{equation} \label{fBB_SM_4}
f_{\rm BB}(y) = 3 (1-|y|) \sum_{m,n=1}^{\infty}\frac{(-1)^{m+n}m^2n^2}{\left[m^2 \,|y|+n^2(1-|y|)\right]^{5/2}} \;.
\end{equation}
Comparing this result (\ref{fBB_SM_4}) with Eq. (\ref{fBB_SM2}) we obtain that the correct normalisation constant in Eq. (\ref{fBB_SM2}) was indeed $B=3$. Moreover, the asymptotic behaviours of $P_{\rm BE}(\tau|T)$ was also computed in \cite{randon-furling08}. As before, due to the $y\to-y$ symmetry it is sufficient to consider only the case $y>0$. Thus, using Eq. (\ref{eq:BB_BE}), one finds that
\begin{eqnarray}\label{summary_asymptotics_BB}
f_{\rm BB}(y) \approx
\begin{cases}
&\frac{\sqrt{2}\,\pi^2}{(1-y)^{\frac{5}{4}}}e^{-\frac{\pi}{\sqrt{1-y}}} \quad \; \quad \quad \quad \; {\rm as} \quad y \to 1 \\
& \\
& \frac{\sqrt{2}\pi^2}{y^{\frac{9}{4}}}e^{-\frac{\pi}{\sqrt{y}}}   \quad \quad \quad \quad\quad \quad {\rm as} \quad  y \to 0^+  \, .
\end{cases}
\end{eqnarray}

\subsection{Moments of $\tau$ for BB}

As in case of BM, the moments of $\tau=t_{\min}-t_{\max}$ can be computed for the BB explicitly. Since $P(\tau|T)$ is
symmetric also for a BB, odd moments vanish and we compute the $k$-th moment of the absolute value $|\tau|$. In this case,
using the scaling form $P(\tau|T)= (1/T)\, f_{BB}(\tau/T)$, we have
\begin{equation}\label{eq:moments_definition}
\langle|\tau|^k\rangle=\int_{-T}^{T}d\tau\,|\tau|^k\,\frac{1}{T}f_{\rm BB}\left(\frac{\tau}{T}\right)=
2T^k\int_{0}^{1}dy\,y^k\,f_{\rm BB}\left(y\right)\,,
\end{equation}
As in the BM case, evaluating the integral on the right hand side using the explicit form of $f_{\rm BB}(y)$ from
Eq. (\ref{fBB_SM_4}) seems cumbersome. We need an integral identity satisfied by $f_{\rm BB}(y)$, just like
Eq. (\ref{eq:int_BM}) for the BM case. Fortunately, such an identity can be derived (see 
Appendix D for the derivation)
\begin{equation}\label{eq:integral_f_bb}
\int_{0}^{1}dy\, \frac{f_{\rm BB}(y)}{\sqrt{1+ u y}}=
\int_{0}^{\infty}dz  \frac{\frac{z}{\sqrt{1+u}}\coth\left(\frac{z}{\sqrt{1+u}}\right)-
1}{\sinh(z)\sinh \left(\frac{z}{\sqrt{1+u}}\right)} \,.
\end{equation}
Again expanding in powers of $u$ on both sides, we can extract the moments. The first four moments read
\begin{eqnarray}\label{eq:moments_BB}
\langle |\tau|\rangle &=&\frac{\pi^2-6}{9}\,T=(0.4299\ldots)\,T\,,\\ \nonumber \langle
\tau^2\rangle &=& \frac{\pi^2-6}{18}\,T^2=(0.2149\ldots)\,T^2\,,\\ \nonumber
\langle |\tau|^3 \rangle &=& \frac{375\pi^2-14\pi^4-1530}{6750}\,T^3=(0.1196\ldots)\,T^3\,,\\
\nonumber \langle \tau^4\rangle &=&\frac{125\pi^2-7\pi^4-390}{2250}\,T^4= (0.0719\ldots)\,T^4\,.\nonumber
\end{eqnarray}
These results are in good agreement with the estimates from numerical simulations.

The second moment $\langle \tau^2\rangle$ can also be derived explicitly using the Veervat construction discussed in
Section \ref{sec:BB} that links a BB to a BE. For a BE, the PDF of $t_{\max}$ has the scaling form
\begin{equation}
P_{\rm BE}(t_{\max}|T) = \frac1T f_{\rm BE}\left(\frac{t_{\max}}{T}\right)\,,
\end{equation}
where $f_{\rm BE}(y)$ is given by \cite{randon-furling08}
\begin{eqnarray} \label{fBE}
f_{\rm BE}(y) = 3 \sum_{m,n=1}^{\infty}\frac{(-1)^{m+n}m^2n^2}{\left(m^2 y+n^2(1-y)\right)^{5/2}} \;.
\end{eqnarray}
Thus the two scaling functions $f_{\rm BB}(y)$ in Eq. (\ref{fBB_SM_4}) and $f_{\rm BE}(y)$ in Eq. (\ref{fBE}) are simply related via
\begin{eqnarray} \label{rel_BB_BE}
f_{\rm BB}(y) = (1-y) f_{\rm BE}(y) \;.
\end{eqnarray}
Consequently, from Eq. (\ref{eq:moments_definition}) using $k=2$ and the relation in Eq. (\ref{rel_BB_BE}) we get
\begin{eqnarray}\label{var_tau_BB2}
\langle \tau^2 \rangle = 2  T^2 \, \Big( \langle y^2\rangle_{\rm BE} - \langle y^3\rangle_{\rm BE} \Big)\,,
\end{eqnarray}
where
\begin{eqnarray}\label{def_mom_BE}
\langle y^m \rangle_{\rm BE} = \int_0^1 dy \, y^m \, f_{\rm BE}(y) \;.
\end{eqnarray}
The first three moments, i.e. $\langle y^m \rangle_{\rm BE}$ for $m=1,2,3$, were computed in Ref. \cite{randon-furling08}
\begin{equation}\label{eq:moments}
\left\langle y\right\rangle _{\rm BE}= \frac{1}{2} ,\textit{        }\textit{        
}\textit{        } \left\langle y^2\right\rangle _{\rm BE}=\frac{15-\pi^2}{18},
\textit{        }\textit{        }\textit{        }
\left\langle y^3\right\rangle _{\rm BE}=1-\frac{\pi^2}{12}.
\end{equation}
Substituting these results in Eq. (\ref{var_tau_BB2}) gives
\begin{equation}
\langle \tau^2\rangle= \frac{\pi^2-6}{18}\,T^2=(0.2149\ldots)\,T^2 \, ,
\label{var_second_BB}
\end{equation}
in perfect agreement with the first derivation in Eq. (\ref{eq:moments_BB}).\\

\noindent {\bf{Covariance of $t_{\min}$ and $t_{\max}$:}} As in the case of BM, the covariance between $t_{\min}$ and
$t_{\max}$ can be computed from the general formula in Eq. (\ref{formula_covariance}), and
using the explicit knowledge of $\langle \tau^2 \rangle$ from Eq. (\ref{eq:moments_BB}). In addition,
we need the first two moments of $t_{\min}$ and $t_{\max}$ for BB. The marginal PDFs of $t_{\min}$ and $t_{\max}$ for BB are 
both uniform over $[0,T]$ \cite{morters10}
\begin{eqnarray}
&&P(t_{\min}|T)=\frac{1}{T} \;, \;\; \quad \; 0\leq t_{\min} \leq T \;, \label{unif_min} \\
&&P(t_{\max}|T)=\frac{1}{T} \;, \; \quad 0\leq t_{\max} \leq T \;. \label{unif_max}
\end{eqnarray}
This gives the first two moments
\begin{eqnarray}
&&\langle t_{\min} \rangle = \langle t_{\max} \rangle = \frac{T}{2} \label{av_tmin_BB} \;, \\
&&\langle t_{\min}^2 \rangle = \langle t_{\max}^2 \rangle = \frac{1}{3}T^2 \;. \label{var_tmin_BB}
\end{eqnarray}
Substituting these results in Eq. (\ref{formula_covariance}) and using $\langle \tau^2\rangle$ from
Eq. (\ref{eq:moments_BB}) we get
\begin{equation} \label{final_cov_BB}
\operatorname{cov}_{\rm BB}(t_{\min},t_{\max})=-\frac{\pi^2-9}{36}T^2 = - (0.0241 \ldots) T^2\;.
\end{equation}
Thus, by comparing Eqs. (\ref{cov}) and (\ref{final_cov_BB}), we see that 
$t_{\min}$ and $t_{\max}$ are more strongly anti-correlated in the BM case than the BB case. \\

\section{Discrete-time random walks} \label{sec:RW}

\begin{figure}
  \centering
\includegraphics[width = 1 \linewidth]{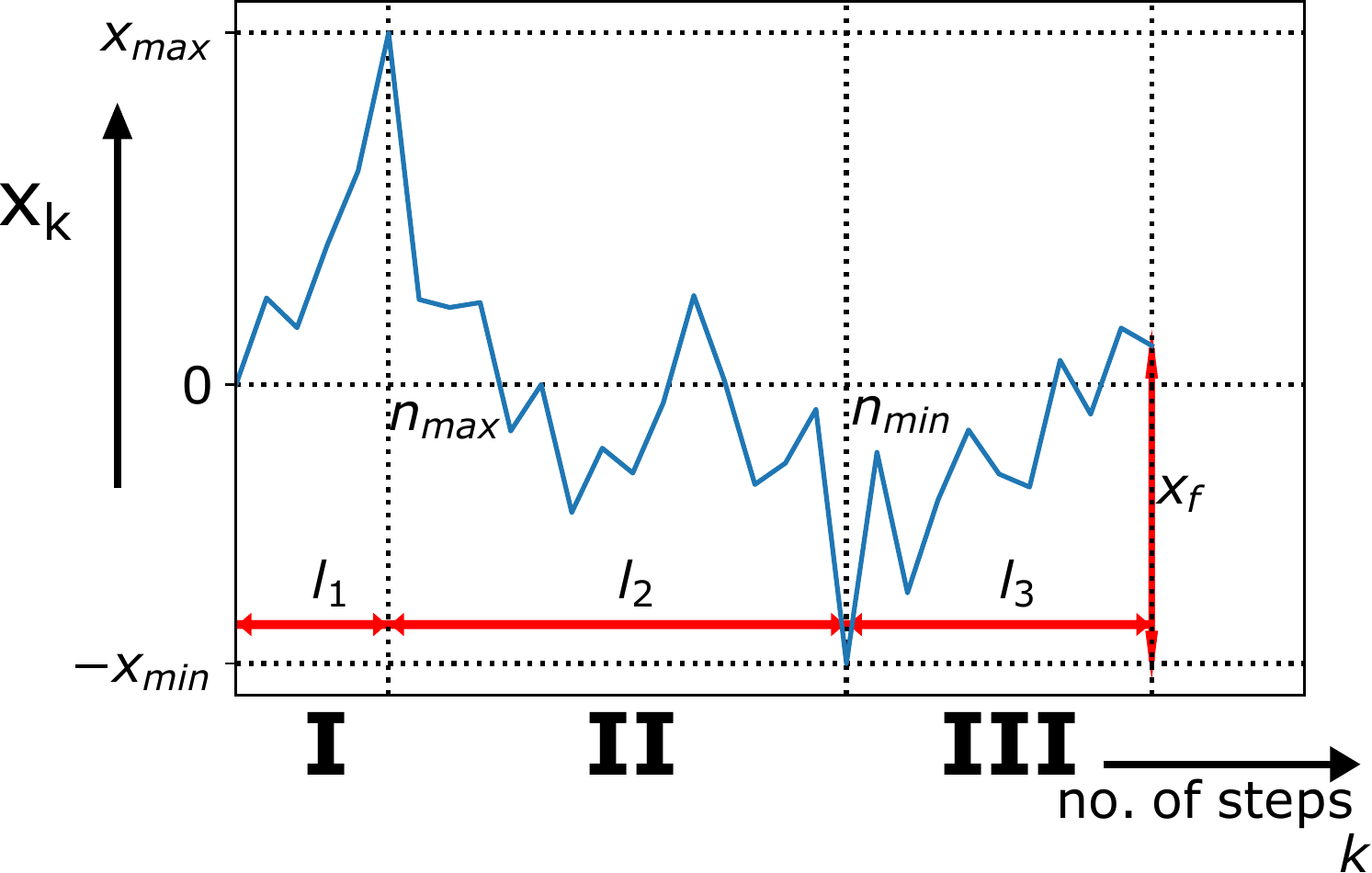}
\caption{A typical trajectory of a discrete-time random walk $x_k$ versus $k$ up to $n$ steps, starting from $x_0=0$. The global maximum $x_{\max}$ occurs at step $n_{\max}$ and the global minimum $-x_{\min} \leq 0$ at step $n_{\min}$. For this trajectory $n_{\min} > n_{\max}$. The final position of the walker at step $n$ is denoted by $x_{\rm f}$, measured with respect to the global minimum $-x_{\min}$. The total duration of $n$ steps has been divided into three segments: $0\leq k \leq n_{\max}$ (I), $n_{\max}\leq k \leq n_{\min}$ (II) and $n_{\min} \leq k \leq n$ (III). The respective durations of these three segments are denoted by $l_1$, $l_2$ and $l_3$.}
\label{fig:exponential} 
\end{figure}
In this section we investigate the time between the maximum and the minimum of discrete-time RWs. Let us consider a time series of $n$ steps generated by the positions of a random walker evolving via the Markov jump process
\begin{eqnarray}
x_{k} = x_{k-1}+\eta_k \;, \label{def_RW}
\end{eqnarray}
starting from $x_0=0$, where $\eta_k$'s are IID random variables, each drawn from a symmetric PDF $p(\eta)$ (for a typical realisation see Fig. \ref{fig:exponential}). Similarly to the continuous-time case described above, we can define the discrete time at which the maximum (minimum) value is reached as $n_{\max}$ ($n_{\min}$), as in Fig. \ref{fig:exponential}. Note that $\tau =n_{\min}-n_{\max}$ is now an integer and that  $\tau \in [-n,n]$. Remarkably, under the additional hypothesis that the distribution $p(\eta)$ is continuous, the probability distribution of $n_{\max}$ is known to be universal, i.e. independent of the PDF $p(\eta)$ for all values of $n$, and not only asymptotically for large $n$. Indeed, Sparre Andersen showed that the probability distribution of $n_{\max}$, given the total number of steps $n$, is~\cite{SA53}
\begin{equation}\label{eq:sa_max}
P(n_{\max}|n)=\binom{2 n_{\max}}{n_{\max}}\,
\binom{2 (n-n_{\max})}{(n-n_{\max})}\,2^{-2n}\,.
\end{equation}
Note that Eq. (\ref{eq:sa_max}) is exact for any $n$ and $n_{\max}\in[0,n]$.
By the symmetry of the jump distribution $p(\eta)$, the time of the minimum $n_{\min}$ is also distributed as (\ref{eq:sa_max}). Note, however, that Eq. (\ref{eq:sa_max}) holds only for continuous jump distributions, thus it is not valid for discrete-space RW, e.g. for lattice walks, which is also discussed below. Notably, using Stirling's formula, in the large $n$ limit the Eq. (\ref{eq:sa_max}) converges to the PDF of $t_{\max}$ for BM given in Eq. (\ref{levy}), with $t_{\max}$ and $T$ replaced by $n_{\max}$ and $n$.
One may ask whether this universality extends also to the time between maximum and minimum, i.e. to $\tau=n_{\min}-n_{\max}$.
In the case of finite jump variance $\sigma^2=\int_{-\infty}^{\infty}d\eta\,\eta^2\,p(\eta)$, the Central Limit Theorem states that the random walk converges, when $n\to\infty$ to a Brownian motion. Thus, one may expect that, in the large $n$ limit, the probability distribution of any observable of the random walk (e.g. $n_{\max}$, $\tau$, $\ldots$) converges to its Brownian counterpart. However, directly verifying this convergence for a generic jump distribution is usually challenging. In this section, we demonstrate that the distribution of $\tau$ converges to the Brownian result (\ref{eq:f_bm}) for two particular jump distributions. Note however that the universal formula for the probability distribution of $n_{\max}$ in Eq. (\ref{eq:sa_max}) is valid even when the variance is not well-defined. Thus a natural question is: ``Does the universality of $\tau$ also extend to jump distributions with a divergent variance?''
To proceed, it is useful to consider separately the cases of finite and divergent jump variance.

\subsection{Finite jump variance} 

\begin{figure}[t]    
    \includegraphics[width=\linewidth]{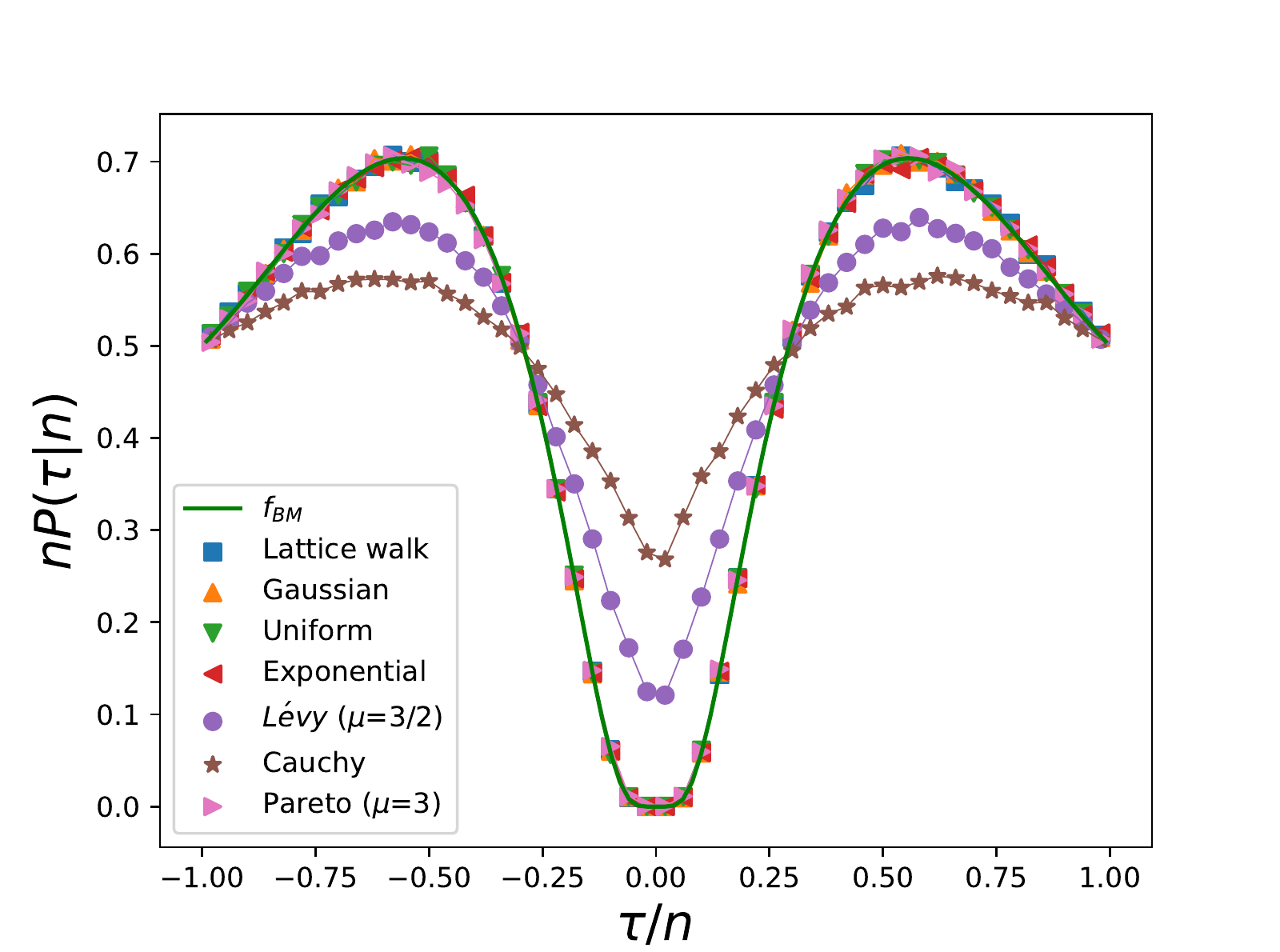} 
    \caption{The distribution $n P(\tau|n)$ as a function of $\tau/n$ for discrete-time random walks (RWs) for different jump distributions. Jump distributions with a finite variance (lattice walks, Gaussian, uniform, double-exponential and Pareto) collapse onto the scaling function $f_{\rm BM}(y)$ shown by the solid (green) line. For distributions with divergent variance, i.e. L\'evy flights with index $\mu = 3/2$ and $\mu = 1$ (Cauchy distribution), the scaling function $f_{\mu}(y)$ depends on $\mu$ (except at the endpoints $y = \pm 1$ where $f_{\mu}(\pm 1) = 1/2$ seems to be universal for all $0<\mu\leq 2$). The empirical curves are obtained by simulating $10^7$ RWs of $n=10^5$ steps for each jump distribution.}
    \label{fig:numeric_rw}
\end{figure}

As stated above, for all jump distributions with a finite variance, one may expect that, for large $n$, the corresponding PDF of the time difference $\tau = t_{\min} - t_{\max}$ would converge for large $n$ to the distribution of $\tau$ for BM. In other words, for $n\to\infty$ we expect
\begin{equation}\label{eq:scaling}
P(\tau | n) \underset{n \to \infty}{\longrightarrow} \frac{1}{n} f_{\rm BM} \left( \frac{\tau}{n}\right) \;.
\end{equation}
In this section, we first verify this universality analytically for the double-exponential distribution $p(\eta)=(1/2)e^{-|\eta|}$. Then, we show that the universality of the distribution of $\tau$ is also valid for discrete-space distributions by computing the asymptotic distribution of $\tau$ in the case of lattice walks.
However, apart from the two special cases discussed below, it turns out that the exact computation of $P(\tau|n)$ is very hard for a generic $p(\eta)$. Therefore, we verify  (\ref{eq:scaling}) numerically for other jump distributions with a finite variance (see Fig. \ref{fig:numeric_rw}). In the case of random walk bridges, i.e. random walks with the additional constraint that they have to go back to the origin at the final step, a result equivalent to the one in Eq. (\ref{eq:scaling}) can be derived. Indeed, in Appendix \ref{app:rw_bridges} we show that, in the case of random walk bridges with double-exponential jumps, in the limit of large $n$
\begin{equation}
P(\tau | n) \underset{n \to \infty}{\longrightarrow} \frac{1}{n} f_{\rm BB} \left( \frac{\tau}{n}\right) \,,
\end{equation}
where $f_{\rm BB}(y)$ is the scaling function in Eq. (\ref{eq:f_bb}).

\subsubsection{Exact result for the double-exponential distribution}
\label{double_exp_jump}

In this section we compute the probability distribution of $\tau$ in the case of RWs with double-exponential jump distribution: $p(\eta)=(1/2)e^{-|\eta|}$. 
The main ingredient needed to compute the probability distribution of $\tau$ is the restricted Green's function $G\left(x,l|M\right)$, defined as the probability that the walker goes from the origin to position $x$ in $l$ steps, while remaining always in the space interval $[0,M]$. Using the Markov property of the process with an arbitrary jump distribution $p(\eta)$, we can easily write down a recursion relation for $G\left(x,l|M\right)$
\begin{equation}
\label{recursive}
G\left(x,l | M \right)=\int_{0}^{M} dx' G\left(x',l-1|M \right)p(\eta = x-x') \;,
\end{equation}
valid for all $l \geq 1$ and starting from the initial condition $G(x,0|M) = \delta(x)$. This equation can be understood as follows. Let the walker arrive at $x' \in [0,M]$ at step $l-1$ (without leaving the box $[0,M]$ up to $l-1$ steps) and then it makes a jump from $x'$ to $x$ at the $l^{\rm th}$ step. 
The probability of this jump is simply $p(\eta = x-x')$. Remarkably, this simple equation (\ref{recursive}) cannot be solved
exactly for arbitrary jump distribution $p(\eta)$. The reason is because the integral is defined over a finite range $[0,M]$. In the semi-infinite case $M \to \infty$, this reduces to Wiener-Hopf equation which can be solved for arbitrary symmetric and continuous
$p(\eta)$ \cite{majumdar10}. Although the solution in this case is not fully explicit for $G\left(x,l | M \to \infty \right)$, one can obtain an explicit expression for
its generating function in terms of the Fourier transform of the jump distribution $p(\eta)$. This is known as the Ivanov formula \cite{ivanov} (see also the Appendix A of Ref. \cite{mounaix} for a transparent derivation). Unfortunately, for finite $M$, no exact solution is known for arbitrary $p(\eta)$. However, for the double-exponential jump distribution, we can obtain an exact solution of Eq. (\ref{recursive}) for finite $l$, as shown below. \\ To proceed, we first consider the generating function
\begin{eqnarray}\label{eq:generating_G}
\tilde{G}(x,s|M)=\sum_{l=1}^{\infty}G(x,l|M)\,s^l.
\end{eqnarray}
By multiplying Eq. (\ref{recursive}) by $s^l$, summing over $l$ and using the initial condition $G(x,0|M) = \delta(x)$, we get
\begin{equation}\label{g_tilde}
\tilde{G}\left(x,s|M\right)=s\int_{0}^{M} dx' \tilde{G}\left(x',s|M \right)\, p(x-x')+s\,p(x).
\end{equation}
The double-exponential distribution $p(\eta)=(1/2)e^{-|\eta|}$ has the special property that if we differentiate it twice, it satisfies a
simple differential equation  
\begin{equation} \label{exp_relation}
p''(x)=p(x)-\delta(x) \;.
\end{equation}
Using this relation, we can then reduce the integral equation in (\ref{g_tilde}) into a differential equation, which then is easier to solve. Differentiating Eq. (\ref{g_tilde}) twice with respect to $x$, and using the identity (\ref{exp_relation}), we get
\begin{equation} \label{differential}
\frac{\partial^2 \tilde{G}\left(x,s|M \right) }{\partial x^2}= (1-s)\tilde{G}\left(x,s|M \right)-s\,\delta(x),
\end{equation}
for $0\leq x\leq M$. For $x>0$, the $\delta$-function in (\ref{differential}) disappears and the general 
solution reads simply
\begin{equation} \label{solution_general}
\tilde{G}\left(x,s|M\right)=A(s,M)\,e^{-\sqrt{1-s}\,x}+B(s,M)\,e^{\sqrt{1-s}\,x} \;,
\end{equation}
where $A(s,M)$ and $B(s,M)$ are two arbitrary constants. In going from the integral (\ref{g_tilde}) to the differential (\ref{differential}) equation, 
we have taken derivatives and hence one has to ensure that the solution of the differential equation also satisfies the integral equation. This condition fixes these unknown constants $A(s,M)$ and $B(s,M)$. Indeed, by substituting Eq. (\ref{solution_general}) into the integral equation (\ref{g_tilde}), it is straightforward to check that Eq. (\ref{g_tilde}) is verified only if
\begin{eqnarray}
\hspace*{-0.5cm}&& A(s,M)=\frac{1-\sqrt{1-s}}{1-\left(\frac{1-\sqrt{1-s}}{1+\sqrt{1-s}}\right)^2 \, e^{-2\sqrt{1-s} \,M}},  \label{A} \\
\hspace*{-0.5cm}&& B(s,M)=-A(s,M)\frac{1-\sqrt{1-s}}{1+\sqrt{1-s}} \, e^{-2\sqrt{1-s}\,M} \;. \label{B}
\end{eqnarray}
Hence, the final exact solution reads
\begin{eqnarray}\label{g_solution}
&& \tilde{G}\left(x,s|M \right)= A(s,M)\\ &\times & \left[e^{-\sqrt{1-s}\,x}-\frac{1-\sqrt{1-s}}{1+\sqrt{1-s}}e^{-\sqrt{1-s}\,(2M-x)}\right] \;\nonumber,
\end{eqnarray}
with the amplitude $A(s,M)$ given in Eq. (\ref{A}). 

We can now proceed with the computation of the probability distribution of $\tau$. As in the case of BM, our strategy will be to first compute the grand joint PDF $P(x_{\min}, x_{\max}, n_{\min}, n_{\max} | n)$ of the four random variables $x_{\min}$, $ x_{\max}$, $n_{\min}$ and $n_{\max}$ and then integrate out $x_{\min}$ and $x_{\max}$ to obtain the joint distribution $P(n_{\min}, n_{\max} | n)$. To compute this grand PDF, we divide the interval $[0,n]$ into three segments of lengths $l_1=n_{\max}$, $l_2=n_{\min}-n_{\max}$ and $l_3=n-n_{\min}$ (see Fig. \ref{fig:exponential}). Here again we consider the case $n_{\max} < n_{\min}$ (the complementary case can be then obtained using the symmetry of the process). The grand PDF can then be written as the product of 
the probabilities $P_{\rm I}$, $P_{\rm II}$ and $P_{\rm III}$ of the three independent segments: $0\leq k \leq n_{\max}$ (${\rm I}$), $n_{\max}\leq k \leq n_{\min}$ (${\rm II}$) and $n_{\min} \leq k \leq n=l_1+l_2+l_3$ (${\rm III}$). To proceed, we first notice that the probability of each segment can be expressed in terms of the restricted Green's function $G\left(x,n| M \right)$. In order to do this, it is crucial to use the fact that the jump distribution $p(\eta)$ is symmetric, which makes the walk reflection-symmetric around the origin. This is best explained with the help of the Fig. \ref{fig:exponential}. Let us first set $M = x_{\min} + x_{\max}$. In segment I, the trajectory has to start at the origin and reach the level $x_{\max}$ at step $l_1$, while staying in the box $[-x_{\min},x_{\max}]$. Using the invariance under the reflection $x \to -x$, followed by a shift of the origin to the level $x_{\max}$, this probability is just 
\begin{eqnarray}\label{PI}
P_{\rm I}=G\left(x_{\max},l_1 | M\right) \;.
\end{eqnarray}
For the second segment, the trajectory starts at $x_{\max}$ and ends at $- x_{\min}$ at step $l_2$, while staying inside the box $[-x_{\min}, x_{\max}]$ (see Fig. \ref{fig:exponential}). Using a similar argument as in the previous case, one gets 
\begin{eqnarray}\label{PII}
P_{\rm II}=G\left(M,l_2 | M \right) \;.
\end{eqnarray}
For the third segment, the trajectory starts at $-x_{\min}$ and stays inside the box $[-x_{\min}, x_{\max}]$ up to $l_3$ steps. Thus this probability is given by  
\begin{eqnarray}\label{PIII}
P_{\rm III}=\int_0^M G\left(x_{\rm f},l_3 | M \right) dx_{\rm f} \;,
\end{eqnarray}
where $x_{\rm f}$ is the final position of the trajectory measured with respect to $-x_{\min}$ (see Fig. \ref{fig:exponential}) and we have integrated over the final position $x_{\rm f}$. The grand joint PDF $P(x_{\min}, x_{\max}, n_{\min}, n_{\max} | n)$ is given by the product of the three segments 
\begin{eqnarray}\label{product}
&& P(x_{\min}, x_{\max}, n_{\min}, n_{\max} | n) = P_{\rm I}  P_{\rm II}  P_{\rm III} \\ &=& G\left(x_{\max},l_1 | M\right)\, G\left(M,l_2 | M \right) \int_0^M G\left(x_{\rm f},l_3 | M \right) dx_{\rm f}  \nonumber,
\end{eqnarray}
where we recall that 
\begin{eqnarray}\label{def_l}
l_1 = n_{\max} \,, \quad l_2 = n_{\min} - n_{\max} \, , \quad l_3 = n-n_{\min} \, ,
\end{eqnarray}
and that $M = x_{\max} + x_{\min}$.
Note that, unlike the BM, for the discrete-time RW, we do not need to put a cut-off $\epsilon$. It is now useful to express this grand PDF in terms of the intervals $l_1, l_2$ and $l_3$ (see Fig. \ref{fig:exponential}). Hence, we write 
\begin{equation}\label{PDF_l1l2l3}
P(x_{\min}, x_{\max}, n_{\min}, n_{\max} | n)\equiv P(x_{\min}, x_{\max}, l_1, l_2, l_3) ,
\end{equation}  
where $l_1, l_2$ and $l_3$ are given in Eq. (\ref{def_l}). It is convenient to take the generating function of the grand PDF $P(x_{\min}, x_{\max}, n_{\min}, n_{\max} | n)$, so we multiply Eq. (\ref{product}) by $s_1^{l_1}s_2^{l_2}s_3^{l_3}$ and sum over $l_1$, $l_2$ and $l_3$ to obtain
\begin{eqnarray}\label{laplace}
&&\sum_{l_1,l_2,l_3=1}^{\infty}
P\left(x_{\min},x_{\max},l_1,l_2,l_3\right)s_1^{l_1}s_2^{l_2}s_3^{l_3}\\ &=&\tilde{G}\left(x_{\max},s_1|M \right)\tilde{G}\left(M,s_2|M \right)\, \int_0^M \, dx_{\rm f}\,  \tilde{G}\left(x_{\rm f},s_3|M \right) \,,\nonumber
\end{eqnarray}
where $\tilde{G}(x,s|M)$ is given in Eq. (\ref{g_solution}).
To obtain the marginal joint distribution of $l_1$, $l_2$, and $l_3$, we still need to integrate over $x_{\min}$ and $x_{\max}$ in Eq. (\ref{laplace}). Let us first define 
\begin{eqnarray}\label{Pl1l2l3}
&& P(l_1, l_2,l_3) \\&=& \int_{0}^\infty dx_{\min} \, \int_{0}^\infty dx_{\max} \,  P\left(x_{\min},x_{\max},l_1,l_2,l_3\right) \;\nonumber.
\end{eqnarray}
We can perform this double integral by making a change of variables $(x_{\min}, x_{\max}) \to (x_{\max}, M = x_{\max} + x_{\min})$. Performing the double integral yields
\begin{eqnarray}\label{eq:gen_fun_1}
&&\sum_{l_1,l_2,l_3=1}^{\infty}
P\left(l_1,l_2,l_3\right)s_1^{l_1}s_2^{l_2}s_3^{l_3}\\&=& \int_{0}^{\infty}dM\left[\int_{0}^{M}dx_{\max} \tilde{G}\left(x_{\max},s_1|M \right)\right]\nonumber\\\nonumber &\times & \tilde{G}\left(M,s_2|M \right)\left[\int_{0}^{M}dx_{\rm f} \tilde{G}\left(x_{\rm f},s_3|M \right)\right]\\\nonumber &=&\int_{0}^{\infty}dM\,I\left(M,s_1\right)\,\tilde{G}\left(M,s_2|M \right)\,I\left(M,s_3\right), \nonumber
\end{eqnarray}
where we have defined
\begin{eqnarray}\label{eq:def_I}
I\left(M,s\right)=\int_{0}^{M}dx \, \tilde{G}\left(x,s|M \right)\,.
\end{eqnarray}
We want to compute the PDF $P(\tau|n)$ of $\tau = n_{\min} - n_{\max}$, for a given total number of steps $n$. We can express $P(\tau|n)$ in terms of the joint PDF $P\left(l_1,l_2,l_3\right)$ computed above as follows
\begin{equation}\label{relation1}
P(\tau|n) = \sum_{l_1, l_3=1}^\infty P(l_1, l_2 = \tau, l_3) \, \delta(l_1+\tau +l_3-n) \;.
\end{equation}
Taking the double generating function of this expression (\ref{relation1}) gives
\begin{equation}\label{relation2}
 \sum_{n=1}^{\infty}\sum_{\tau=1}^{n}P(\tau|n) s_2^\tau \, s^n= \sum_{l_1,\tau,l_3=1}^\infty P(l_1,  \tau, l_3) s^{l_1} (s\, s_2)^{\tau}\, s^{l_3}\, .
\end{equation}
Notice that the right-hand side of Eq. (\ref{relation2}) can be read off Eq. (\ref{eq:gen_fun_1}) by setting $s_1 \to s$, $s_2 \to s\, s_2$ and $s_3 \to s$.  This yields
\begin{equation}\label{eq:relation3}
\sum_{\tau, n} P(\tau|n) s_2^\tau \, s^n =\int_{0}^{\infty}dM\,\tilde{G}\left(M,s\,s_2|M \right)I\left(M,s\right)^2.
\end{equation}
To use the relation (\ref{eq:relation3}), we also need to compute $I\left(M,s\right)$. Substituting Eq. (\ref{g_solution}) in Eq. (\ref{eq:def_I}), we get
\begin{eqnarray}\label{eq:I_explicit}
&& I\left(M,s\right)=\int_{0}^{M}dx \, \tilde{G}\left(x,s|M \right)
=\frac{A(s,M)}{\sqrt{1-s}}\\&\times & \left[1-\frac{2}{1+\sqrt{1-s}}
e^{-\sqrt{1-s}M}  + \frac{1-\sqrt{1-s}}{1+\sqrt{1-s}}e^{-2\sqrt{1-s}M}\right]\,,\nonumber
\end{eqnarray}
where $A(s,M)$ is given in Eq. (\ref{A}). Using Eqs. (\ref{g_solution}) and (\ref{eq:I_explicit}), we can now write an explicit expression for the right-hand side of Eq. (\ref{eq:relation3}). This yields

\begin{eqnarray}\label{eq:relation4}
&& \sum_{ n,\tau} P(\tau|n) s_2^\tau \, s^n = \int_{0}^{\infty}dM\,A(s\, s_2,M) e^{-\sqrt{1-ss_2}M} \nonumber \\ & \times & \Big( 1-\frac{2}{1+\sqrt{1-s}}
e^{-\sqrt{1-s}M}  + \frac{1-\sqrt{1-s}}{1+\sqrt{1-s}}e^{-2\sqrt{1-s}M} \Big)^2\,\nonumber \\ &\times &\Big(\frac{2\sqrt{1-ss_2}}{1+\sqrt{1-ss_2}}\Big)\Big(\frac{A(s,M)}{\sqrt{1-s}}\Big)^2 
 ,
\end{eqnarray}
where $A(s,M)$ is given in Eq. (\ref{A}).
We now want to extract the asymptotic behavior of $P\left(\tau |n\right)$ for large $n$. In this limit, we expect that $P(\tau | n)$ approaches a scaling form 
\begin{eqnarray}\label{scaling_form}
P(\tau | n) \underset{n \to \infty}{\longrightarrow} \frac{1}{n} f_{\rm exp} \left( \frac{\tau}{n}\right) \;.
\end{eqnarray}
Our goal now is to extract this scaling function $f_{\rm exp}(y)$ from the exact formula (\ref{eq:relation4}) and show that $f_{\rm exp}(y) = f_{\rm BM}(y)$ given in Eq. (\ref{eq:f_bm}). Since we are interested in the scaling limit $\tau, n \to \infty$ keeping the ratio $y = \tau/n$ fixed, we also need to investigate the generating function in Eq. (\ref{eq:relation4}) in the corresponding scaling limit. It is convenient to first parametrise the Laplace variables as $s=e^{-\lambda}$ and  $s_2=e^{-\lambda_2}$. In these new variables, the scaling limit corresponds to  
$\lambda,\lambda_2\rightarrow 0 $ with $\lambda_2/\lambda$ fixed. In this limit, the double sum in the left-hand side of Eq. (\ref{eq:relation4}) can be replaced by a double integral. Thus, taking the limit $\lambda,\lambda_2\rightarrow 0$ keeping the ratio $\lambda_2/\lambda$ fixed on both sides of Eq. (\ref{eq:relation4}), we get 
\begin{eqnarray} \label{scaling1}
&& \int_{0}^{\infty}dn \int_{0}^{n}d \tau \, P\left(\tau | n \right)\,e^{-\lambda_2 \tau}e^{-\lambda \, n} \approx
\frac{2\sqrt{\lambda+\lambda_2}}{\lambda}\nonumber \\ &\times &\int_{0}^{\infty}dM  
\frac{e^{-\sqrt{\lambda+\lambda_2} M}\left(1-e^{-\sqrt{\lambda}M}\right)^2}{\left(1+e^{-\sqrt{\lambda}M}\right)^2\left(1-e^{-2\sqrt{\lambda+\lambda_2}M}\right)}.
\end{eqnarray}
Rescaling $z=\sqrt{\lambda+\lambda_2}\,M$ in the integral on the right-hand side leads to
\begin{eqnarray}\label{integral1}
&& \int_{0}^{\infty}dn\int_{0}^{n}d \tau P\left(\tau | n\right)e^{-\lambda_2 \tau}e^{-\lambda n}\\ &\approx &
\frac{1}{\lambda} \int_{0}^{\infty}dz  
\frac{\tanh^2\left(\sqrt{\frac{\lambda}{\lambda+\lambda_2}}\frac{z}{2}\right)}{\sinh(z)}\,.\nonumber
\end{eqnarray}
Substituting the scaling form (\ref{scaling_form}) on the left-hand side of Eq. (\ref{integral1}) gives 
\begin{eqnarray} \label{integral1bis}
&&\int_{0}^{\infty}dn\int_{0}^{n}d \tau \frac{1}{n}f_{\rm exp} \left(\frac{\tau}{n}\right)\, e^{-\lambda_2 \tau}e^{-\lambda n}\\&=&
\int_{0}^{\infty}dn\int_{0}^{1}dy\, f_{\rm exp}(y)\,e^{-\lambda_2 y n}e^{-\lambda n}=\int_{0}^{1}dy\frac{f_{\rm exp}(y)}{\lambda+\lambda_2 y}\,. \nonumber
\end{eqnarray}
Comparing this left-hand side (\ref{integral1bis}) with the right-hand side of Eq. (\ref{integral1}), with $u=\frac{\lambda_2}{\lambda}$ fixed, we get the identity
\begin{equation}\label{integral2}
\int_{0}^{1}dy\frac{f_{\exp}(y)}{1+uy}=\int_{0}^{\infty}dz \frac{1}{\sinh(z)}\tanh^2\left(\frac{z}{2\sqrt{1+u}}\right).
\end{equation}
This representation of $f_{\rm exp}(y)$ in Eq. (\ref{integral2}) turns out to be useful to compute the moments of $\tau$ explicitly (see Section \ref{sec:moments}).
The next step is to invert this integral equation (\ref{integral2}) to obtain $f_{\rm exp}(y)$ explicitly. For this, it is convenient to first rewrite Eq. (\ref{integral2}) in terms of the variables $u=-\frac{1}{w}$ on the left-hand side and $t=\frac{z}{2\sqrt{1+u}}$ on the right-hand side. This gives
\begin{equation}\label{stieltjes}
\int_{0}^{1}dy\frac{f_{\rm exp}(y)}{w-y}=\frac{2}{w}\sqrt{1-\frac{1}{w}}\int_{0}^{\infty}dt \frac{\tanh^2(t)}{\sinh \left(2t\sqrt{1-\frac{1}{w}}\right)} \;.
\end{equation}
We now recognise the left-hand side of Eq. (\ref{stieltjes}) as the Stieltjes transform of the function $f_{\rm exp}(y)$. A Stieltjes transform of this type can be inverted using the so-called Sochocki-Plemelj formula (see for instance the book \cite{mushk_book}). Using this inversion formula we get (see Appendix \ref{app:integral} for the details of the computation):
\begin{equation}\label{eq:scaling_fexp}
f_{\rm exp}(y)=\frac{1}{y}\sum_{n=1}^{\infty}(-1)^{n-1}\tanh^2\left(\frac{n\pi}{2}\sqrt{\frac{y}{1-y}}\right) \,.
\end{equation}
This result above has been derived assuming $\tau = n_{\min} - n_{\max} > 0$, i.e., when the minimum occurs after the maximum. In the complementary case $\tau < 0$, i.e. when the maximum occurs after the minimum, it is clear that $P(\tau|n) = P(-\tau|n)$ and this follows simply from the $x\to-x$ symmetry of the process. Hence, we get, for $\tau \in [-n, n]$, and in the scaling limit $\tau, n \to \infty$ keeping the ratio $y = \tau/n$ fixed
\begin{eqnarray}\label{scaling_form2}
P(\tau | n) \underset{n \to \infty}{\longrightarrow} \frac{1}{n} f_{\rm exp} \left( \frac{\tau}{n}\right) \;,
\end{eqnarray}
where the scaling function $f_{\rm exp}(y)$ is given exactly by
\begin{equation}\label{scaling_fexp2}
f_{\rm exp}(y)=\frac{1}{|y|}\sum_{n=1}^{\infty}(-1)^{n-1}\tanh^2\left(\frac{n\pi}{2}\sqrt{\frac{|y|}{1-|y|}}\right) \,.
\end{equation}
Comparing with the Brownian case in Eq. (\ref{eq:f_bm}), we see that $f_{\rm exp}(y) = f_{\rm BM}(y)$. This exact computation for the double-exponential jump distribution confirms explicitly the expectation based on the Central Limit Theorem. The asymptotics of $f_{\rm exp}(y)$ is thus given in Eq. (\ref{summary_asymptotics}). In particular, in the limit $y\to 1$, we get that $f(y)\to 1/2$. This limit value can be also computed directly for the double-exponential distribution $p(\eta)=(1/2)e^{-|\eta|}$ (see Section \ref{sec:universal}). This result is also confirmed by numerical simulations (see Fig. \ref{fig:numeric_rw}). Moreover, the fact that $f_{\rm exp}(y)=f_{\rm BM}(y)$ implies that Eq. (\ref{integral2}) is also satisfied by the Brownian scaling function $f_{\rm BM}(y)$. This thus provides the derivation of the integral identity announced in Eq. (\ref{eq:int_BM}). 

\subsubsection{Exact result for lattice walks}

We now consider a one-dimensional unbiased lattice walk. This corresponds to the discrete PDF 
\begin{eqnarray}\label{eq:pdf_lattice}
p(\eta)=\frac{1}{2}\left(\delta\left(\eta-1\right)+\delta\left(\eta+1\right)\right).
\end{eqnarray}
At each step $k$, the position $x_k$ of the walker is increased or decreased by $1$ with equal probability. Contrary to the BM and double-exponential case, here we need to be careful when defining the times $n_{\max}$ and $n_{\min}$. Indeed, for such discrete-space walks there is a finite probability that the global maximum or the global minimum are not unique. Therefore, for simplicity we define $n_{\max}$ ($n_{\min}$) as the time at which the maximal (minimal) value is reached \emph{for the first time}. Even if this choice may seem arbitrary, we expect that, in the large $n$ limit, the final result will be independent of which particular global maximum (minimum) we consider.\\
As before, the main ingredient to compute the distribution of $\tau=n_{\min}-n_{\max}$ is the restricted Green's function $G(x,l|M)$, defined as the probability that the walker goes from the origin to position $x$ in exactly $l$ steps, without leaving the space interval $[0,M]$. Note that now $x$ and $M$ are integers. One can easily write a recursion relation for $G(x,l|M)$:
\begin{equation}\label{eq:recursive_G}
 G(x,l|M) = \frac{1}{2}G(x+1,l-1|M)+\frac{1}{2}G(x-1,l-1|M) \;,
\end{equation}
valid for $l\geq 1$. Eq. (\ref{eq:recursive_G}) means that if the walker is at position $x\in(0,M)$ at time $l$, it either was at position $x-1$ at time $l-1$ and then, with probability $1/2$, it jumped up, or it was at position $x+1$ at time $l-1$ and then, with probability $1/2$, it jumped down. Note that this Eq. (\ref{eq:recursive_G}) can be also obtained substituting the expression for $p(\eta)$ given in Eq. (\ref{eq:pdf_lattice}) into Eq. (\ref{recursive}). One needs to impose appropriate boundary conditions. Since we are forcing the walker to remain always in the interval $[0,M]$, we impose:
\begin{eqnarray}\label{eq:boundary_conditions}
G(-1,l|M)=0\,,\quad G(M+1,l|M)=0\,.
\end{eqnarray}
We recall that $G(x,l|M)$ is by definition the probability that the walker goes from the origin to position $x$ in $l$ steps, while remaining inside the interval $[0,M]$. Thus, the initial condition is
 \begin{eqnarray}\label{eq:initial_condition}
G(x,0|M)=\delta_{x,\,0}\,,
\end{eqnarray}
where $\delta_{a,\,b}$ is the Kronecker delta function: $\delta_{a,\,b}=1$ if $a=b$ and $\delta_{a,\,b}=0$ otherwise.
To solve Eq. (\ref{eq:recursive_G}) we first multiply both sides by $s^l$ and sum over $l\geq 1$. This yields
\begin{eqnarray}\label{eq:recursive_G2}
&& \tilde{G}(x,s|M)\\&=& \delta_{x\,,0}+\frac{s}{2}\left(\tilde{G}(x+1,s|M)+\tilde{G}(x-1,s|M)\right)\,,\nonumber
\end{eqnarray}
where the generating function $\tilde{G}(x,s|M)$ is defined in Eq. (\ref{eq:generating_G}).
The boundary conditions for $\tilde{G}(x,s|M)$ are
\begin{eqnarray}\label{eq:boundary_conditions2}
\tilde{G}(-1,s|M)=0\,,\quad \tilde{G}(M+1,s|M)=0\,.
\end{eqnarray}
Note that the recursion (\ref{eq:recursive_G2}) is non-homogeneous due to the term $\delta_{x\,,0}$. However, one can easily include this term in the lower boundary condition. In this way, we obtain the homogeneous relation
\begin{equation}\label{eq:recursive_G3}
\tilde{G}(x,s|M)=\frac{s}{2}\left(\tilde{G}(x+1,s|M)+\tilde{G}(x-1,s|M)\right),
\end{equation}
with modified boundary conditions
\begin{eqnarray}\label{eq:boundary_conditions3}
\tilde{G}(-1,s|M)=\frac{2}{s}\,,\quad \tilde{G}(M+1,s|M)=0\,.
\end{eqnarray}
This homogeneous recursion relation can now be solved for $x\in [-1,M+1]$, yielding, after a few steps of algebra
\begin{eqnarray}\label{eq:solution_G_tilde}
\tilde{G}(x,s|M)&=&\frac{2}{s}\Big(\frac{w(s)^{x+1}}{1-w(s)^{2(M+2)}}\\ &+& \frac{w(s)^{-(x+1)}}{1-w(s)^{-2(M+2)}}\Big)\,,\nonumber
\end{eqnarray}
where 
\begin{eqnarray}\label{eq:definition_w}
w(s)=\frac{1}{s}\left(1-\sqrt{1-s^2}\right).
\end{eqnarray}
Since the global maximum and the global minimum are in general not unique, we need to use a slightly different method with respect to the one presented at the beginning of this section. Indeed, we need to include the information that $n_{\max}$ and $n_{\min}$ are the times at which the global maximum and the global minimum are attained for the first time. Hence, we need to impose that $x_k < x_{\max}$ for $k\leq n_{\max}-1$ and that $x_k > x_{\min}$ for $k\leq n_{\min}-1$. As we will see, this additional condition slightly modifies the procedure described above. Considering the case $n_{\max}<n_{\min}$ (the complementary case $n_{\min}<n_{\max}$ can be studied analogously), we can factorize the grand probability distribution $P(x_{\min},x_{\max},n_{\min},n_{\max}|n)$ as the product of three factors: $P_{\rm I}$, $P_{\rm II}$ and $P_{\rm III}$, corresponding to the three segments defined above (see Fig. \ref{fig:exponential}). 
In segment $I$ ($0\leq k\leq n_{\max}$), the walker needs to attain the maximum value $x_{\max}$ at time $n_{\max}$ for the first time, while remaining always above $-x_{\min}$. Thus, it has first to arrive at $x_{\max}-1$ at time $n_{\max}-1$ without leaving the interval $[-x_{\min}+1,x_{\max}-1]$ and then to jump to $x_{\max}$ at time $n_{\max}$. The probability weight of this last jump is $1/2$. Hence, using the reflection invariance $x\rightarrow-x$ the probability of the first segment can be written as
\begin{eqnarray}\label{eq:P_I}
P_{\rm I}=G(x_{\max}-1,l_1-1|M-2) \frac{1}{2},
\end{eqnarray}
where $l_1=n_{\max}$ and $M=x_{\min}+x_{\max}$. Note that after time $n_{\max}$ the walker is free to reach $x_{\max}$ again. For the second segment ($n_{\max}\leq k\leq n_{\min}$) we apply a similar reasoning: to reach the global minimum $-x_{\min}$ at time $n_{\min}$ for the first time, the RW first has to arrive at $-x_{\min}+1$ while remaining in the interval $[-x_{\min}+1,x_{\max}]$ and then to jump to $-x_{\min}$. Thus, the probability of the second segment is
\begin{eqnarray}\label{eq:P_II}
P_{\rm II}=G(M-1,l_2-1|M-1) \frac{1}{2}\,,
\end{eqnarray}
where $l_2=\tau=n_{\min}-n_{\max}$.
Finally, the probability of the last segment is the probability to remain in the interval $[-x_{\min},x_{\max}]$ up to time $n$ starting from $-x_{\min}$. This is simply given by:
\begin{eqnarray}\label{eq:P_III}
P_{\rm III}=\sum_{x_{\rm f}=0}^{ M}G(x_{\rm f},l_3|M),
\end{eqnarray}
where $l_3=n-n_{\min}$ and $x_{\rm f}$ is the final position measured with respect to $-x_{\min}$.
Thus the grand joint probability $P(x_{\min},x_{\max},t_{\min},t_{\max})$ is given by the product of the three factors:
\begin{eqnarray}\label{eq:grand_proba}
&& P(x_{\min},x_{\max},t_{\min},t_{\max}|n)=P_{\rm I}\,P_{ \rm II}\,P_{\rm III}\\&=&\frac{1}{4}G(x_{\max}-1,l_1-1|M-2)\,
 G(M-1,l_2-1|M-1)\, \nonumber \\&\times & \sum_{x_{\rm f}=0}^{ M}G(x_{\rm f},l_3|M)\,.\nonumber
\end{eqnarray}
Again it is convenient to express this probability in terms of the intervals $l_1$, $l_2$, $l_3$, as in Eq. (\ref{PDF_l1l2l3}). Summing over the position variables $x_{\max}$ and $x_{\min}$ we obtain the joint probability:
\begin{eqnarray}\label{Proba_l_1_l_2_l_3}
&& P(l_1, l_2, l_3)\\ &=& \sum_{x_{\max}\,,x_{\min}=1}^\infty G(x_{\max}-1,l_1-1|M-2) \nonumber \\&
 \times &\frac{1}{4} G(M-1,l_2-1|M-1)\, \sum_{x_{\rm f}=0}^{ M}G(x_{\rm f},l_3|M)\,.\nonumber
\end{eqnarray} 
Performing the change of variables $(x_{\max},\,x_{\min})\to(x_{\max},\,M)$ in the summations above, we obtain 
\begin{eqnarray}\label{Proba_l_1_l_2_l_3_2}
&& P(l_1, l_2, l_3) \\ &=&\frac{1}{4} \sum_{M=2}^\infty\sum_{x_{\max}=1}^{ M-1}G(x_{\max}-1,l_1-1|M-2) \nonumber \\
 &\times &  G(M-1,l_2-1|M-1)\, \sum_{x_{\rm f}=0}^{ M}G(x_{\rm f},s_3|M)\,.\nonumber 
\end{eqnarray} 
We multiply both sides by $s^{l_1}\,s^{l_2}\,s^{l_3}$ and we sum over $l_1$, $l_2$ and $l_3$:
\begin{eqnarray}\label{Proba_l_1_l_2_l_3_3}&& \sum_{l_1,l_2,l_3=1}^\infty P(l_1, l_2, l_3) s_1^{l_1} s_2^{l_2} s_3^{l_3} =\frac{1}{4} \sum_{M=2}^\infty \sum_{x_{\max}=1}^{ M-1} \\& \times & s_1 \,\tilde{G}(x_{\max}-1,s_1|M-2) 
\,s_2\,\tilde{G}(M-1,s_2|M-1)\nonumber \\ & \times &\sum_{x_{\rm f}=0}^{ M}\tilde{G}(x_{\rm f},s_3|M)\nonumber\,,
\end{eqnarray}
where $\tilde{G}(x,s|M)$ is the generating function of $G(s,l|M)$ given in Eq. (\ref{eq:solution_G_tilde}). Note that in Eq. (\ref{Proba_l_1_l_2_l_3_3}) we have used the initial condition $G(x,0|M)=\delta_{x\,,0}$. Rearranging the terms in Eq. (\ref{Proba_l_1_l_2_l_3_3}), we get
\begin{eqnarray}\label{Proba_l_1_l_2_l_3_3_bis}&& \sum_{l_1,l_2,l_3=1}^\infty P(l_1, l_2, l_3) s_1^{l_1} s_2^{l_2} s_3^{l_3} =\frac{s_1\,s_2}{4} \sum_{M=2}^\infty \\ &\times & \tilde{I}(M-2,s_1)\tilde{G}(M-1,s_2|M-1) \tilde{I}(M,s_3)\nonumber\,,
\end{eqnarray}
where
\begin{eqnarray}\label{eq:I_lattice}
\tilde{I}(M,s)=\sum_{x=0}^{ M}\tilde{G}(x,s|M)\,.
\end{eqnarray} 
Plugging the expression for $\tilde{G}(x,s|M)$ given in Eq. (\ref{eq:solution_G_tilde}), we get
\begin{eqnarray}\label{eq:I_lattice_2}
\tilde{I}(M,s)&=&\frac{2}{s}\Big(\frac{1}{1-w(s)^{2(M+2)}}\frac{w(s)^{M+1}-1}{1-\omega(s)^{-1}}\\ &+& \frac{1}{1-w(s)^{-2(M+2)}}\frac{w(s)^{-(M+1)}-1}{1-\omega(s)}\Big)\,. \nonumber
\end{eqnarray} 
It is now useful to recall the expression (\ref{relation2}), which relates the probability distribution $P(\tau|n)$ to the distribution $P(l_1, l_2 = \tau, l_3)$ and is given by 
\begin{eqnarray}\label{relation2bis}
&& \sum_{\tau, n=1}^{\infty}P(\tau|n) s_2^\tau \, s^n\\& =& \sum_{l_1,\tau,l_3=1}^\infty P(l_1, l_2 = \tau, l_3) s^{l_1} (s\, s_2)^{\tau}\, s^{l_3}\, .\nonumber
\end{eqnarray}
It is easy to show that Eq. (\ref{relation2bis}) is still valid for a discrete-space RW. Thus, using Eqs. (\ref{Proba_l_1_l_2_l_3_3_bis}) and (\ref{relation2bis}), we obtain:
\begin{eqnarray}\label{Proba_l_1_l_2_l_3_4}
\sum_{\tau,n=1}^\infty P(\tau|n)s_2^{\tau}s^n\, &=& \frac{s^2\,s_2}{4}\sum_{M=2}^\infty \tilde{I}(M-2,s)\\ &\times & \,\tilde{I}(M,s)\,G(M-1,s\,s_2|M-1)\,. \nonumber
\end{eqnarray}
Note that this relation (\ref{Proba_l_1_l_2_l_3_4}) is exact and valid even for finite $n$. As for the case of double-exponential jumps, we are interested in the large $n$ limit. In this limit we expect the probability $P(\tau|n)$ to have a scaling form
\begin{eqnarray}\label{scaling_form_lattice}
P(\tau | n) \underset{n \to \infty}{\longrightarrow} \frac{1}{n} f_{\rm LW} \left( \frac{\tau}{n}\right) \;.
\end{eqnarray}
Moreover, to investigate the limit $\tau,n\rightarrow\infty$ with $\tau/n$ fixed it is useful, as before, to parametrise the Laplace variables as $s\,=\,e^{-\lambda}$ and $s_2=e^{-\lambda_2}$ and to consider the limit $\lambda,\lambda_2 \rightarrow 0$, with $\lambda/\lambda_2$ fixed. In this limit, similarly to the case of the double-exponential distribution (see Eq. (\ref{integral1bis})), the left-hand side of Eq. (\ref{Proba_l_1_l_2_l_3_4}) can be approximated, using Eq. (\ref{scaling_form_lattice}) and the replacing the sums by integrals, as
\begin{eqnarray}\label{eq:approximation1}
\sum_{\tau,n}P(\tau|n)s_2^{\tau}s^n\,\approx \int_{0}^{1}\,dy\,\frac{f_{\rm LW}(y)}{\lambda+\lambda_2 \, y}\,.
\end{eqnarray}
We now consider the right-hand side of Eq. (\ref{Proba_l_1_l_2_l_3_4}) and we set $s\,=\,e^{-\lambda}$ and $s_2=e^{-\lambda_2}$. When $n$ is large, we expect the sum over $M$ to be dominated by terms with $M\gg 1$. Thus, we can approximate $M-2\approx M-1 \approx M$ and substitute the sum in Eq. (\ref{Proba_l_1_l_2_l_3_4}) with an integral. This gives
\begin{eqnarray}\label{eq:rhs_approx}
&&\frac{s^2\,s_2}{4}\sum_{M=2}^\infty \tilde{I}(M-2,e^{-\lambda}) \,\tilde{I}(M,e^{-\lambda})\,\\ &\times & G(M-1,e^{-(\lambda+\lambda_2)}|M-1)\nonumber \\ &\approx & \frac{e^{-(2\lambda+\lambda_2)}}{4} \int_{0}^\infty dM\,  \tilde{I}(M,e^{-\lambda})^2 G(M,e^{-(\lambda+\lambda_2)}|M)\,.\nonumber
\end{eqnarray}
Expanding $\tilde{I}(M,e^{-\lambda})$ in Eq. (\ref{eq:I_lattice_2}) for small $\lambda$, we obtain:
\begin{eqnarray}\label{eq:I_expanded}
\tilde{I}(M,e^{-\lambda})\approx\sqrt{\frac{2}{\lambda}}\tanh \left(M\sqrt{\frac{\lambda}{2}}\right).
\end{eqnarray}
The Green's function $G(M,e^{-(\lambda+\lambda_2)}|M)$ in Eq. (\ref{eq:solution_G_tilde}) can be approximated, for $\lambda,\,\lambda_2\to 0$, as
\begin{eqnarray}\label{eq:G_expanded}
G(M,e^{-(\lambda+\lambda_2)}|M)\approx \frac{2\sqrt{2 (\lambda+\lambda_2)}}{\sinh\left(M\sqrt{2(\lambda+\lambda_2)}\right)}.
\end{eqnarray}
Substituting the expressions (\ref{eq:I_expanded}) and (\ref{eq:G_expanded}) into Eq. (\ref{eq:rhs_approx}) and approximating $e^{-(2\lambda+\lambda_2)}\approx 1$, we get that the right-hand side of Eq. (\ref{Proba_l_1_l_2_l_3_4}) can be approximated by
\begin{eqnarray}\label{eq:rhs_approx_2}
\int_0^\infty \, dM\, \frac{\sqrt{2(\lambda+\lambda_2)}}{\lambda}\frac{\tanh^2\left(M\sqrt{\lambda/2}\right)}{\sinh(M\sqrt{2(\lambda+\lambda_2)})}\,.
\end{eqnarray}
Using this approximation (\ref{eq:rhs_approx_2}) together with Eq. (\ref{eq:approximation1}) to approximate the left-hand side of Eq. (\ref{Proba_l_1_l_2_l_3_4}), we get that in the limit $n,\tau\to\infty$, with $\tau/n$ fixed, Eq. (\ref{Proba_l_1_l_2_l_3_4}) becomes\begin{eqnarray}
 && \int_{0}^{1}\,dy\,\frac{f_{\rm LW}(y)}{\lambda+\lambda_2 \, y}\\ &=& \int_0^\infty \, dM\, \frac{\sqrt{2(\lambda+\lambda_2)}}{\lambda}\frac{\tanh^2\left(M\sqrt{\lambda/2}\right)}{\sinh(M\sqrt{2(\lambda+\lambda_2)})}\,.\nonumber
\end{eqnarray}
Setting $u=\lambda_2 /\lambda$ and performing the change of variables $z=\sqrt{2(\lambda+\lambda_2)}M$ we find the relation:
\begin{equation}\label{integral2_new}
\int_{0}^{1}dy\frac{f_{\rm LW}(y)}{1+uy}=\int_{0}^{\infty}dz \frac{1}{\sinh(z)}\tanh^2\left(\frac{z}{2\sqrt{1+u}}\right)\,.
\end{equation}
Comparing this result with Eq. (\ref{integral2}) we notice that $f_{\rm LW}(y)$ and $f_{\exp}(y)$ satisfy the same integral relation. As explained in the previous section, this integral relation can be exactly solved (see Appendix \ref{app:integral}) and one obtains that
\begin{eqnarray}
f_{\rm LW}(y)=f_{\rm BM}(y)\,,
\end{eqnarray} 
where $f_{\rm BM}(y)$ is given by Eq. (\ref{eq:f_bm}). Hence, also in the case of lattice walks we have explicitly verified the expectations based on the Central Limit Theorem.
See Fig. \ref{fig:numeric_rw} for a numerical verification of this result.

\subsection{Divergent jump variance}

We now consider a class of RWs characterised by jump distributions with divergent variance $\sigma^2=\int_{-\infty}^{\infty}\,d\eta\,\eta^2 \,p(\eta)$. In particular, we consider L\'evy flights for which the jump probability $p(\eta)$ has heavy tails: 
\begin{equation}
p(\eta)\sim \frac{1}{|\eta|^{-(\mu+1)}}
\end{equation}
for $\eta\rightarrow \pm \infty$, with $0< \mu < 2$. Note that for such values of the L\'evy index $\mu$ the variance $\sigma^2$ is divergent. Thus, for these RWs the Central Limit Theorem does not hold. Indeed, we verify numerically (see Fig. \ref{fig:numeric_rw}) that in the limit of large number of steps $n$, the PDF of $\tau$ takes a scaling form
\begin{eqnarray}\label{eq:f_mu}
P(\tau|n)\approx\frac{1}{n}\,f_{\mu}\left(\frac{\tau}{n}\right),
\end{eqnarray}
where the scaling function $f_{\mu}(y)$ depends on the L\'evy index $\mu$. Our general result is thus less universal with respect to the distribution of the time of the maximum $n_{\max}$, which is the same for any symmetric jump distribution. However, we observe from simulations that $f_{\mu}(y)$ goes to the value $1/2$ when $y\rightarrow \pm 1$ , independently of $\mu$. Thus, the PDF of the event ``$\tau = n$'' is universal in the limit of large $n$ for any symmetric distribution $p(\eta)$:
\begin{equation}\label{eq:res_n}
P(\tau=n|n)=1/(2n)
\end{equation}
In Section \ref{sec:BM}, we have verified that the scaling function $f_{\rm BM}(y)$ approaches the limit value $1/2$ linearly with a negative slope $-1/2$. In the case of L\'evy flights, this linear behaviour in the vicinity of $y=\tau/n=1$ seems to remain valid, but with a slope which depends on the L\'evy exponent $\mu$ (see Fig. \ref{fig:numeric_rw}).
The asymptotic result in Eq. (\ref{eq:res_n}) can be directly obtained in the case of lattice walks (see Appendix \ref{app:limit}). As we will see in the next section, this result in Eq. (\ref{eq:res_n}) is valid, in the case of continuous jump distributions, for any finite $n$. However, rigorously proving this fact for L\'evy flights appears to be a challenging task.

\section{Universal probability of the event $\tau=n_{\min}-n_{\max}=n$}
\label{sec:universal}

\begin{figure}[t]
  \centering
\includegraphics[width=1\linewidth]{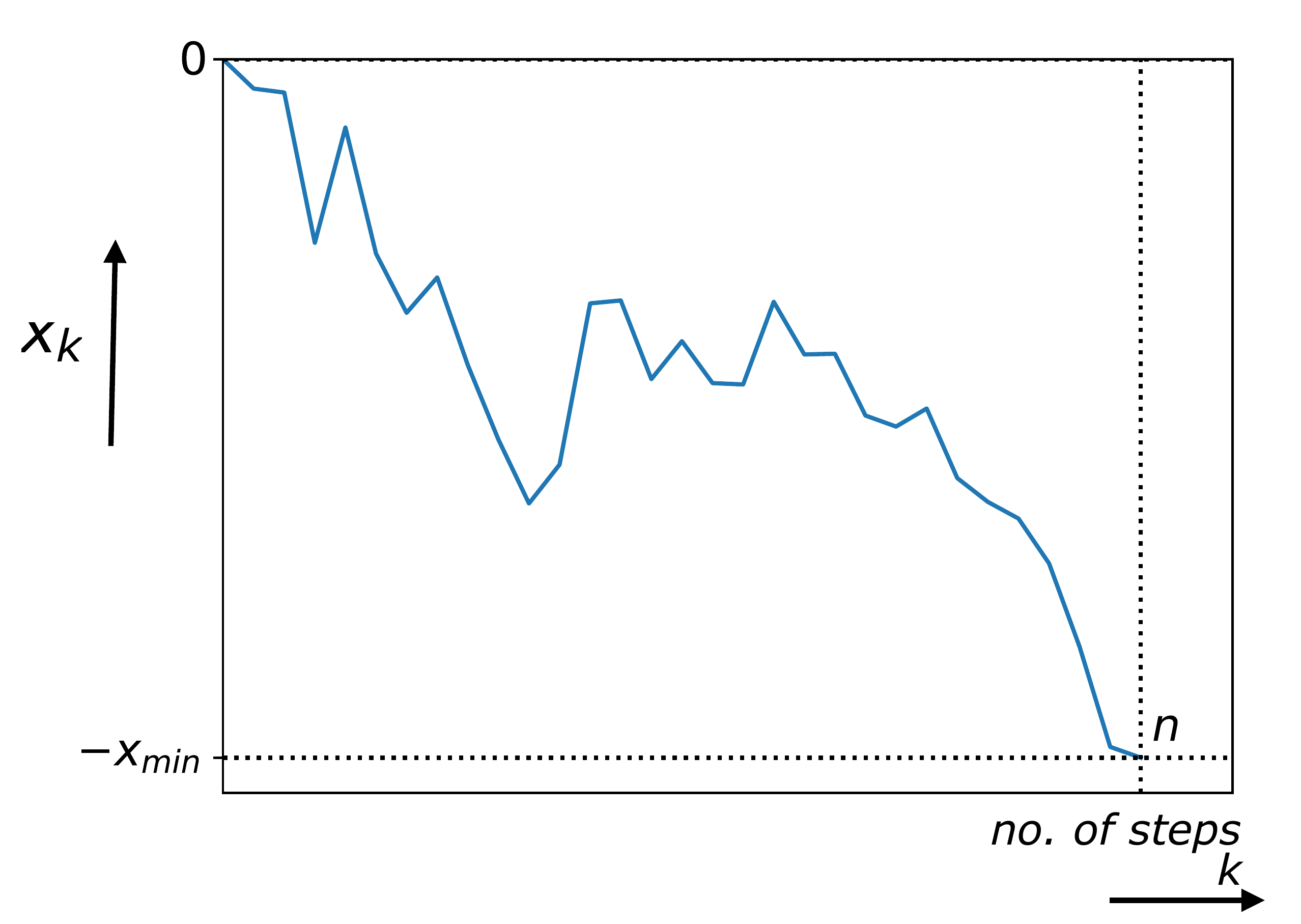}
\caption{A typical trajectory of a discrete-time random walk that contributes to the probability $P(\tau = n|n)$ where $\tau = n_{\min} - n_{\max}$. The event ``$\tau = n$'' can only happen when $n_{\max} = 0$ and $n_{\min} = n$. Consequently, the trajectories that contribute to this event start at $x_0 = 0$ and arrive at $x_n = -x_{\min}$, while staying inside the box $[-x_{\min},0]$.}
\label{fig:appendix} 
\end{figure}

Let us consider a discrete-time random walk $x_k$ generated by the Markov process in Eq. (\ref{def_RW}) with a generic jump 
distribution $p(\eta)$. In this section, we want to investigate the probability $p_n$ that the number of steps 
$\tau=n_{\min}-n_{\max}$ between the global maximum and the global minimum is exactly equal to the total number of steps $n$. 
Since $\tau$ is bounded by construction between $-n$ and $n$, the event ``$\tau = n$'' can only happen when $n_{\max} = 0$ and 
$n_{\min} = n$, which corresponds to configurations as the one in Fig. \ref{fig:appendix}. Thus, this probability 
$p_n=P(\tau=n|n)$ has a simple and nice interpretation. Indeed, using the $x\to-x$ symmetry of the process, $p_n$ is the 
probability that a RW starting from the origin remains positive up to step $n$ and that the last value $x_n$ is a record, 
meaning that $x_n>x_k$ for $k=0,\ldots, n-1$. \\

In Section \ref{sec:RW}, we have observed from numerical simulations that, in the limit of large number $n$ of steps, the 
probability $p_n$ of the event ``$\tau = n$'' appears to be completely independent of the distribution $p(\eta)$ of the jumps 
(see Fig. \ref{fig:numeric_rw}). Indeed, in the limit of large $n$,
\begin{equation}\label{eq:result_asymptotic}
p_n=P(\tau=n|n)\simeq \frac{1}{2n}
\end{equation} 
for any symmetric jump distribution $p(\eta)$. Notably, this universality appears to be valid for a variety of distributions, 
including discrete distributions and distributions with divergent first moment, e.g. Cauchy distribution. 
Moreover, in the case of the double-exponential jump distribution and in the case of lattice walks, 
we analytically showed that the probability distribution of $\tau$ converges, in the large $n$ limit, to the scaling form
\begin{equation}\label{eq:scaling_general}
P(\tau|n)\simeq \frac{1}{n}f_{\rm BM}\left(\frac{\tau}{n}\right)\,,
\end{equation}
where the scaling function $f_{\rm BM}(y)$ is given in Eq. (\ref{eq:f_bm}). 
In Section (\ref{sec:BM}) we have observed that when $y=\tau/n\to 1$, i.e. when $\tau \to n$, this scaling function 
$f_{\rm BM}(y)$ converges to the asymptotic value $1/2$ (see Eq. (\ref{summary_asymptotics})). Thus, taking the 
limit $\tau\to n$ in Eq. (\ref{eq:scaling_general}) we obtain the result in Eq. (\ref{eq:result_asymptotic}). 
Moreover, Eq. (\ref{eq:result_asymptotic}) can be also derived directly in the case of 
lattice walks (see Appendix \ref{app:limit}).

\begin{figure}[t]
  \centering
\includegraphics[width=1\linewidth]{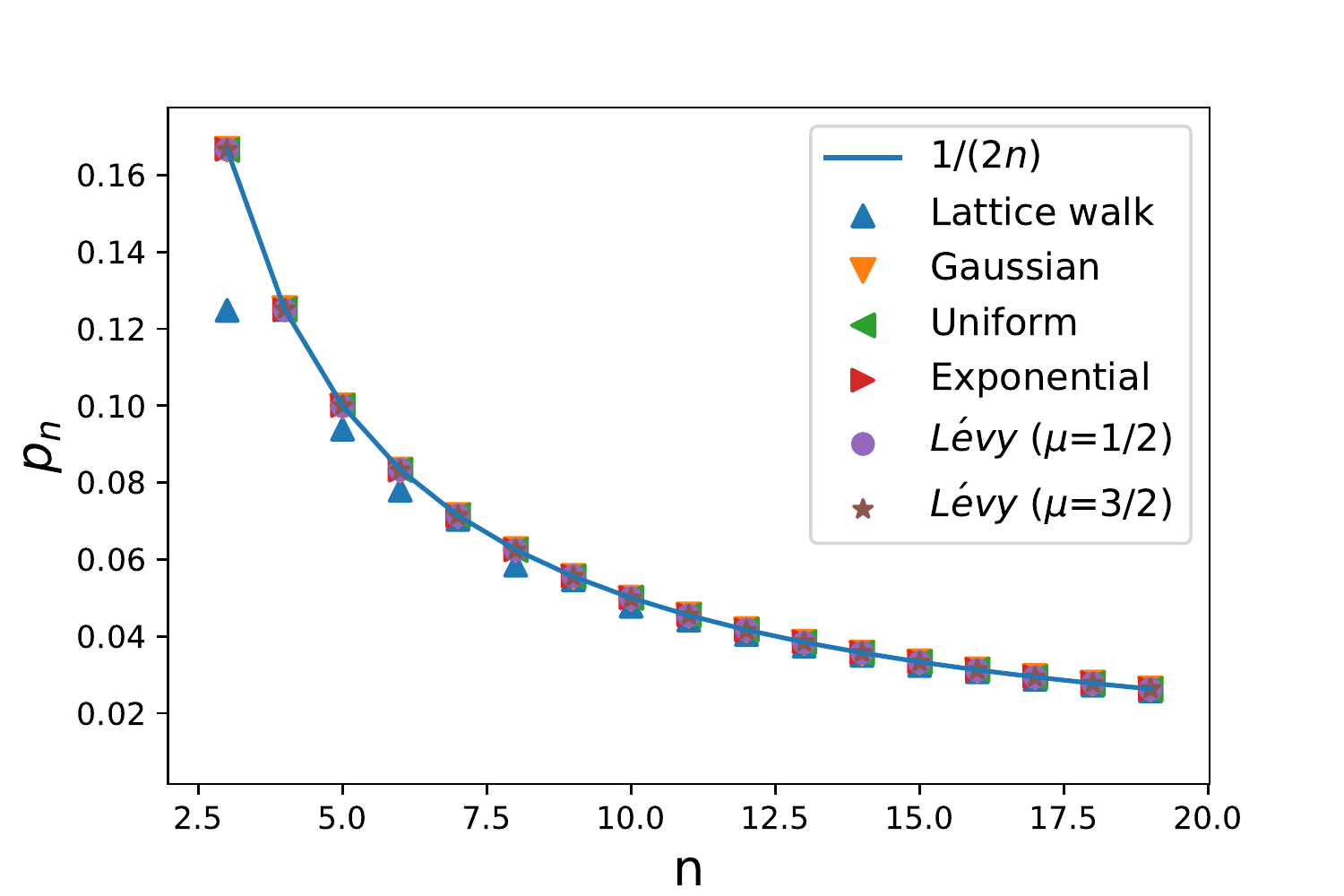}
\caption{The probability $p_n=P(\tau=n|n)$ as a function of $n$ for discrete-time random walks (RWs) for different jump distributions. Continuous jump distributions collapse onto the universal result $1/(2n)$ shown by the solid (blue) line. In the case of lattice walks, which are discrete in space, the result $1/(2n)$ is only reached asymptotically for large $n$. The empirical curves are obtained by simulating $10^7$ RWs for each jump distribution.}
\label{fig:numerics_universality} 
\end{figure}

We now want to show that, in the case of continuous jump distributions, the universal result in Eq. 
(\ref{eq:result_asymptotic}) is exactly valid even for finite $n$. In Fig. \ref{fig:numerics_universality}, we verify this 
universality numerically for several different continuous jump distributions. Even if finding a theoretical explanation of 
this universality appears to be non-trivial, it is possible to analytically derive the result $p_n=1/(2n)$ in the special case 
of double-exponential jumps. Moreover, besides the trivial cases $n=1$ and $n=2$, in Appendix~\ref{app:n3}, we provide a proof 
of this universality in the case $n = 3$. However, generalising the method presented in Appendix~\ref{app:n3} to $n>3$ appears 
to be challenging. Thus, proving this conjecture $p_n=1/(2n)$ for symmetric and continuous jump distribution
remains an interesting open problem.

It is easy to verify that this result $p_n=1/(2n)$ for any finite $n$ does not hold for RWs with discrete jump 
distributions (see Fig. 
\ref{fig:numerics_universality}). For instance, if we consider the discrete distribution $p(\eta)=\delta(|\eta|-1)/2$, 
corresponding to lattice walks, it is easy to check for $n=3$ that
\begin{equation}
p_3=\frac{1}{8}\neq \frac{1}{2n}\,.
\end{equation}
Note that for RWs with discrete distributions the global maximum and the global minimum are degenerate with finite 
probability. Thus, here we define $n_{\max}$ ($n_{\min}$) as the step at which the global maximum (minimum) is reached for the 
first time.

\begin{figure*}[t]   
 \includegraphics[angle=0,width=\linewidth]{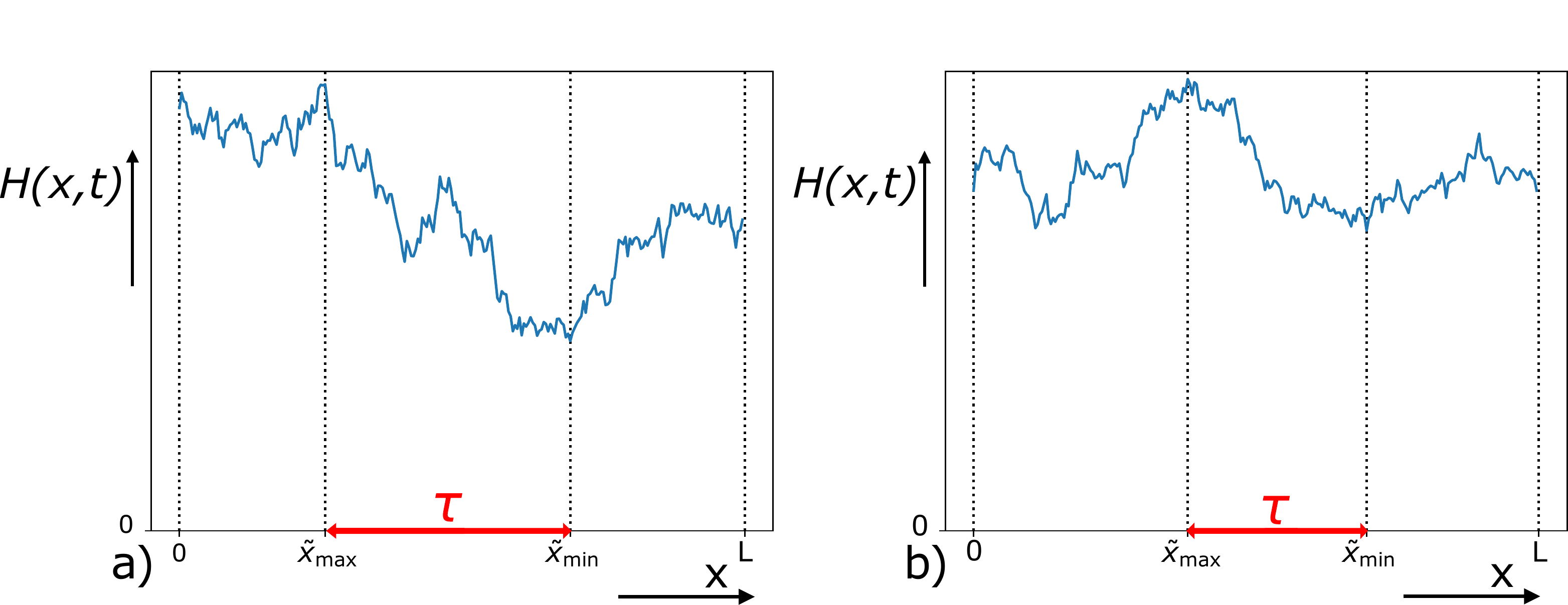} 
    \caption{Typical realizations of the height $H(x,t)$ of $(1+1)-$dimensional  
Kardar-Parisi-Zhang fluctuating interfaces, evolving on a substrate of size $L$, as a function of the 
position $x$ and at fixed time $t$: a) for the free boundary conditions and b) for the periodic boundary condition. 
The maximal height is reached at position $\tilde{x}_{\max}$ and the minimal height at 
position $\tilde{x}_{\min}$. The distance between the points of maximal and minimal height is denoted as $\tau$.}
    \label{fig:kpz_realization}
\end{figure*}

In the case of RWs with double exponential jumps, i.e. when $p(\eta) = (1/2) e^{-|\eta|}$, we can analytically show that $p_n=1/(2n)$ for any finite $n$. Since $-n \leq \tau \leq n$, it follows that the event ``$\tau = n_{\min} - n_{\max} = n$'' corresponds, as explained above, to having the maximum at step $n_{\max}=0$ and the minimum at step $n_{\min}=n$ (see Fig. \ref{fig:appendix}). This corresponds to a trajectory that starts at the origin at step $0$, reaches $-x_{\min}$ at step $n$, stays in the box $[-x_{\min},0]$ for all intermediate steps. To compute the probability of such a trajectory, it is useful first to reflect the trajectory $x \to -x$, so that we just need to compute the probability that the walker starting at $0$ arrives at $x_{\min}\geq 0$ at step $n$, while staying in the box $[0,x_{\min}]$, with $x_{\min}$ integrated over $[0, +\infty)$. This probability can be conveniently expressed in terms of our basic building block $G(x,n|M)$, defined as the probability that the walker goes from the origin to position $x$ in $n$ steps, always remaining in the box $[0,M]$. Indeed, after integrating over $x_{\min}$, we get
\begin{eqnarray}\label{P1/2_2}
p_n=P(\tau  = n|n) = \int_0^\infty G(x_{\min}, n |x_{\min}) \, d x_{\min} \;.
\end{eqnarray}
Hence, to prove that $p_n = 1/(2n)$, we need to evaluate the integral on the right-hand side of Eq. (\ref{P1/2_2}). 
Actually, for the double-exponential jump distribution, the generating
function of $G(x,n|M)$ was exactly computed in Section \ref{sec:RW}. Hence, 
multiplying both terms of Eq. (\ref{P1/2_2}) by $s^n$ and summing over $n$, we obtain,
\begin{equation}\label{eq:app_lapl}
\sum_{n=1}^{\infty}p_n\,s^n = \int_0^\infty \tilde{G}(x_{\min}, s |x_{\min}) \, d x_{\min} \;.
\end{equation}
Using the expression for $\tilde{G}(x_{\min},s|M)$ in Eqs. (\ref{g_solution}) and (\ref{A}), with $M=x_{\min}$, we obtain
\begin{eqnarray}\label{eq:app_lap2}
&&\sum_{n=1}^{\infty}p_n\,s^n = \int_0^\infty  
d x_{\min} \, \frac{1-\sqrt{1-s}}{1-(\frac{1-\sqrt{1-s}}{1+\sqrt{1-s}})^2 
e^{-2\sqrt{1-s}\,x_{\min}}}\nonumber \\ &\times & 
\left[e^{-\sqrt{1-s}\,x_{\min}}-\frac{1-\sqrt{1-s}}{1+\sqrt{1-s}}e^{-\sqrt{1-s}\,x_{\min}}\right]\;.
\end{eqnarray}
Computing the integral on the right-hand side, we obtain, after few steps of algebra,
\begin{equation}
\sum_{n=1}^{\infty}p_n\,s^n=-\frac{1}{2}\log\left(1-s\right)\,.
\end{equation}
The right-hand side can be rewritten in Taylor series for $0<s<1$ as follows
\begin{equation}
\sum_{n=1}^{\infty}p_n\,s^n=\frac{1}{2}\sum_{n=1}^{\infty}\frac{1}{n}s^n\,,
\end{equation}
which implies that for any finite $n\geq 1$
\begin{equation}
p_n=P(\tau  = n|n)=\frac{1}{2n}\,.
\end{equation}

\section{Fluctuating interfaces} \label{sec:fluctuating}

\begin{figure*}[t]
 \includegraphics[angle=0,width=\linewidth]{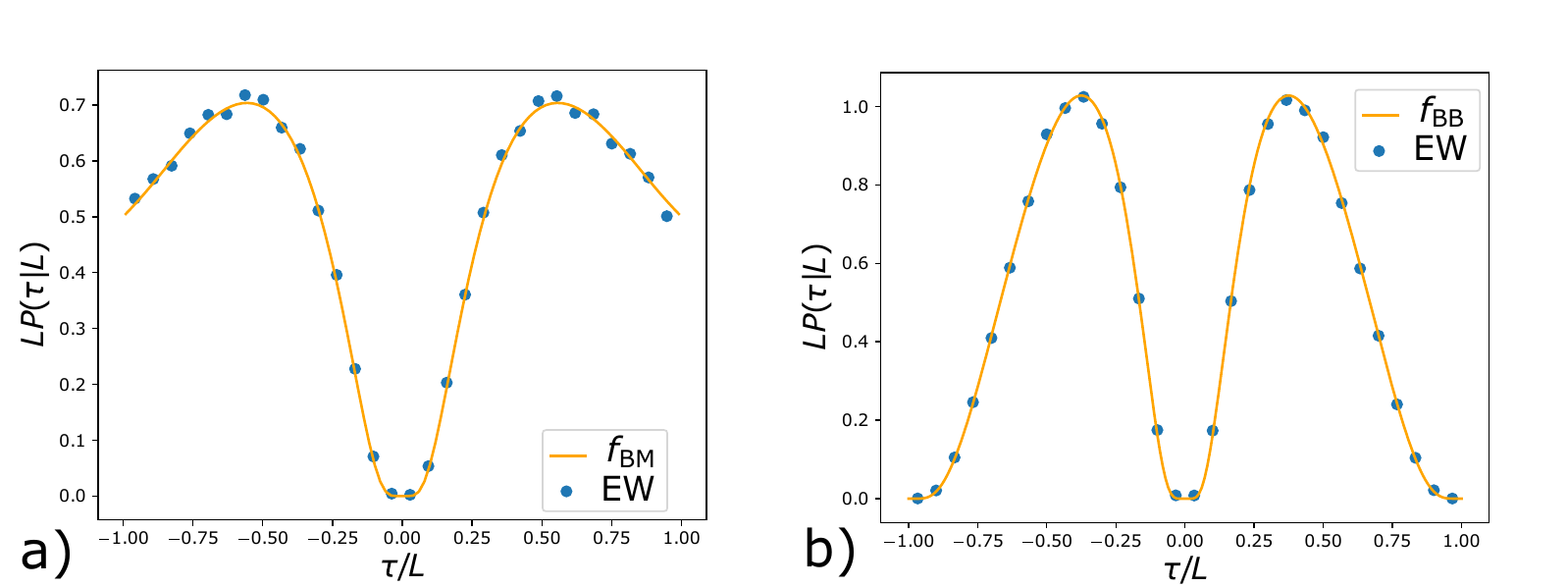} 
    \caption{
Scaling plot of $P(\tau = \tilde x_{\min} - \tilde x_{\max}|L)$ for the Edwards-Wilkinson interface obtained from the numerical integration of Eq. (\ref{discrete_EW}) with $\Delta t=0.01$ and $L=512$: a) for the free boundary conditions and b) for the periodic boundary conditions. The solid line in a) represents the analytical scaling function $f_{\rm BM}(y)$ given in Eq. (\ref{eq:f_bm}) while the filled dots represent simulation data. In b), the solid line represents the analytical scaling function $f_{\rm BB}(y)$ given in Eq. (\ref{eq:f_bb}), while the filled dots represent simulation data. The numerical data are obtained by averaging over $10^6$ samples.}
    \label{fig:numeric_ew}
\end{figure*}

A remarkable application of our results is to KPZ/EW fluctuating interfaces. We consider a $(1+1)-$dimensional fluctuating interface evolving over a substrate of finite size $L$. Let $H(x,t)$ be the height of the interface at position $x$ at time $t$, with $0\leq x \leq L$, as in Fig. \ref{fig:kpz_realization} \cite{review_kpz1,review_kpz2,spohn_houches}. We describe the evolution of the height field in time using the KPZ equation~\cite{kpz86} 
\begin{equation} \label{eq:kpz}
\frac{\partial H(x,t)}{\partial t}=\frac{\partial^2 H(x,t)}{\partial x^2}+\lambda\left(\frac{\partial H(x,t)}{\partial x}\right)^2+\eta\left(x,t\right),
\end{equation}
where $\lambda \geq 0$ and $\eta(x,t)$ is a Gaussian white noise with zero mean and correlator $\langle\eta\left(x,t\right)\eta\left(x',t'\right)\rangle=2\delta(x-x')\delta(t-t')$. The linear case $\lambda=0$ corresponds to EW equation \cite{edwads82}. We consider both free boundary conditions, where the endpoints $H(0,t)$ and $H(L,t)$ evolve freely, and periodic boundary condition, where the constraint $H(0,t)=H(L,t)$ is present (see Fig. \ref{fig:kpz_realization}). Here we are interested in describing this system in the large time limit. However, since the zero mode, characterised by the average height 
\begin{eqnarray}\label{eq:avg_H}
\overline{H(t)}= \frac{1}{L}\int_{0}^{L}H(x,t)\,dx\,,
\end{eqnarray}
typically grows with time, the height $H(x,t)$ will never reach a stationary state, even for a finite system. Thus, it is useful to define the displacement from the average height, i.e. the relative height
\begin{eqnarray}\label{eq:h_def}
h(x,t)=H(x,t)-\overline{H}(t)\,.
\end{eqnarray}
In this way, we are fixing the zero mode to be exactly zero. Indeed, note that $h(x,t)$ satisfies by construction
\begin{eqnarray}\label{eq:condition_on_h}
\int_0^L \, dx \,h(x,t)\,=\,0\,.
\end{eqnarray}
For finite $L$, it turns out that $h(x,t)$ reaches a stationary state $h(x)$ for late times. In this stationary state, we define the position at which the height is minimal as 
\begin{eqnarray}\label{eq:x_min_def}
\tilde{x}_{\min}=\operatorname{argmin}_{0\leq x\leq L}\left(h(x)\right),
\end{eqnarray}
and the position of maximal height as
\begin{eqnarray}\label{eq:x_max_def}
\tilde{x}_{\max}=\operatorname{argmax}_{0\leq x\leq L}\left(h(x)\right)\,.
\end{eqnarray}
We are mainly interested in computing the joint PDF $P(\tilde{x}_{\max}\,,\tilde{x}_{\min}|L)$ of $\tilde{x}_{\max}$ and $\tilde{x}_{\min}$  and the PDF of the position distance $\tau=\tilde{x}_{\min}-\tilde{x}_{\max}$ between maximum and minimum, which we denote as $P(\tau|L)$. We also denote the maximal and the minimal relative height as $h_{\max}=h(\tilde{x}_{\max})$ and $h_{\min}=h(\tilde{x}_{\min})$.

\subsection{Edwards-Wilkinson case}

We start by considering the simpler case of EW interfaces, corresponding to $\lambda=0$ in Eq. (\ref{eq:kpz}). 
First, we consider FBC, i.e. we assume that the height values $h(0)$ and $h(L)$ at the extremes of the interval evolve freely according to Eq. (\ref{eq:kpz}). In this case, the PDF of the stationary state $h(x)$ of EW equation is given by~\cite{majumdar04,comtet05,schehr06}
\begin{equation}\label{eq:stationary_FBC}
P_{\rm st}\left(\{h\}\right)=A_L\,e^{-\frac{1}{2}\int_{0}^{L}dx (\partial_{x}h)^2}\delta\left[\int_{0}^{L}h(x)dx\right] \;,
\end{equation}
where the delta function enforces the constraint (\ref{eq:condition_on_h}) and $A_L$ is the normalisation constant 
\begin{equation}
A_L=\sqrt{2\pi}L^{3/2}\,.
\end{equation}
On the other hand, in the case of PBC $h(0)$ and $h(L)$ evolve freely but with $h(0)=h(L)$. Thus, the stationary distribution of $h(x)$ contains the additional factor $\delta\left(h(0)-h(L)\right)$:
\begin{eqnarray}\label{eq:stationary_PBC}
P_{\rm st}\left(\{h\}\right) &=& B_L\,e^{-\frac{1}{2}\int_{0}^{L}dx (\partial_{x}h)^2}\delta\left[\int_{0}^{L}h(x)dx\right]\\ & \times\, &\delta  \left(h(0)-h(L)\right) \;\nonumber ,
\end{eqnarray}
with $B_L$ corresponding to the normalisation constant
\begin{equation}
B_L=L\,.
\end{equation}
From the expressions of the stationary probabilities (\ref{eq:stationary_FBC}) and (\ref{eq:stationary_PBC}), we observe that, for both FBC and PBC, the stationary height $h(x)$ behaves locally as a BM, apart from a global zero area constraint. Indeed if we identify (a) space with time, i.e. $x \Leftrightarrow t$, (b) the total substrate length $L$ with the total duration $T$, i.e. $L \Leftrightarrow T$ and (c) the stationary relative height $h(x)$ with the position $x(t)$ of a BM, i.e. $h(x) \Leftrightarrow x(t)$, we find a one-to-one mapping between the stationary EW interface and the positions of a BM (for the case of the FBC). In the case of the PBC, the stationary interface corresponds to a BB. Note however that due to the zero area constraint (\ref{eq:condition_on_h}) the process $x(t)$ obtained through the mapping is not exactly a BM/BB. Indeed, $x(t)$ has to satisfy an equivalent constraint:
\begin{eqnarray}\label{eq:condition_x}
\int_{0}^{T}\,dt\,x(t)\,=\,0\,.
\end{eqnarray}
Hence, the statistical properties of this process $x(t)$ will in general differ from those of a usual BM/BB.
For instance, the PDF of the maximal (minimal) height $h_{\max}$ ($h_{\min}$) is known for both boundary conditions to be different from the PDF of the maximum (minimum) value of an usual BM/BB \cite{majumdar04,comtet05}. Indeed, the zero area constraint affects the value of the maximum (minimum). On the other hand, this constraint just corresponds to a global shift by the zero mode. Thus, it is clear that, due to the locally Brownian nature of $h(x)$, the positions at which the extrema occur are not affected for both FBC and PBC. Hence, for FBC the joint PDF of $\tilde{x}_{\max}$ and $\tilde{x}_{\min}$ $P(\tilde{x}_{\max}\,,\,\tilde{x}_{\min}|L)$ will coincide with that of the times $t_{\max}$ and $t_{\min}$ for a BM of total duration $T=L$. This implies that the stationary probability distribution of the position difference $\tau$ between minimum and maximum is given by
\begin{eqnarray}\label{eq:P_tau_FBC}
P(\tau=\tilde{x}_{\min}-\tilde{x}_{\max}|L)=\frac{1}{L}f_{\rm BM}(\frac{\tau}{L})\,,
\end{eqnarray}
where the scaling function $f_{\rm BM}(y)$ is given by Eq. (\ref{eq:f_bm}). In the case of PBC, exploiting the mapping to a BB, we have that
\begin{eqnarray}\label{eq:P_tau_PBC}
P(\tau=\tilde{x}_{\min}-\tilde{x}_{\max}|L)=\frac{1}{L}f_{\rm BB}(\frac{\tau}{L})\,,
\end{eqnarray}
where the scaling function $f_{\rm BB}(y)$ is given in Eq. (\ref{eq:f_bb}).
To check this prediction for $P(\tau|L)$ in Eqs. (\ref{eq:P_tau_FBC}) and (\ref{eq:P_tau_PBC}) for the EW interface, we numerically integrated the space-time discretised form of Eq. (\ref{eq:kpz}) with $\lambda = 0$
\begin{eqnarray}\label{discrete_EW}
H(i,t+ \Delta t) - H(i,t)&=&  \Delta t \big[H(i+1,t) + H(i-1,t)\nonumber \\&-& 2 H(i,t) \big] + \eta_i(t) \sqrt{2 \Delta t} \;,
\end{eqnarray}
where $\eta_i(t)$'s are IID random variables for each $i$ and $t$, each drawn from a Gaussian distribution of zero mean and unit variance. We considered both the FBC and the PBC with $\Delta t = 0.01$ and $L = 512$. We have run the simulation for a sufficiently large time to ensure that the system has reached the stationary state and then measured the PDF $P(\tau|L)$. Even though we expect the results in  Eqs. (\ref{eq:P_tau_FBC}) and (\ref{eq:P_tau_PBC}) to be valid for all values of $L$, this expectation is only for the continuum version of the EW equation (\ref{eq:kpz}) with $\lambda=0$. Since for the simulation we have use the discrete version (\ref{discrete_EW}) of this equation, we expect these results in Eqs. (\ref{eq:P_tau_FBC}) and (\ref{eq:P_tau_PBC}) to hold only for large $L$. Actually, for $L=512$, we already see an excellent agreement between simulations and analytical results. In Fig. \ref{fig:numeric_ew} a) we compare the simulations with the analytical prediction for the FBC in Eq. (\ref{eq:P_tau_FBC}). The corresponding simulation results for the PBC are shown in Fig. \ref{fig:numeric_ew} b) and compared with the analytical prediction in Eq. (\ref{eq:P_tau_PBC}).

\subsection{Kardar-Parisi-Zhang case}  
At variance with the case of EW equation, for the KPZ equation (\ref{eq:kpz}) with $\lambda >0$, the stationary state for the relative heights is expected to converge to the same measures (\ref{eq:stationary_FBC}) and (\ref{eq:stationary_PBC}) (respectively for the FBC and the PBC), but only in the limit $L \to \infty$. Therefore, we expect that the results for $P(\tau|L)$ in Eqs. (\ref{eq:P_tau_FBC}) and (\ref{eq:P_tau_PBC}) to hold also for the KPZ equation for large $L$. However, verifying these analytical predictions numerically for the KPZ equation is challenging because the non-linear term is not easy to discretise \cite{lam_shin,lam_shin2}. 
\begin{figure}[t]    
\includegraphics[width=1\linewidth]{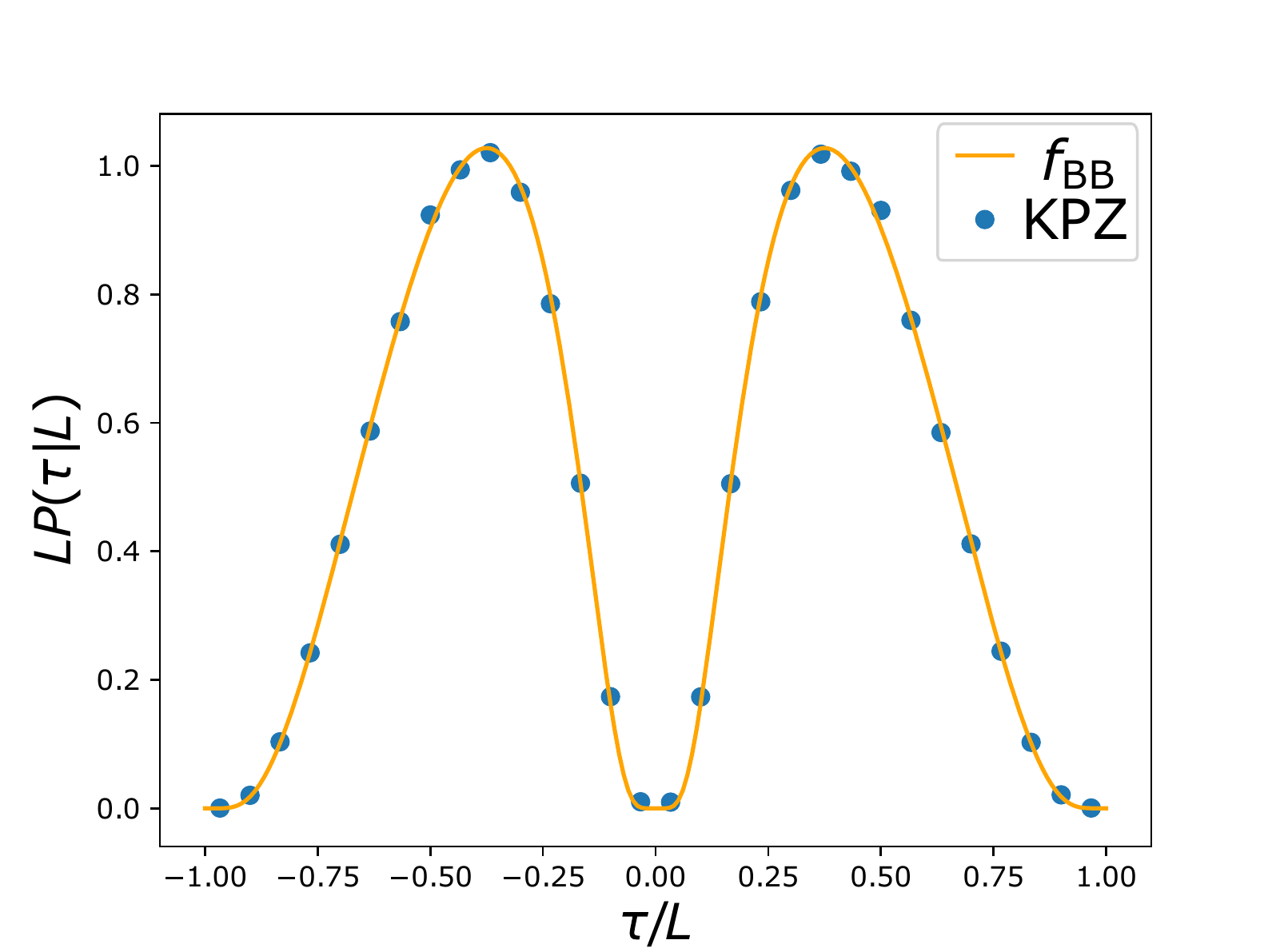} 
\caption{Scaling plot of $P(\tau = \tilde x_{\min} - \tilde x_{\max}|L)$ for the KPZ interface with PBC obtained using the discretisation scheme (\ref{lam_shin}) with $\Delta t=0.01$ and $L=512$. The solid line represents the analytical scaling function $f_{\rm BB}(y)$ given in Eq. (\ref{eq:f_bb}), while the filled dots represent the simulation data. The numerical data are obtained by averaging over $10^6$ samples.}\label{Fig_KPZ}
\end{figure} 
Several discretisation schemes have been proposed in the literature and we found it suitable to use the scheme proposed by Lam and Shin \cite{lam_shin2}, where the non-linear term $\lambda (\partial_x H(x,t))^2$ is discretised as follows
\begin{eqnarray}\label{lam_shin}
&&\frac{\lambda}{3} \Big[ (H(i+1,t)-H(i,t))^2+ (H(i+1,t)- H(i,t))\nonumber \\ &\times & (H(i,t) - H(i-1,t)) \\ &+& (H(i,t) - H(i-1,t))^2  )\Big]\nonumber \;.
\end{eqnarray}  
The advantage of this scheme is that one can prove analytically that, for the PBC, the Fokker-Planck equation associated with this discrete model admits a stationary solution,
\begin{equation}
P_{\rm st}(\{ H\}) \propto \exp \left[-\frac{1}{2} \sum_{i=1}^L (H(i+1,t)-H(i,t))^2 \right]\,,
\end{equation} 
independently of $\lambda$. In the $L \to \infty$ limit, the stationary measure converges to the Brownian measure  $P_{\rm st}(\{ H\}) \propto \exp \left[-\frac{1}{2} \int_0^L (\partial_x H)^2 \, dx \right]$. Therefore, with this discretisation scheme (\ref{lam_shin}) and PBC, we expect to recover the BB result for $P(\tau|L)$ as in Eq. (\ref{eq:P_tau_PBC}). In Fig. \ref{Fig_KPZ} we compare the simulation results for $P(\tau|L)$ for the KPZ equation with PBC and $\lambda = 1$ (with parameters $\Delta t = 0.01$ and $L=512$), with the analytical scaling function in Eq. (\ref{eq:P_tau_PBC}) for the BB -- the agreement is excellent. Unfortunately, for the KPZ equation with the FBC, there is no convenient discretisation scheme for the non-linear term that correctly produces the stationary measure for finite $L$. Of course, we still expect that, in this case, the results for $P(\tau|L)$ for the KPZ equation in the stationary state will again converge to the Brownian prediction given in Eq. (\ref{eq:P_tau_FBC}) in the large $L$ limit. However, numerically verifying this for finite but large $L$ seems challenging, due to the absence of a good discretisation scheme for the non-linear term in the FBC case.   
\section{Conclusions} \label{sec:conclusions}
In summary, we have presented an exact solution for the probability distribution of the time $\tau$ between the maximum and the minimum for a class of stochastic processes. First, we have considered a one-dimensional BM of duration $T$. In this case, we have used a path-integral method to show that the PDF of $\tau$ has a scaling form for any $\tau$ and $T$, i.e. $P(\tau|T)=(1/T)f_{\rm BM}(\tau/T)$, and we have exactly computed the scaling function $f_{\rm BM}(y)$. In particular we find that $f_{\rm BM}(y)\sim e^{-\pi/\sqrt{|y|}}$ when $y\to 0$, while $f_{\rm BM}(y)\sim 1/2$ when $y\to \pm 1$. We have generalised our result to a one-dimensional BB, finding a different scaling function $f_{\rm BB}(y)$. We have verified numerically that the PDF of $\tau$ for BM is universal in the sense of the Central Limit Theorem,  i.e. the scaling function $f_{\rm BM}(y)$ is also valid for discrete-time RWs with finite-variance jumps in the limit of large number of steps $n$. For two particular RW models, namely RW with double-exponential jumps and lattice walks, we have proved analytically this universality. In the case of L\'evy flights with a divergent jump variance we have observed from numerical simulations that the PDF of $\tau$ differs from the Brownian case. Indeed, for L\'evy walks the precise shape of $P(\tau|n)$ depends on the tail behaviour of the jump distribution.

For discrete-time RWs with symmetric and continuous jump distribution, we found numerically  that 
the probability $p_n=P(\tau=n|n)=1/(2n)$ for any finite $n$, completely independent of the jump distribution. We could
prove this result analytically for the double exponential jump distribution. For general symmetric and continuous jump distribution, we 
could prove this super-universality only for $n\leq 3$. We believe that there must be an elegant combinatorial proof of this result, but
it has eluded us so far. Proving this conjecture for $n >3$ remains a challenging open problem.

Finally, we have also observed that the distribution of $\tau$ for BM and BB emerges in the statistical description of fluctuating interfaces. Indeed, the space distance between the maximal and minimal height of a $(1+1)$-dimensional stationary KPZ interface growing over a substrate of size $L$ has the same probability distribution as $\tau$ for BM, if one considers FBC, or for BB, in the case of PBC.

For further studies it would be interesting to compute the distribution of $\tau$ in the case of L\'evy flights. Moreover, in this paper we have only investigated processes with symmetric increments. It would be relevant to study how the PDF of $\tau$ gets modified when one considers an additional drift in the process, such as for drifted BM. This could be useful to describe financial data, which have the tendency to increase or decrease persistently in time.
Finally, it would be also interesting to compute the distribution of $\tau$ for stochastic processes with correlated noise, such as run-and-tumble particles or active BM. 
\appendix
\section{Computation of the integral $J(\alpha,\beta)$}
\label{app:Iab}
In this appendix we explicitly compute the integral 
\begin{eqnarray}\label{appendix_Iab1}
J(\alpha, \beta)& =&\int_0^\infty dx_{\min}  \int_0^\infty dx_{\max} e^{-\frac{\beta}{(x_{\min} + x_{\max})^2}}\\ &\times & \frac{1}{(x_{\min} + x_{\max})^6}\sin \left(\frac{\alpha \, x_{\min}}{x_{\min}  + x_{\max}} \right)  \, . \nonumber
\end{eqnarray}
First of all, we perform the change of variable $(x_{\min},x_{\max})\to(m=x_{\min},M=x_{\min}+x_{\max})$
\begin{eqnarray}\label{appendix_Iab2}
J(\alpha, \beta)& =&\int_0^\infty dm \int_{m}^\infty \frac{dM}{M^6}\, e^{-\frac{\beta}{M^2}}\sin \left(\frac{\alpha \,m}{M} \right)  \, . 
\end{eqnarray}
In the integral over $M$ we make the change of variable $M\to z=M/m$, this yields, after inverting the order of the integrals,
\begin{eqnarray}\label{appendix_Iab3}
J(\alpha, \beta)& =&\int_{1}^\infty \frac{dz}{z^6} \int_0^\infty \frac{dm}{m^5} e^{-\frac{\beta}{(mz)^2}}\sin \left(\frac{\alpha }{z} \right)  \, . 
\end{eqnarray}
In order to decouple the two integrals we change variable $m\to y=zm$ in the integral over $m$. We get that
\begin{eqnarray}\label{appendix_Iab4}
J(\alpha, \beta)& =&\int_{1}^\infty \frac{dz}{z^2} \sin \left(\frac{\alpha }{z} \right) \int_0^\infty \frac{dy}{y^5} e^{-\frac{\beta}{y^2}} \, . 
\end{eqnarray}
The two integrals can now easily computed and one finds the final result
\begin{eqnarray}  
J(\alpha, \beta) = \frac{1-\cos \alpha}{2 \alpha \, \beta^2 } \,,
\end{eqnarray} 
which is given in Eq. (\ref{Iab_2}).

\section{Solution of the integral relation} \label{app:integral}
In this appendix we show that the solution of Eq. (\ref{stieltjes})
\begin{equation}\label{stieltjes_app}
\int_{0}^{1}dy\frac{f(y)}{w-y}=\frac{2}{w}\sqrt{1-\frac{1}{w}}\int_{0}^{\infty}dt \frac{\tanh^2(t)}{\sinh \left(2t\sqrt{1-\frac{1}{w}}\right)} \;.
\end{equation}
is
\begin{equation}\label{solution_appendix}
f(y)=\frac{1}{y}\sum_{n=1}^{\infty}(-1)^{n-1}\tanh^2\left(\frac{n\pi}{2}\sqrt{\frac{y}{1-y}}\right) \;, 
\end{equation}
for $0\leq y\leq 1$.
Moreover, using Eqs. (\ref{stieltjes_app}) and (\ref{solution_appendix}), we show that
\begin{equation}\label{eq:integral_normalization}
\int_{0}^{1}dy\,\sum_{n=1}^{\infty}(-1)^{n+1}\frac{1}{y}\tanh^2\left(\frac{n\pi}{2}\sqrt{\frac{y}{1-y}}\right)=\frac{1}{2}\,,
\end{equation}
which is used in Section \ref{sec:BM} to determine the normalisation constant $A$ of the scaling function $f_{\rm BM}(y)$.\\
First of all, we recognise the left-hand side of Eq. (\ref{stieltjes_app}) as the Stieltjes transform of the function $f(y)$. To invert this Stieltjes transform we use the Sochocki-Plemelj formula \cite{mushk_book}. Setting $w = y + i \epsilon$ with $y$ real, in our case this formula reads
\begin{eqnarray}\label{antitransform_app}
f(y)&=&-\frac{1}{\pi}\lim_{\epsilon\rightarrow 0}\operatorname{Im}\Bigg[ \frac{2}{(y+i\epsilon)}\sqrt{1-\frac{1}{(y+i\epsilon)}}\\ &\times & \int_{0}^{\infty}dt \frac{\tanh^2(t)}{\sinh\left(2t\sqrt{1-\frac{1}{(y+i\epsilon)}}\right)} \Bigg]\,.\nonumber
\end{eqnarray}
We first expand the integrand of the right-hand side of Eq. (\ref{antitransform_app}) for small $\epsilon$ and take the imaginary part
\begin{eqnarray} \label{antitransform2_app}
\operatorname{Im}\left[\frac{1}{(y+i\epsilon)}\sqrt{1-\frac{1}{(y+i\epsilon)}}\frac{1}{\sinh\left(2t\sqrt{1-\frac{1}{(y+i\epsilon)}}\right)}\right]\\ \simeq
\frac{\epsilon}{y^3}\frac{t \cos\left(2t\sqrt{\frac{1-y}{y}}\right)-\frac{3-2y}{2}\sqrt{\frac{y}{1-y}}\sin\left(2t\sqrt{\frac{1-y}{y}}\right)}{\sin^2\left(2t\sqrt{\frac{1-y}{y}}\right)+\frac{\epsilon^2t^2}{y^3(1-y)}\cos^2\left(2t\sqrt{\frac{1-y}{y}}\right)}.\nonumber
\end{eqnarray}
Note that we have kept the leading term of order ${\cal O}(\epsilon^2)$ in the denominator in the second line of Eq. (\ref{antitransform2_app}), so that the integral over $t$ does not diverge. Substituting Eq. (\ref{antitransform2_app}) in Eq. (\ref{antitransform_app}) and making the change of variable $v=2t\sqrt{\frac{1-y}{y}}$, we get
\begin{eqnarray}\label{antitransform3_app}
&&f(y)= \lim_{\epsilon \to 0}\Bigg[-\frac{\epsilon}{2\pi y^2(1-y)}\int_{0}^{\infty}dv \\&\times &\tanh^2\left(\frac{v}{2}\sqrt{\frac{y}{1-y}}\right) \frac{v\cos(v)-(3-2y)\sin(v)}{\sin^2(v)+\frac{\epsilon^2 v^2}{\left(2y\left(1-y\right)\right)^2} \cos^2(y)} \Bigg] \;.\nonumber
\end{eqnarray}
To compute the integral on the right-hand side, we split it as a sum of integrals over $v \in [0,\pi/2]$ and $ v \in [n\pi-\pi/2,n\pi+\pi/2]$ for $n\geq1$. The integral over $[0,\pi/2]$ is convergent (since there is no divergence of the integrand even when $\epsilon \to 0$ in the denominator) and is of order ${\cal O}(\epsilon)$. Thus it vanishes in the limit $\epsilon \to 0$. Hence
\begin{equation}\label{formula_app}
f(y)=\lim_{\epsilon \to 0} \left[ -\frac{\epsilon}{2\pi y^2(1-y)}\sum_{n=1}^{\infty}I_n(y) \right] \;,
\end{equation}
where
\begin{eqnarray}\label{In_app}
I_n(y) &=&\int_{n\pi-\pi/2}^{n\pi+\pi/2}dv \tanh^2\left(\frac{v}{2}\sqrt{\frac{y}{1-y}}\right)\\ &\times & \frac{v\cos(v)-(3-2y)\sin(v)}{\sin^2(v)+\frac{\epsilon^2 v^2}{\left(2y\left(1-y\right)\right)^2} \cos^2(v)} \;.\nonumber
\end{eqnarray}
For $n \geq 1$, we need to keep the ${\cal O}(\epsilon^2)$ regulator in the denominator of the right-hand side of Eq. (\ref{In_app}) since there is a double pole at $v=n \pi$. Therefore, in the $\epsilon \to 0$ limit, the dominant contribution to $I_n(y)$ comes from the neighbourhood of $v = n \pi$. Indeed, setting $v  =n \pi + \epsilon\, z$, we find  to leading order in the small $\epsilon$ limit 
\begin{eqnarray}\label{In_2_app}
I_n(y)&\simeq & \epsilon\int_{-\infty}^{+\infty}dz\tanh^2\left(\frac{n\pi+\epsilon z}{2}\sqrt{\frac{y}{1-y}}\right)\\ & \times & \frac{(n\pi)(-1)^n}{\epsilon^2(z^2+\frac{(n\pi)^2}{(2y(1-y))^2})}\nonumber\\ &=& \frac{2\pi y(1-y)(-1)^n}{\epsilon} \tanh^2\left(\sqrt{\frac{y}{1-y}}\frac{n\pi}{2}\right).\nonumber
\end{eqnarray}
Substituting this result in Eq. (\ref{formula_app}), we see that the limit $\epsilon \to 0$ clearly exists and is given, for $0\leq y \leq 1$, by 
\begin{equation}\label{scaling_fexp_app}
f(y)=\frac{1}{y}\sum_{n=1}^{\infty}(-1)^{n-1}\tanh^2\left(\frac{n\pi}{2}\sqrt{\frac{y}{1-y}}\right) \;.
\end{equation}
We can now show the validity of Eq. (\ref{eq:integral_normalization}). Setting $u=-1/w$ and changing variable $t\to z=2\sqrt{1+u}t$ in Eq. (\ref{stieltjes_app}), we get
\begin{equation}\label{eq:integral_app}
\int_{0}^{1}dy\frac{f(y)}{1+uy}=\int_{0}^{\infty}dz \frac{1}{\sinh(z)}\tanh^2\left(\frac{z}{2\sqrt{1+u}}\right).
\end{equation}
Setting $u=0$ in Eq. (\ref{eq:integral_app}) and plugging the expression for $f$ given in Eq. (\ref{scaling_fexp_app}), we obtain that
\begin{eqnarray}
&&\int_{0}^{1}dy\,\frac{1}{y}\sum_{n=1}^{\infty}(-1)^{n-1}\tanh^2\left(\frac{n\pi}{2}\sqrt{\frac{y}{1-y}}\right)\\ &=&\int_{0}^{\infty}dz \frac{1}{\sinh(z)}\tanh^2\left(\frac{z}{2}\right)\nonumber \,.
\end{eqnarray}
The integral in the second line is equal to $\frac{1}{2}$ and hence we obtain Eq. (\ref{eq:integral_normalization}).

\section{Direct proof of $P(\tau=T|T)=1/(2T)$ for lattice walks} \label{app:limit}
In Section \ref{sec:RW} we show that for RWs with finite jump variance the probability distribution $P(\tau|n)$, in the limit of large $n$, approaches the scaling form 
\begin{equation}
P(\tau|n)\rightarrow \frac{1}{n}f_{\rm BM}\left(\frac{\tau}{n}\right)\,,
\end{equation}
where the scaling function $f_{\rm BM}(y)$ is given in Eq. (\ref{eq:f_bm}). The asymptotics of this function $f_{\rm BM}(y)$ have been derived in Section \ref{sec:BM}. In particular, in the limit $y\to 1$, we obtain that (see Eq. (\ref{asymptote_bm}))
\begin{equation}
f_{\rm BM}(y)\to \frac{1}{2}\,.
\end{equation}
This result indicates that, for large $n$,  
\begin{eqnarray} \label{P1/2}
P(\tau =  n|n) \approx \frac{1}{2\,n} \;. 
\end{eqnarray}
Remarkably, in Section (\ref{sec:universal}) we show that this result is valid even for finite $n$ in the case of continuous-space RWs.\\
In this appendix we show that this result (\ref{P1/2_2}) can be proved directly in the case of lattice walks, corresponding to the discrete jump distribution
\begin{equation}
p(\eta)=(1/2)\delta(|\eta|-1)\,.
\end{equation} 
Due to the discrete-space nature of these lattice walks, the result in Eq. (\ref{P1/2_2}) is only valid for large $n$. As explained in Section \ref{sec:RW}, since we are considering a discrete-space random walk, we need to be careful when defining the global minimum and the global maximum. Indeed, the global extrema of a lattice walk will be degenerate with finite probability. Hence, for simplicity, we define $n_{\min}$ ($n_{\max}$) as the time at which the global minimum (maximum) is reached for the first time. In order to reach the global minimum $-x_{\min}$ at step $n$ for the first time, the random walk has first to arrive at position $-x_{\min}+1$ at time $n-1$ and then, with probability $1/2$, to jump down. To enforce that the global maximum is attained for the first time at step $n_{\max}=0$, we need to impose that the walker starts at the origin and that $x_k\leq 0$ for any $k$. Overall, we need to compute the probability that the walker starts from the origin and arrives at position $-x_{\min}+1$ after $n-1$ steps, remaining always in the space-interval $[-x_{\min}+1,0]$, and then it jumps to position $-x_{\min}$. Using, the $x\to-x$ symmetry and summing over $x_{\min}$, we obtain that
\begin{equation}
P(\tau=n|n)=\sum_{x_{\min}=1}^{\infty}G(x_{\min}-1,n|x_{\min}-1)\,\times \frac{1}{2}\,,
\end{equation}
where $G(x,l|M)$ is the probability to go from the origin to position $x$ in $l$ steps, always remaining inside the interval $[0,M]$, and the factor $1/2$ is the probability of the last jump. Multiplying both terms by $s^n$ and we summing over $n$, we obtain
\begin{equation}
\sum_{n=1}^{\infty}P(\tau=n|n)s^n=\frac{1}{2}\sum_{x_{\min}=1}^{\infty}\tilde{G}(x_{\min}-1,s|x_{\min}-1)\,,
\end{equation}
where the generating function $\tilde{G}(x_{\min}-1,s|x_{\min}-1)$ is given in Eqs. (\ref{eq:solution_G_tilde}) and (\ref{eq:definition_w}). We want to study the large $n$ limit, which corresponds to the $s\to 1$ limit. Thus, we set $s=e^{-p}$ and we take the limit $p\rightarrow 0$:
\begin{equation}
\sum_{n=1}^{\infty}P(\tau=n|n)e^{-pn}=\frac{1}{2}\sum_{x_{\min}=1}^{\infty}\tilde{G}(x_{\min}-1,e^{-p}|x_{\min}-1)\,.
\end{equation}
In the limit $p\to 0$ we expect the sum on the right-hand side to be dominated by large values of $x_{\min}$, hence we can approximate $x_{\min}-1\simeq x_{\min}$. Moreover, when $p$ is small it is reasonable to approximate the sums in both terms with integrals. This yields
\begin{equation}\label{eq:appendix_lw}
\int_{0}^{\infty}dn\,P(\tau=n|n)e^{-pn}=\frac{1}{2}\int_{0}^{\infty}dx_{\min}\,\tilde{G}(x_{\min},e^{-p}|x_{\min})\,.
\end{equation}
Expanding the expression for $\tilde{G}(x_{\min},e^{-p}|x_{\min})$ in Eqs. (\ref{eq:solution_G_tilde}) and (\ref{eq:definition_w}) for small $p$, we get
\begin{equation}\label{eq:appendix_tilde_G}
\tilde{G}(x_{\min},e^{-p}|x_{\min})\simeq \frac{2\sqrt{2p}}{\sinh(x_{\min}\sqrt{2p})}\,.
\end{equation}
Plugging this expression (\ref{eq:appendix_tilde_G}) into Eq. (\ref{eq:appendix_lw}) and changing variable $x_{\min}\to x=\sqrt{2}x_{\min}$, we get
\begin{equation}\label{eq:appendix_lw_2}
\int_{0}^{\infty}dn\,P(\tau=n|n)e^{-pn}=\int_{0}^{\infty}dx\,\frac{\sqrt{p}}{\sinh(x \sqrt{p})}\,.
\end{equation}
We next invert this Laplace transform with respect to $p$ using the identity \cite{prudnikov}
\begin{equation}\label{eq_sinh}
\frac{\sqrt{p}}{\sinh\left(\sqrt{p}\right)}=\sum_{m=1}^{\infty}\frac{2m^2\pi^2(-1)^{m+1}}{p+m^2\pi^2} \;,
\end{equation}
and noting that each term on the right-hand side corresponds to a simple pole in the complex $p$-plane. Hence the
inversion of the Laplace transform becomes simple and we get
\begin{eqnarray}\label{eq_sinh_2}
G\left(x_{\min},n|x_{\min} \right) &=& \frac{2\pi^2}{x_{\min}^3}\sum_{m=0}^{\infty}(-1)^{m+1}m^2e^{-\frac{m^2\pi^2}{x_{\min}^2}n}\\ &=& \frac{1}{n}\frac{d}{dx_{\min}}\left[\sum_{m=0}^{\infty}(-1)^{m+1}e^{-\frac{m^2\pi^2}{x_{\min}^2}n}\right] \;.\nonumber
\end{eqnarray}
Integrating over $x_{\min}$, Eq. (\ref{P1/2_2}) gives
\begin{eqnarray}\label{P1/2_final}
P(\tau  = n|n) &=& \int_0^\infty G(x_{\min}, n |x_{\min}) d x_{\min}\\ & \approx & \frac{1}{n}\left(\sum_{m=0}^{\infty}(-1)^{m+1}+1\right)=\frac{1}{2n}\;.\nonumber
\end{eqnarray}
Note that, in the last line, we have used the regularisation as in Eq. (\ref{yto1_2}) to evaluate the sum on the second line.  \\

\section{Probability distribution of $\tau=n_{\min}-n_{\max}$ for random walk bridges.}
\label{app:rw_bridges}

In Section \ref{sec:RW} we have shown that, in the case of discrete-time random walks with finite jump variance, the probability distribution of $\tau=n_{\min}-n_{\max}$ converges, in the limit of large number of steps $n$, to the Brownian result in Eq. (\ref{eq:f_bm}). One may wander whether a similar result holds also in the case of random walk bridges, i.e. random walks with the additional constraint that they have to go back to the origin at the final step. More precisely, we consider a time series $x_k$, with $k=0,1,\ldots,n$, generated by the Markov rule
\begin{equation}
x_k=x_{k-1}+\eta_k\,,
\end{equation}
with initial condition $x_0=0$ and with the constraint  $x_n=0$. The jumps $\eta_k$ are IID random variables with PDF $p(\eta)$, which is assumed to be symmetric around zero. When the jump variance $\sigma^2=\int_{-\infty}^{\infty}d\eta\,p(\eta)\eta^2$ is finite, the Central Limit Theorem states that, in the limit of large $n$, the stochastic process $x_k$ converges to a Brownian bridge. Thus, we expect that also the probability distribution of the time $\tau$ between the global maximum and the global minimum converges to the result in Eq. (\ref{eq:f_bb}), obtained in the case of a Brownian bridge. Here we directly verify this convergence in the case of the double-exponential jump distribution $p(\eta)=(1/2)e^{-|\eta|}$. A similar result can be easily obtained also in the case of lattice walks. First of all, we notice that, due to the bridge constraint, the probability distribution $P(\tau|n)$ is now implicitly conditioned to the fact that the final position is zero. Thus, using Bayes' theorem, we obtain that
\begin{equation}\label{eq:beyes}
P(\tau|n)=\frac{P(\tau,x_n=0|n)}{P(x_n=0|n)}\,.
\end{equation}
We first compute the denominator $P(x_n=0|n)$, which is the probability that an unconstrained random walk goes back to the origin at step $n$. We define the propagator $G(x,n)$ as the probability that the walker is at position $x$ after $n$ steps. Note that $P(x_n=0|n)=G(0,n)$ and that the initial condition is $G(x,0)=\delta(x)$. Using the Markov property, we can write down a recursion relation for $G(x,n)$:
\begin{equation}\label{eq:integral_recursion_app}
G(x,n)=\int_{-\infty}^{\infty}dx'\,G(x',n-1)p(x-x')\,,
\end{equation}
for any $n\geq 1$. This equation means that in order to arrive at position $x$ at step $n$, the walker must have been at some position $x'$ at step $n-1$ and then it must have jumped, with probability $p(x-x')$, to position $x$. In the case of the double-exponential distribution $p(\eta)=(1/2)e^{-|\eta|}$, as explained in Section \ref{sec:RW}, one can solve this kind of integral equations using the fact that 
\begin{equation}\label{eq:trick}
p''(x)=p(x)-\delta(x)\,,
\end{equation}
It is convenient to consider the generating function of $G(x,n)$:
\begin{equation}
\tilde{G}(x,s)=\sum_{n=1}^{\infty}G(x,n)s^n\,.
\end{equation}
Multiplying both terms of Eq. (\ref{eq:integral_recursion_app}) and summing over $n\geq1$, we obtain 
\begin{equation}\label{eq:integral_app_lt}
\tilde{G}(x,s)=s\int_{-\infty}^{\infty}dx'\,\tilde{G}(x',s)p(x-x')+s\,p(x)\,,
\end{equation}
where we have used the initial condition $G(x,0)=\delta(x)$. Differentiating Eq. (\ref{eq:integral_app_lt}) twice with respect to $x$ and using Eq. (\ref{eq:trick}), we obtain
\begin{equation}\label{eq:differential_app}
\frac{\partial^2\tilde{G}(x,s)}{\partial x^2}=(1-s)\tilde{G}(x,s)-s\delta(x)\,.
\end{equation}
When $x>0$, the $\delta$-function disappears and the most general solution of the differential equation (\ref{eq:differential_app}) is
\begin{equation}\label{eq:tilde_G_+}
\tilde{G}(x,s)=A_{+}(s)e^{-\sqrt{1-s}x}+B_{+}(s)e^{\sqrt{1-s}x}\,,
\end{equation}
where $A_{+}(s)$ and $B_{+}(s)$ are two arbitrary constants. Similarly, for $x<0$, we obtain
\begin{equation}\label{eq:tilde_G_-}
\tilde{G}(x,s)=A_{-}(s)e^{\sqrt{1-s}x}+B_{-}(s)e^{-\sqrt{1-s}x}\,,
\end{equation}
where $A_{-}$ and $B_{-}$ are again arbitrary constants. First of all, in the limit $x\to\infty$ we know that the propagator $G(x,n)$ goes asymptotically to zero. This implies that $\tilde{G}(x,s)$ cannot diverge when $x\to\infty$ and hence we obtain that $B_{+}=0$. Similarly, considering the limit $x\to -\infty$, we obtain that $B_{-}=0$. Moreover, imposing that $\tilde{G}(x,s)$ is continuous at the origin, we obtain that
\begin{equation}
A_{+}=A_{-}\equiv A\,.
\end{equation}
Finally, to determine the constant $A$, we integrate Eq. (\ref{eq:differential_app}) for $x\in(-\epsilon,\epsilon)$. This yields
\begin{equation}
\frac{\partial\tilde{G}(\epsilon,s)}{\partial x}-\frac{\partial\tilde{G}(-\epsilon,s)}{\partial x}=\int_{-\epsilon}^{\epsilon}dx\,(1-s)\tilde{G}(x,s)-s\,.
\end{equation}
Taking the limit $\epsilon\to 0$ the integral on the right-hand side vanishes and we obtain the condition
\begin{equation}
\frac{\partial\tilde{G}(0^+,s)}{\partial x}-\frac{\partial\tilde{G}(0^-,s)}{\partial x}=-s\,.
\end{equation}
Using Eqs. (\ref{eq:tilde_G_+}) and (\ref{eq:tilde_G_-}) and setting $A_+=A_-=A$, we obtain
\begin{equation}
A=\frac{s}{2\sqrt{1-s}}\,.
\end{equation}
Thus, the generating function of $G(x,n)$ is given by
\begin{equation}\label{eq:tilde_G_2}
\tilde{G}(x,s)=\frac{s}{2\sqrt{1-s}}e^{-\sqrt{1-s}|x|}\,.
\end{equation}
We are interested in the large $n$ limit, which corresponds to the limit $s\to 1$. Thus, it is convenient to parametrise $s=e^{-p}$ and to consider the limit $p\to 0$. In this limit, the sum over $n$ in the definition of $\tilde{G}(x,s)$ can be approximated with an integral. Hence, using $s=e^{-p}$ and expanding
the right-hand side of Eq. (\ref{eq:tilde_G_2}) for small $p$, we obtain
\begin{equation}
\int_{0}^{\infty}dn	\, G(x,n)\,e^{-pn}\simeq\frac{1}{2\sqrt{p}}e^{-\sqrt{p}|x|}\,.
\end{equation}
Inverting the Laplace transform, we obtain that in the large $n$ limit
\begin{equation}
G(x,n)\simeq \frac{1}{\sqrt{4\pi n}}e^{-\frac{x^2}{4n}}\,.
\end{equation}
Setting $x=0$ and using $P(x_n=0|n)=G(0,n)$, we obtain that for large $n$
\begin{equation}\label{eq:prob_bridge}
P(x_n=0|n)\simeq \frac{1}{\sqrt{4\pi n}}\,.
\end{equation}
To proceed, we need to determine the probability $P(\tau,x_n=0|n)$. The method to compute this probability is similar to the one presented in Section \ref{sec:RW}, with the only difference that we do not need to integrate over the final position $x_n$, which is instead fixed. We will first write the joint probability $P(x_{\min},x_{\max},n_{\min},n_{\max},x_n=0|n)$ of the global minimum $x_{\min}$, the global maximum $x_{\max}$, the time of the minimum $n_{\min}$, the time of the maximum $n_{\max}$ and the event ``$x_n=0$''. Considering the case $n_{\min}>n_{\max}$, this probability can be computed as a product of three factors $P_{\rm I}$, $P_{\rm II}$, and $P_{\rm III}$, corresponding to the three segments in Fig. \ref{fig:exponential}): $0\leq k\leq n_{\max}$ (I), $n_{\max}\leq k\leq n_{\min}$ (II), and $n_{\min}\leq k\leq n$ (III). Each of these probability factors can be expressed in terms of the restricted Greens' function $G(x,n|M)$, defined as the probability that the walker goes from the origin to position $x$ in $n$ steps, without leaving the space-interval $[0,M]$. We recall that in our case $M=x_{\min}+x_{\max}$. The generating function of $G(x,n|M)$ has been computed in Section \ref{sec:RW} and is given by (see Eqs. (\ref{A}) and (\ref{g_solution})):
\begin{eqnarray}\label{eq:g_solution_app}
&& \tilde{G}\left(x,s|M \right)= A(s,M)\\ &\times & \left[e^{-\sqrt{1-s}\,x}-\frac{1-\sqrt{1-s}}{1+\sqrt{1-s}}e^{-\sqrt{1-s}\,(2M-x)}\right] \,,\nonumber
\end{eqnarray}
where
\begin{eqnarray}\label{eq:A_app}
&& A(s,M)=\frac{1-\sqrt{1-s}}{1-\left(\frac{1-\sqrt{1-s}}{1+\sqrt{1-s}}\right)^2 \, e^{-2\sqrt{1-s} \,M}}\,.
\end{eqnarray}
The probabilities of the first two segments are exactly identical to the ones computed in the case of random walks. Thus,
\begin{equation}
P_{\rm I}=G(x_{\max},l_1|M)\,
\end{equation}
where $l_1=n_{\max}$ and
\begin{equation}
P_{\rm II}=G(M,l_2|M)\,,
\end{equation}
where $l_2=n_{\min}-n_{\max}$. 
The probability of the third segment is modified as follows to take into account the constraint $x_n=0$
\begin{equation}
P_{\rm III}=G(x_{\min},l_3|M)\,,
\end{equation}
where $l_3=n-n_{\max}$. The grand joint PDF $P(x_{\min},x_{\max},n_{\min},n_{\max},x_n=0|n)$ is given by the product of the three factors above
\begin{eqnarray}\label{eq:product_appendix}
&&P(x_{\min},x_{\max},n_{\min},n_{\max},x_n=0|n)=P_{\rm I}P_{\rm II } P_{\rm III} \nonumber\\
&=&G(x_{\max},l_1|M)G(M,l_2|M)G(x_{\min},l_3|M)\,, 
\end{eqnarray}
where $M=x_{\min}+x_{\max}$. It is useful to express the left-hand side in terms of the intervals $l_1$, $l_2$ and $l_3$, hence we define
\begin{eqnarray}
&&P(x_{\min},x_{\max},n_{\min},n_{\max},x_n=0|n)\\ &\equiv &P(x_{\min},x_{\max},l_1,l_2,l_3,x_n=0)\,.\nonumber
\end{eqnarray}
Multiplying both terms of Eq. (\ref{eq:product_appendix}) by $s_1^{l_1}s_2^{l_2}s_3^{l_3}$ and summing over $l_1$, $l_2$ and $l_3$, we get
\begin{eqnarray}\label{eq:laplace_app}
&&\sum_{l_1,l_2,l_3}
P\left(x_{\min},x_{\max},l_1,l_2,l_3,x_n=0\right)s_1^{l_1}s_2^{l_2}s_3^{l_3}\\ &=&\tilde{G}\left(x_{\max},s_1|M \right)\tilde{G}\left(M,s_2|M \right)\,   \tilde{G}\left(M-x_{\max},s_3|M \right) \,,\nonumber
\end{eqnarray}
where $\tilde{G}(x,s|M)$ is given in Eq. (\ref{eq:g_solution_app}).
We now integrate both terms of Eq. (\ref{eq:laplace_app}) over $x_{\min}$ and $x_{\max}$ in order to obtain the marginal probability $P(l_1,l_2,l_3,x_n=0)$ of $l_1$, $l_2$ and $l_3$.
Making a change of variables $(x_{\min}, x_{\max}) \to (x_{\max}, M = x_{\max} + x_{\min})$, we obtain
\begin{eqnarray}\label{eq:gen_fun_1_app}
&&\sum_{l_1,l_2,l_3}
P\left(l_1,l_2,l_3,x_n=0\right)s_1^{l_1}s_2^{l_2}s_3^{l_3}\\ &= & \int_{0}^{\infty}dM\,\tilde{G}\left(M,s_2|M \right)\nonumber 
\int_{0}^{M}dx_{\max}\,\tilde{G}\left(x_{\max},s_1|M \right)\,\\ &\times &\tilde{G}\left(M-x_{\max},s_3|M \right)
 \,, \nonumber
\end{eqnarray}
To compute the marginal probability $P(\tau,x_n=0|n)$ of $\tau=n_{\min}-n_{\max}$ and of the event ``$x_n=0$'', we write it in terms of the joint PDF $P(l_1,l_2,l_3,x_n=0)$ as follows
\begin{eqnarray}
&& P(\tau,x_n=0|n)\\
&=& \sum_{l_1,l_2=1}^{\infty}  P(l_1,l_2=\tau,l_3,x_n=0)\delta(l_1+\tau+l_3-n)\nonumber
\end{eqnarray}
Multiplying both terms by $s_2^\tau s^n$ and summing over $n$ and $\tau$, we obtain
\begin{eqnarray}\label{eq:relation2_app}
&& \sum_{ n=1}^{\infty}\sum_{ \tau=1}^{n}P(\tau,x_n=0|n) s_2^\tau \, s^n\\& =& \sum_{l_1,\tau,l_3=1}^{\infty} P(l_1, l_2 = \tau, l_3,x_n=0) s^{l_1} (s\, s_2)^{\tau}\, s^{l_3}\, .\nonumber
\end{eqnarray}
Notice that the right-hand side of Eq. (\ref{eq:relation2_app}) can be read off Eq. (\ref{eq:gen_fun_1_app}) by setting $s_1 \to s$, $s_2 \to s\, s_2$ and $s_3 \to s$.  This yields
\begin{eqnarray}\label{eq:int_p_tau}
&&\sum_{ n,\tau} P(\tau,x_n=0|n) s_2^\tau \, s^n 
\\ &=& \int_{0}^{\infty}dM\,\tilde{G}\left(M,s\,s_2|M \right) 
\int_{0}^{M}dx_{\max}\,\tilde{G}\left(x_{\max},s|M \right)\nonumber \\ &\times &\,\tilde{G}\left(M-x_{\max},s|M \right)
 \,, \nonumber
\end{eqnarray}
We are interested in the limit $\tau,n\to\infty$ with $y=\tau/n$ fixed. Thus, it is useful to parametrise the variables $s$ and $s_2$ as $s=e^{-\lambda}$ and $s_2=e^{-\lambda_2}$ and to take the limit $\lambda,\lambda_2\to 0$ with $u=\lambda_2/\lambda$ fixed. In this limit the double sum on left-hand side of Eq. (\ref{eq:int_p_tau}) can be approximated with a double integral. Expanding the right-hand side of Eq. (\ref{eq:int_p_tau}) to leading order in $\lambda$ and $\lambda_2$, we obtain
\begin{eqnarray}
&&\int_{0}^{\infty}dn\,\int_{0}^{n}d\tau\, P(\tau,x_n=0|n) e^{-\lambda_2 \tau} \, e^{-\lambda n}
\\ &=& \int_{0}^{\infty}dM \frac{\sqrt{\lambda+\lambda_2}}{\sinh(\sqrt{\lambda+\lambda_2}M)} \nonumber \\ &\times &\int_{0}^{M}dx_{\max}\frac{\sinh(\sqrt{\lambda }x_{\max})\sinh(\sqrt{\lambda}(M- x_{\max}))}{\sinh^2 (\sqrt{\lambda}M)} \,. \nonumber
\end{eqnarray}
Performing the integral over $x_{\max}$ we obtain
\begin{eqnarray}\label{eq:double_laplace_app}
&&\int_{0}^{\infty}dn\,\int_{0}^{n}d\tau\, P(\tau,x_n=0|n) e^{-\lambda_2 \tau} \, e^{-\lambda n}
\\ &=& \int_{0}^{\infty}dM \frac{\sqrt{\lambda+\lambda_2}}{2\sqrt{\lambda}} \frac{M\sqrt{\lambda}\coth(M\sqrt{\lambda})-1}{\sinh(\sqrt{\lambda+\lambda_2}M)\sinh (\sqrt{\lambda}M)} \,. \nonumber
\end{eqnarray}
In order to invert the double Laplace transform in Eq. (\ref{eq:double_laplace_app}) we use the following identities \cite{prudnikov}:
\begin{equation}\label{eq:identity1_app}
\frac{\sqrt{p}}{\sinh(\sqrt{p})}=2\pi^2\sum_{m=1}^{\infty}\frac{(-1)^{m+1}\,m^2}{p+m^2 \pi^2}\,,
\end{equation}
\begin{equation}\label{eq:identity2_app}
\frac{\sqrt{q}\coth(\sqrt{q})-1}{\sqrt{q}\sinh(\sqrt{q})}=4\pi^2\sum_{k=1}^{\infty}\frac{(-1)^{k+1}\,k^2}{(q+k^2 \pi^2)^2}\,.
\end{equation}
Using these relations (\ref{eq:identity1_app}) and (\ref{eq:identity2_app}) with $p=(\lambda+\lambda_2)M^2$ and $q=\lambda\,M^2$ in Eq. (\ref{eq:double_laplace_app}), we obtain
\begin{eqnarray}\label{eq:double_laplace_app_2}
&&\int_{0}^{\infty}dn\,\int_{0}^{n}d\tau\, P(\tau,x_n=0|n) e^{-\lambda_2 \tau} \, e^{-\lambda n}
\\ &=& \int_{0}^{\infty}dM \frac{4 \pi^4}{M^6}\sum_{m,k=1}^{\infty}(-1)^{m+k}\frac{m^2}{\lambda+\lambda_2+\frac{m^2\pi^2}{M^2}}\nonumber \\ &\times& \frac{k^2}{\left(\lambda+\frac{k^2\pi^2}{M^2}\right)^2} \,. \nonumber
\end{eqnarray}
To invert the double Laplace transform on the left-hand side of Eq. (\ref{eq:double_laplace_app_2}) we perform the change of variables $(\tau,n)\to(\tau,\bar{\tau}=n-\tau)$. Thus, defining $\lambda_3=\lambda+\lambda_2$, one obtains
\begin{eqnarray}\label{eq:double_laplace_app_3}
&&\int_{0}^{\infty}d\bar{\tau}\,\int_{0}^{\infty}d\tau\, P(\tau,x_n=0|\tau+\bar{\tau})\\ &\times & e^{-\lambda_3 \tau} \, e^{-\lambda \bar{\tau}}
 = \int_{0}^{\infty}dM \frac{4 \pi^4}{M^6}\sum_{m,k=1}^{\infty}(-1)^{m+k}\frac{m^2}{\lambda_3+\frac{m^2\pi^2}{M^2}}\nonumber \\ &\times& \frac{k^2}{\left(\lambda+\frac{k^2\pi^2}{M^2}\right)^2} \,. \nonumber
\end{eqnarray}
We can now invert the double Laplace transform noticing that the term in $\lambda$ on the right-hand side corresponds to a single pole in the complex plane, while the term in $\lambda_3$ corresponds to a double pole. This yields, using $\bar{\tau}=n-\tau$,
\begin{eqnarray}
&&P(\tau,x_n=0|n) = \int_{0}^{\infty}dM \frac{4 \pi^4}{M^6}\sum_{m,k=1}^{\infty}\\ &\times & (-1)^{m+k} m^2 k^2 (n-\tau) e^{-\frac{m^2\pi^2}{M^2}\tau-\frac{k^2\pi^2}{M^2}(n-\tau)}\nonumber
\end{eqnarray}
Performing the integral over $M$, we obtain, after few steps of algebra
\begin{eqnarray}
&&P(\tau,x_n=0|n) \\ &=&  \frac{3 }{2 \sqrt{\pi}}\sum_{m,k=1}^{\infty}\frac{(-1)^{m+k} m^2 k^2 (n-\tau) }{\left(m^2 \tau+k^2 (n-\tau)\right)^{5/2}}\nonumber
\end{eqnarray}
Finally, using Eqs. (\ref{eq:beyes}) and (\ref{eq:prob_bridge}) we obtain that, in the large $n$ limit,
\begin{eqnarray}
&&P(\tau|n) \\ &=&  3\sqrt{n}\sum_{m,k=1}^{\infty}\frac{(-1)^{m+k} m^2 k^2 (n-\tau) }{\left(m^2 \tau+k^2 (n-\tau)\right)^{5/2}}\nonumber
\end{eqnarray}
This expression can be rewritten in the following scaling form
\begin{equation}\label{eq:scal_f_bb}
P(\tau|n)=\frac{1}{n}f_{\rm BB}\left(\frac{\tau}{n}\right)\,,
\end{equation}
where the scaling function $f_{\rm BB}(y)$ is the scaling function in Eq. (\ref{eq:f_bb}), which was obtained in the case of continuous-time Brownian bridges. Thus, we have directly verified the prediction of the Central Limit Theorem.\\
Eq. (\ref{eq:double_laplace_app}) can also be used to compute a useful integral relation for $f_{\rm BB}(y)$, which allows to determine the moments of $\tau$ for a Brownian bridge. Indeed, using Eq. (\ref{eq:beyes}) we can rewrite the left-hand side of Eq. (\ref{eq:double_laplace_app}) as
\begin{eqnarray}
&&\int_{0}^{\infty}dn\,\int_{0}^{n}d\tau\,  P(\tau,x_n=0|n) e^{-\lambda_2 \tau} \, e^{-\lambda n}\\ &=&\int_{0}^{\infty}dn\,\int_{0}^{n}d\tau\, P(\tau|n)P(x_n=0|n) e^{-\lambda_2 \tau} \, e^{-\lambda n}\, .\nonumber
\end{eqnarray}
Using the expression for $P(x_n=0|n)$ in Eq. (\ref{eq:prob_bridge}) and using the scaling form in Eq. (\ref{eq:scal_f_bb}), we get
\begin{eqnarray}
&&\int_{0}^{\infty}dn\,\int_{0}^{n}d\tau\, P(\tau|n)P(x_n=0|n) e^{-\lambda_2 \tau} \, e^{-\lambda n} \nonumber
\\ &=&\int_{0}^{\infty}dn\,\int_{0}^{n}d\tau\, \frac{1}{n}f_{\rm BB}\left(\frac{\tau}{n}\right)\frac{1}{2\sqrt{\pi\,n}} e^{-\lambda_2 \tau} \, e^{-\lambda n}\nonumber \\&=& \int_{0}^{\infty}dn\,\frac{1}{2\sqrt{\pi\,n}} \int_{0}^{1}dy\, f_{\rm BB}(y)  \, e^{-(\lambda+\lambda_2 y) n}
\\ &=& \frac{1}{2}\int_{0}^{1}dy\, \frac{f_{\rm BB}(y)}{\sqrt{\lambda + \lambda_2 y}}\, ,\nonumber
\end{eqnarray}
where we have performed the change of variable $y=\tau/n$ in going from the second to the third line above. Changing variable $z=M\sqrt{\lambda+\lambda_2}$ in the right-hand side of Eq. (\ref{eq:double_laplace_app}), we get
\begin{eqnarray}
&&\frac{1}{2}\int_{0}^{1}dy\, \frac{f_{\rm BB}(y)}{\sqrt{\lambda + \lambda_2 y}}\\ &=&
\int_{0}^{\infty}dz \frac{1}{2\sqrt{\lambda}} \frac{z\frac{\sqrt{\lambda}}{\sqrt{\lambda_2+\lambda}}\coth\left(z\frac{\sqrt{\lambda}}{\sqrt{\lambda_2+\lambda}}\right)-1}{\sinh(z)\sinh \left(z\frac{\sqrt{\lambda}}{\sqrt{\lambda_2+\lambda}}\right)} \,. \nonumber
\end{eqnarray} 
Defining $u=\lambda_2/\lambda$, after few steps of algebra, we get
\begin{equation}
\int_{0}^{1}dy\, \frac{f_{\rm BB}(y)}{\sqrt{1+ u y}}=
\int_{0}^{\infty}dz  \frac{\frac{z}{\sqrt{1+u}}\coth\left(\frac{z}{\sqrt{1+u}}\right)-1}{\sinh(z)\sinh \left(\frac{z}{\sqrt{1+u}}\right)} \,. 
\end{equation} 
As explained in the main text, this integral relation is useful to compute the moments of $\tau$ for a Brownian bridge.
\section{Proof of $P(\tau=n|n)=1/(2n)$ for $n\le 3$ for discrete-time random walks}
\label{app:n3}
We consider a discrete-time RW on the line generated by the Markov jump process
\begin{equation}
\label{evol_rw.1}
x_k=x_{k-1}+\eta_k\,; \quad x_0=0\, ,
\end{equation}
where $\eta_k$'s IID variables each drawn from a symmetric and continuous PDF $p(\eta)$. We are interested
in computing the probability $p_n$ of the event that $``\tau=n_{\min}-n_{\max}=n''$, where $n_{\min}$ and $n_{\max}$
denote respectively the time of the global minimum and the global maximum and $n$ is the total number of steps. This event then corresponds to trajectories where for the $n$-step walk, the maximum occurs at $x_0=0$ and the minimum occurs at the
$n$-step. This then corresponds to trajectories that start at the origin $x_0=0$, stay non-positive
up to $n$-steps and, in addition, the position $x_n$ is the global minimum. By symmetry of the walk, $p_n$
also counts the probability that $n_{\min}=0$ and $n_{\max}=n$, i.e., the probability that the walker
starts at the origin, stays non-negative up to step $n$ and additionally, the position $x_n$ is the global maximum.
Mathematically, the latter event can be expressed as
\begin{equation}
p_n= \Big\langle \theta(x_1)\theta(x_2)\ldots \theta(x_n)\, \mathbbm{1} \left(x_n>M_{n-1}\right) 
\Big\rangle\, ,
\label{pn_def.1}
\end{equation}
where 
\begin{equation}\label{Mn}
M_{n-1}=\max_{0\leq k \leq n-1}x_k
\end{equation}
is the global maximum up to step $n-1$ and $\theta(x)$ is the Heaviside theta function: $\theta(x)=1$ if $x>0$ and $\theta(x)=0$ if $x<0$. The indicator function
$\mathbbm{1}$ denotes the event that $x_n$ is bigger than all previous values, so that $x_n$ is the global maximum. The average
$\langle\rangle$ is over the joint distribution of the IID noises $\{\eta_1,\eta_2,\,\ldots, \eta_n\}$ 
\begin{equation}
P(\eta_1,\eta_2,\ldots, \eta_n)= \prod_{i=1}^n p(\eta_i)\, .
\label{factorised_jpdf}
\end{equation}
Our conjecture is that for symmetric and continuous $p(\eta)$
\begin{equation}
p_n= \frac{1}{2n}\, \quad {\rm for}\,\, {\rm all}\,\, n\ge 1\, .
\label{conjecture.1}
\end{equation}
In this appendix, we prove this conjecture for $n\le 3$.

\vskip 0.4cm

\noindent {\bf {The case $n=1$:}} This case is trivial, because one gets from Eq. (\ref{pn_def.1})
\begin{equation}
p_1= \langle \theta(x_1)\rangle= \langle \eta_1\rangle =\frac{1}{2}\, ,
\label{p1.1}
\end{equation}
where we used $x_1=\eta_1$ from Eq. (\ref{evol_rw.1}).
\vskip 0.4cm

\noindent {\bf {The case $n=2$:}} For a $2$-step walk, Eq. (\ref{pn_def.1}) reads, using Eq. (\ref{evol_rw.1})
\begin{equation}
p_2= \langle \theta(x_1)\theta(x_2) \theta(x_2-x_1)\rangle= \langle \theta(\eta_1)\theta(\eta_1+\eta_2)\theta(\eta_2)\rangle\, .
\label{p2.1}
\end{equation}
However, if $\eta_1>0$ and $\eta_2> 0$, one automatically has $\eta_1+\eta_2>0$. Hence, Eq. (\ref{p2.1}) simply reduces to
\begin{equation}
p_2= \langle \theta(\eta_1)\theta(\eta_2)\rangle= \langle \theta(\eta_1)\rangle \langle \theta(\eta_2)\rangle=\frac{1}{4}\, .
\label{p2.2}
\end{equation}

\noindent {\bf {The case $n=3$:}} Already the case $n=3$ starts to be nontrivial. In this case, $p_3$ in 
Eq. (\ref{pn_def.1}) counts the events that $x_1=\eta_1>0$, $x_2=\eta_1+\eta_2>0$, $x_3=\eta_1+\eta_2+\eta_3>0$,
and in addition, $x_3-x_2=\eta_3>0$, $x_3-x_1=\eta_2+\eta_3>0$. Thus one can write $p_3$ in Eq. (\ref{pn_def.1})
in terms of $\eta_i$'s as
\begin{equation}
p_3= \langle \theta(\eta_1)\theta(\eta_3) \theta(\eta_1+\eta_2)\theta(\eta_2+\eta_3)\rangle \, .
\label{p3.1}
\end{equation}
Furthermore, for given $\eta_1$ and $\eta_2$, the event $\eta_2+\eta_1>0$ and $\eta_2+\eta_3>0$ is equivalent
to the event $\eta_2> \max(-\eta_1,-\eta_3)$. Hence Eq. (\ref{p3.1}), using Eq. (\ref{factorised_jpdf}), can be
expressed as 
\begin{equation}
p_3 = \int_0^{\infty} d\eta_1 \,p(\eta_1)\int_{0}^{\infty} d\eta_3 \,p(\eta_3)\,
\int_{\max(-\eta_1,-\eta_3)}^\infty d\eta_2\, p(\eta_2)\, .
\label{p3.2}
\end{equation}
Consider the integrand in the $(\eta_1\ge 0, \eta_3\ge 0)$ quadrant where it is symmetric under the exchange of $\eta_1$ and $\eta_3$.
Hence, we restrict the integral in the region $\eta_1>\eta_3>0$ where $\max(-\eta_1,-\eta_3)=-\eta_3$ and we get
\begin{equation}
p_3 = 2 \int_0^{\infty} d\eta_1 \,p(\eta_1)\int_{0}^{\eta_1} d\eta_3 \,p(\eta_3)\,
\int_{-\eta_3}^\infty d\eta_2\, p(\eta_2)\, .
\label{p3.3}
\end{equation} 
where the factor $2$ comes from the symmetric contribution from the region $\eta_3>\eta_1>0$. Furthermore, using
the fact that $p(\eta)$ is symmetric, we can write
\begin{equation}
p_3= 2 \int_0^{\infty} d\eta_1 \,p(\eta_1)\int_{0}^{\eta_1} d\eta_3 \,p(\eta_3)\, \left[\frac{1}{2}+ \int_0^{\eta_3} 
d\eta_2\, p(\eta_2)\right]\, .
\label{p3.4}
\end{equation}
To proceed further, we make the change of variables
\begin{equation}
z_i= \int_0^{\eta_i} p(\eta)\, d\eta\, .
\label{zeta.1}
\end{equation}
With this change of variables the integral in Eq. (\ref{p3.4}) reduces magically to
\begin{equation}
p_3 =  2 \int_0^{1/2} dz_1 \int_0^{z_1} dz_3\, \left[\frac{1}{2}+ z_3\right] =  \frac{1}{6}\, .
\label{p3.5}
\end{equation}
Essentially, the change of variables in Eq. (\ref{zeta.1}) transform the $\eta_k$'s to $z_k$'s and each $z_k$
is uniformly distributed over $z\in [-1,1]$ and thus the dependence on $p(\eta)$ completely drops out. This is the
key mechanism behind the super-universality. 

One would like to continue for $n>3$, but it becomes rather cumbersome quickly and we haven't found a simple
recursive pattern to compute these multiple integrals for $n>3$. There ought to exist an elegant combinatorial
proof of this beautiful universal result for all $n$, which unfortunately eludes us for the moment.
We thus leave this as a challenging open problem.

\end{document}